\documentclass[traditabstract,a4paper]{aa}
\usepackage{amsmath}
\usepackage{graphicx}
\usepackage{color}
\usepackage{natbib}

\bibpunct{(}{)}{;}{a}{}{,}

\begin{document}

\author{Edo van Uitert \inst{\ref{inst1}} \and Henk Hoekstra \inst{\ref{inst1}} \and Malin Velander \inst{\ref{inst1}} \and David G. Gilbank \inst{\ref{inst2},\ref{inst3}} \and Michael D. Gladders \inst{\ref{inst4}} \and H.K.C. Yee \inst{\ref{inst3}} }
\institute{Leiden Observatory, Leiden University, Niels Bohrweg 2, NL-2333 CA Leiden, The Netherlands, email: vuitert@strw.leidenuniv.nl \label{inst1} \and Department of Physics and Astronomy, University of Waterloo, Waterloo, Ontario, N2L 3G1, Canada \label{inst2} \and Department of Astronomy and Astrophysics, University of Toronto, 50 St. George Street, Toronto, Ontario, M5S 3H4, Canada \label{inst3} \and Department of Astronomy and Astrophysics, University of Chicago, 5640 S. Ellis Ave., Chicago, IL 60637, USA \label{inst4} }

\title{Galaxy-galaxy lensing constraints on the relation between baryons and dark matter in galaxies in the Red Sequence Cluster Survey 2}

\definecolor{grijs}{gray}{0.5}

\titlerunning{The relation between baryons and dark matter}

\abstract {We present the results of a study of weak gravitational lensing by galaxies using imaging data that were obtained as part of the second Red Sequence Cluster Survey (RCS2). In order to compare to the baryonic properties of the lenses we focus here on the $\sim$300 square degrees that overlap with the data release 7 (DR7) of the Sloan Digital Sky Survey (SDSS). The depth and image quality of the RCS2 enables us to significantly improve upon earlier work for luminous galaxies at $z\geq0.3$. To model the lensing signal we employ a halo model which accounts for the clustering of the lenses and distinguishes between satellite and central galaxies. Comparison with dynamical masses from the SDSS shows a good correlation with the lensing mass for early-type galaxies. The correlation is less clear for late-type galaxies, possibly due to rotation. For low luminosity (stellar mass) early-type galaxies we find a satellite fraction of $\sim$40\% which rapidly decreases to $<10$\% with increasing luminosity (stellar mass). The satellite fraction of the late-types has a value in the range 0-15\%, independent of luminosity or stellar mass. At high masses the satellite fraction is not well constrained, which we partly attribute to the modelling assumptions. To infer virial masses we apply simple models based on an independent satellite kinematics analysis to account for intrinsic scatter in the scaling relations. We find that early-types in the range $10^{10}<L_r<10^{11.5} L_{\odot}$ have virial masses that are about five times higher than those of late-type galaxies and that the mass scales as $M_{200} \propto L^{2.34^{+0.09}_{-0.16}}$. For an early-type galaxy with a fiducal luminosity of $10^{11}L_{r,\odot}$, we obtain a mass $M_{200}=(1.93^{+0.13}_{-0.14})\times10^{13}h^{-1}M_{\odot}$. We also measure the virial mass-to-light ratio, and find for $L_{200}<10^{11}L_{\odot}$ a value of $M_{200}/L_{200}=42\pm10$ for early-types, which increases for higher luminosities to values that are consistent with those observed for groups and clusters of galaxies. For late-type galaxies we find a lower value of $M_{200}/L_{200}= 17\pm9$. Our measurements also show that early- and late-type galaxies have comparable halo masses for stellar masses $M_*<10^{11}M_{\odot}$, whereas the virial masses of early-type galaxies are higher for higher stellar masses. To compare the efficiency with which baryons have been converted into stars, we determine the total stellar mass within $r_{200}$. Our results for early-type galaxies suggest a variation in efficiency with a minimum of $\sim$10\% for a stellar mass $M_{*,200}=10^{12}M_{\odot}$. The results for the late-type galaxies are not well constrained, but do suggest a larger value.}


\maketitle

\section{INTRODUCTION} 
\hspace{4mm} There is now overwhelming evidence that galaxies are surrounded by dark matter haloes. Studying the global properties of the haloes, such as their virial masses or density profiles, however, has proven difficult due to a lack of reliable tracers of the gravitational potential at large distances. Improving observational constraints is important because the details of galaxy formation are not completely clear, even though significant progress has been made in recent years \citep[e.g.][]{Bower10,Kim09}. The relation between the baryons and the dark matter in galaxies has been studied using numerical simulations \citep[e.g.][]{Wang06,Croton06,Somerville08,Moster10} and it is important to confront the predictions with observations. This requires reliable estimates of both the dark matter and the baryonic content of galaxies. \\
\indent Several observables can be used to trace the baryons, such as the luminosity of a galaxy, which is readily available. It is also possible to derive stellar masses by fitting stellar synthesis models to either the spectral features of a galaxy \citep{Kauffmann03,Gallazzi05} or to its colours \citep{BelldJ01,Salim07}. The stellar mass estimates are tightly correlated to various other important global properties of galaxies \citep[colour, metallicity, luminosity, environment, see e.g.][and references therein]{Grutzbauch11} and they are therefore considered a useful tracer of the baryonic content of a galaxy. \\
\indent Numerical simulations suggest that the dark matter haloes of massive galaxies extend out to hundreds of kiloparsecs \citep[e.g.][]{Springel05}, which is supported by observations \citep[e.g.][]{Hoekstra04}. For nearby galaxies it is possible to study the dark matter distribution using the dynamics of planetary nebulae \citep[e.g.][]{Napolitano09}. In addition, studies of satellite galaxies around central galaxies \citep[e.g.][]{More11,Conroy07} have provided constraints on the relation between baryons and dark matter. Unfortunately these studies require spectroscopy of large numbers of objects, which makes them rather expensive. Furthermore, the observations are limited to small scales due to the requirement of having optical tracers, which complicates the determination of the virial mass of the haloes galaxies reside in, unless one is willing to extrapolate the measurements. \\
\indent Fortunately it is possible to probe the matter distribution on large scales, thanks to an effect called weak gravitational lensing; we can measure the distortion of the shapes of faint background galaxies (sources) caused by the bending of light rays by intervening mass concentrations (lenses). The distortion is independent of the type of matter in the lenses, and so the projected mass of the lens is measured without any assumption on the physical state of the matter at scales from a few kiloparsec to a few megaparsec.  \\
\indent The weak lensing signal around a single galaxy is too weak to detect since it is 10-100 times smaller than the intrinsic ellipticities of galaxies. Therefore the galaxy-galaxy signal has to be averaged over many lenses to decrease the shape noise. Although individual galaxies cannot be studied in this way, their average properties can be determined \citep[e.g.][]{Brainerd96,Fischer00,Hoekstra04}. Only more recently has it become possible to study lenses as a function of properties such as type, luminosity, stellar mass, etc., because early studies lacked the ancillary data needed to subdivide the lenses into subsamples. For instance \citet{Hoekstra05} used nearly 34 square degrees of the Red Sequence Cluster Survey (RCS) \citep{Gladders05} for which photometric redshifts were available \citep{Hsieh05}, to study the relation between the virial mass and baryonic contents of isolated galaxies in the redshift range $0.2<z<0.4$, and derived star formation efficiencies for early- and late-type galaxies. Thanks to the wealth of ancillary data, the Sloan Digital Sky Survey \citep[SDSS;][]{York00} has had a major impact on galaxy-galaxy lensing studies \citep[e.g.][]{GuzikS02,Mandelbaum06}. This is evidenced by \citet{Mandelbaum06} who used nearly 5000 square degrees of the SDSS DR4 \citep{Adelman06} to study galaxies in the redshift range $0.02<z<0.35$ as a function of galaxy type and environment, and constrained the stellar mass to virial mass relation, the luminosity to virial mass relation and the satellite fractions of the lens samples. \\
\indent Currently no survey can surpass the precision that can be achieved by the SDSS at low redshift ($z<0.3$) because of the large survey area and the availability of spectroscopic data. We note, however, that complementing the SDSS data with deeper imaging by the Panoramic Survey Telescope \& Rapid Response System\footnotemark 
\footnotetext[1]{http://pan-starrs.ifa.hawaii.edu/public/}
\citep[Pan-STARRS;][]{Kaiser02} will provide a major improvement, as is demonstrated by the results we present here. For lenses with $z>0.3$ it is possible to achieve a significant improvement over the SDSS results by surveying a smaller area with deeper data and good image quality; it allows us to use sources at higher redshifts. This is important because the amplitude of the lensing signal scales proportionally to the ratio of the angular diameter distance between the lens and the source and the distance between the observer and the source. The signal decreases rapidly when the lens redshift approaches the peak of the source redshift distribution, which occurs around $z\sim 0.35$ for the SDSS. \\
\indent In this paper we use data from the second generation Red Sequence Cluster Survey (RCS2) to measure the weak lensing signal around galaxies that are observed in the SDSS. The RCS2 is a nearly 900 square degree imaging survey carried out by the Canada-France-Hawaii-Telescope (CFHT), and is $\sim$2 magnitudes deeper than the SDSS in $r'$. The increase in depth combined with a median seeing of 0.7", which is a factor of two smaller than the seeing in the SDSS, results in a source galaxy number density that is about five times higher, and a source redshift distribution that peaks at z$\sim$0.7.  \\
\indent We use the overlapping area between the two surveys, which amounts to approximately 300 square degrees, in order to assign the spectroscopic redshifts, luminosities, stellar masses and dynamical masses from the SDSS to the lenses. The lensing analysis itself is performed on the RCS2 data. Even though the overlap between the surveys is modest, the loss in survey area is outweighed by the gain in the number density of source galaxies and the improvement of the lensing efficiency. This enables us to improve the measurements of the lensing signal around the most massive galaxies, which mostly reside at redshifts where the SDSS is not very sensitive. \\
\indent In this paper we describe the lenses in Section \ref{sec_lenssamp}. The weak lensing analysis is discussed in Section \ref{sec_lensing}. The halo model that we have implemented is introduced in Section \ref{sec_halo}. In Section \ref{sec_mdyn} we compare the weak lensing mass to the dynamical mass. We describe the luminosity results in Section \ref{sec_lumi}, and the stellar mass results in Section \ref{sec_mstel}. We summarize our conclusions in Section \ref{sec_conc}. Throughout the paper we assume a WMAP5 cosmology \citep{Komatsu09} with $\sigma_8=0.8$, $\Omega_{\Lambda}=0.73$, $\Omega_M=0.27$, $\Omega_b=0.045$ and the dimensionless Hubble parameter $h=0.7$. All distances quoted are in physical (rather than comoving) units unless explicitly stated otherwise.


\section{LENS SAMPLE \label{sec_lenssamp}}
\hspace{4mm} The SDSS has imaged roughly a quarter of the entire sky, and has observed the spectra for about one million galaxies \citep{Eisenstein01,Strauss02}. The combination of spectroscopic coverage and photometry in five optical bands ($u,g,r,i,z$) in the SDSS provides a wealth of galaxy information that is not available for the RCS2. To use this information, but also benefit from the improved lensing quality of the RCS2, we use the 300 square degrees overlap between the surveys for our analysis. We match the RCS2 catalogues to the DR7 \citep{Abazajian09} spectroscopic catalogue, to the MPA-JHU DR7\footnotemark  
\footnotetext[2]{http://www.mpa-garching.mpg.de/SDSS/DR7/}
stellar mass catalogue and to the NYU Value Added Galaxy Catalogue (NYU-VAGC)\footnotemark
\footnotetext[3]{http://sdss.physics.nyu.edu/vagc/}
\citep{Blanton05,Adelman08,Padmanabhan08} which yields the spectroscopic redshift, luminosity, stellar mass, and the dynamical mass of $1.7\times10^4$ galaxies. These form the lens sample of this work; we study the distortion these galaxies imprint as a function of their baryonic content on the shapes of the background galaxies. \\
\indent As the relation between dark matter and baryons depends on galaxy type, we split the lens sample into early- and late-type galaxies using the $frac\_deV$ parameter included in the SDSS photometric catalogues. This parameter is determined by simultaneously fitting $frac\_deV$ times the best-fitting De Vaucouleur profile plus (1-$frac\_deV$) times the best-fitting exponential profile to an object's brightness profile. This has been done in the $g$, $r$ and $i$ band, and we use the average value. We classify galaxies with $frac\_deV>0.5$ as early-types, and galaxies with $frac\_deV<0.5$ as late-types. The classification of early-types is at least 96\% complete and 76\% reliable (96\% of all early-type galaxies are in the early-type sample, while 76\% of all the galaxies in the early-type sample are actually early-types), and the classification of late-types is at least 55\% complete and 90\% reliable \citep{Strateva01,Mandelbaum06}. \\
\indent We visually inspect the brightest and most massive early- and late-type galaxies of our lens sample using our RCS2 imaging data. We find that about 30 of the 100 most massive late-types (with a stellar mass in the range $10^{11.4}-10^{12.5}M_{\odot}$) actually consists of multiple objects with small separations. These galaxies reside at a redshift of $\sim0.4$, and are not well resolved in the SDSS. They are not removed from the analysis as that may introduce a selection bias. More importantly, including them facilitates a comparison to the literature. As a test, we excluded these lenses, and found that the results did not significantly change (note, however, that due to the low number of massive late-type lenses, the errors are large).


\subsection{Luminosities \& Stellar Masses}
\begin{figure}
  \resizebox{\hsize}{!}{\includegraphics[angle=-90]{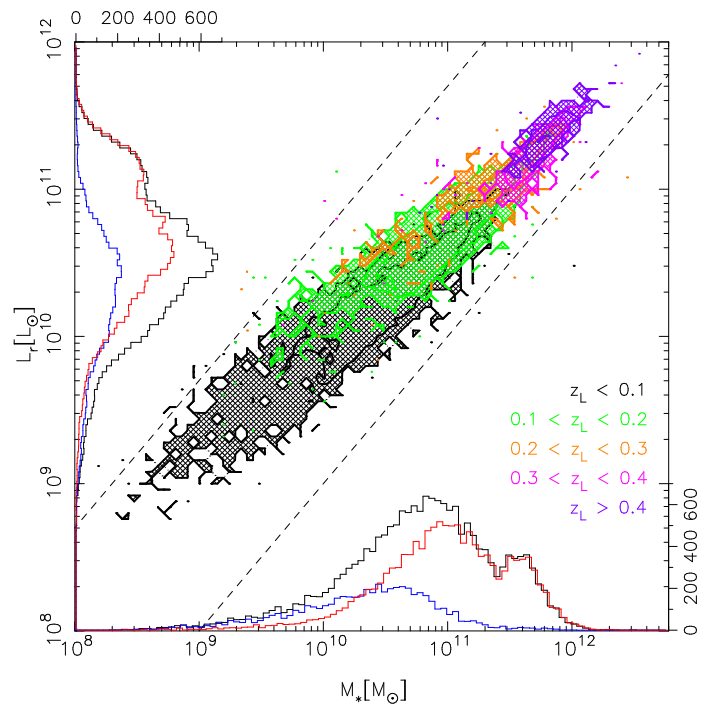}}
  \caption{Stellar mass versus luminosity of the lens sample. The colour coding represents the redshift of the galaxies as denoted in the lower right-hand corner. The histograms for all galaxies (black line), the early-types (red line) and the late-types (blue line), as a function of stellar mass and luminosity are also shown, and are drawn slightly offset for clarity. The dashed diagonal lines indicate the additional mass-to-light ratio cuts we have applied (objects with $M_*/L_r$ between 0.2 and 10 have been selected) to remove outliers that may contaminate the lensing signal.}
  \label{plot_mstellumi}
\end{figure}
\hspace{4mm} The MPA-JHU stellar mass catalogue contains about $7 \times 10^5$ unique galaxies, and provides the $r$-band absolute magnitudes and the stellar mass estimates of our lenses. The absolute magnitudes that are used to compute the luminosities and stellar masses are based on the Petrosian apparent magnitudes from the SDSS. The Petrosian apparent magnitude measures the flux within a circular aperture whose radius depends on the azimuthally averaged brightness profile in the $r$-band. It does not include the flux at very large radii from a galaxy, and therefore underestimates the total flux by typically a few tenths of a magnitude \citep{Blanton01}. Although we do not correct for the missing flux as it would complicate a comparison with previous observational work, this should be kept in mind when comparing our results to predictions from numerical simulations. \\
\indent The absolute magnitudes have been corrected for extinction using the dust maps from \citet{Schlegel98}, the k-corrections have been calculated to $z=0.0$ using the {\tt KCORRECT} v4\_2 code \citep{Blanton03,BlantonR07}, and the distance modulus is determined with $h=0.7$. We convert the absolute magnitudes into solar luminosities using the absolute AB magnitude in the SDDS $r$-band of $M_{solar}=4.65$ for $z=0.0$. We account for passive evolution by dividing the luminosities of the early-type galaxies by $(1+z)$. The luminosity evolution of late-type galaxies can in principle be computed if the star formation histories (SFHs) are accurately known. The SFHs are generally uncertain, however, since they depend on many parameters such as the stellar mass, environment, assembly history, and AGN activity of a galaxy. Hence the luminosity evolution is difficult to determine and the correction highly uncertain. We therefore do not correct the luminosites of late-type galaxies for evolution. \\
\indent The stellar masses have been estimated by fitting a library of \citet{BruzualC03} stellar population models to the $u,g,r,i,z$ photometry of the galaxies in the SDSS. The initial mass function (IMF) was taken to be a \citet{Kroupa01} IMF and the modelling methodology follows \citet{Salim07}. \\
\indent Nearly all galaxies with a spectroscopic redshift from DR7 are present in the stellar mass catalogue. Figure \ref{plot_mstellumi} shows the stellar mass versus luminosity for the matched galaxies. The different colours represent galaxies at different redshifts. The most massive and luminous galaxies in our sample reside in the highest redshift range, and are almost exclusively early-type galaxies. Also shown are the histograms of the stellar masses and of the luminosities on respectively the horizontal and vertical axis. The dashed lines indicate the additional $0.2<M_*/L_r<10$ cut we apply to minimize the outlier contamination of the lensing bins.


\subsection{Dynamical Masses \label{sec_lensmdyn}}
\hspace{4mm} The motions of stars in a galaxy provide an alternative way to estimate the mass of a galaxy at small radii, and constrain the scaling relations between baryons and dark matter. Spectroscopic observations are required to measure the velocity dispersion, which is converted into a dynamical mass estimate via the scalar virial theorem, taking into account projection effects and assumptions on the structure of the stellar orbits: 
\begin{equation}
  GM_{\mathrm{dyn}} = K_V(n) \sigma_{\mathrm{los}}^2 R_e,
  \label{eq_mdyn}
\end{equation}
with $\sigma_{\mathrm{los}}$ the line-of-sight velocity dispersion of the galaxy, $R_e$ the effective radius (containing 50\% of the light of the best fit S\'{e}rsic model), and $K_V(n)$ a term that includes the effects of structure on stellar dynamics, which can be approximated by \citep{Bertin02}:
\begin{equation}
  K_V(n) \cong \frac{73.32}{10.465+(n-0.94)^2}+0.954,
\end{equation}
with $n$ the S\'{e}rsic index \citep{Sersic68}. \\
\indent Using the dynamical mass as a tracer for the total mass of a galaxy has various complications. Firstly, it is implicitly assumed that the velocity dispersion in Equation \ref{eq_mdyn} is only generated by the radial motions of the stars, and the $K_V(n)$ term is derived under the assumption that the mass distribution is spherical, dynamically isotropic, and non-rotating. In reality, however, the rotation of a galaxy contributes to the measured velocity dispersion as well, and this effect is particularly important in late-type galaxies. The majority of the early-type galaxies in our study are massive and luminous. They are expected to rotate slowly \citep[e.g.][]{Emsellem07}, so their dynamical mass estimates are less affected. The dynamical masses of late-type galaxies, however, are potentially overestimated. A second complication arises from the fact that the spectroscopic fibre within which the velocity dispersion is measured has a fixed size. Therefore, the physical region over which the velocity dispersion is averaged depends on the redshift of a galaxy, and hence it probes different regions for galaxies at different redshifts. If the velocity dispersion changes with radius, we would effectively assign different dynamical masses to the same galaxy depending on its redshift. Thirdly, the dynamical mass is measured within the effective radius. The effective radius is a rather arbitrary point, as it depends on parameters such as the shape, the brightness profile and the orientation of a galaxy, and the distribution of dust within the galaxy.  Even if a galaxy is spherical and isotropic, it is not clear whether the effective radius marks a special point in relation to the total mass content of a galaxy, given that the dark matter does not follow the distribution of stars. This is most obvious in the outer regions of a galaxy, where most of the matter is dark. \\
\indent To calculate the dynamical mass of our lenses, we retrieve the velocity dispersions from the SDSS spectroscopic catalogue. As it is complex to estimate the velocity dispersion of galaxies whose spectra are dominated by multiple components, e.g. galaxies with different stellar populations or different kinematic components, the SDSS only provides estimates for spheroidal systems whose spectra are dominated by red stars. At low redshift, the selection also includes the bulges of late-type galaxies because their spectra are similar to the spectra of early-type galaxies. The S\'{e}rsic index and the effective radius are obtained from the NYU-VAGC. The sizes and fluxes are underestimated 10\% and 15\% respectively for large galaxies and galaxies with high S\'{e}rsic indices \citep{Blanton05}, whereas the S\'{e}rsic index itself is underestimated by $\sim0.5$ to $\sim1.3$ for galaxies with high S\'{e}rsic indices. It is shown in \citet{Guo09} that these biases arise from background overestimation and subtraction. 
As a result, the dynamical mass estimates for these galaxies may be slightly biased, but we do not account for it since we do not know the correction for each galaxy. To ensure that the dynamical mass is computed in approximately the rest-frame $r$-band, we split the sample according to redshift. For galaxies at $z<0.2$ we use the S\'{e}rsic index and effective radius in the $r$-band, for galaxies between $0.2<z<0.4$ we use the values in the $i$-band, and for galaxies at $z>0.4$ we average the values of the $i$- and $z$-band. \\


\section{LENSING ANALYSIS \label{sec_lensing}}

\subsection{The RCS2}

\hspace{4mm} The lensing signal can be detected with high significance at low redshifts ($z<0.3$) using SDSS data only. At higher redshifts, the significance decreases rapidly, because of the limited imaging depth and image quality of the SDSS. To improve the lensing signal-to-noise ratio at $z\ge0.3$, we use the deep imaging data from the Red Sequence Cluster Survey 2 (RCS2) \citep{Gilbank10} instead. The RCS2 is a nearly 900 square degree imaging survey in three bands ({\it g', r'} and {\it z'}) carried out with the Canada-France-Hawaii Telescope (CFHT) using the 1 square degree camera MegaCam. The primary survey area is divided into 13 well-separated patches on the sky (including the uncompleted patch 1303), each with an area ranging from 20 to 100 square degrees \footnotemark.
\footnotetext[4] {The CFHT Legacy Survey Wide, comprising of 171 square degrees of imaging data in $u^{\star}$, {\it g', r', i'} and {\it z'}, is also included in the RCS2, but is not used in this study.}
Since the RCS2 consists of single exposures only, it is difficult to identify cosmic rays, especially those that hit stars and galaxies. However, only a small fraction of objects is hit by a cosmic ray, and the affected objects do not bias the measurements, but act as a negligible source of noise \citep{Hoekstra04}. We perform the weak lensing analysis in the SDSS and RCS2 overlap using the 8 minute exposures of the {\it r'}-band ({\it $r'_{lim}\sim$}24.8), which is best suited for lensing as it has a median seeing of 0.7".


\subsection{Image processing}
\begin{figure}
  \resizebox{\hsize}{!}{\includegraphics[angle=-90]{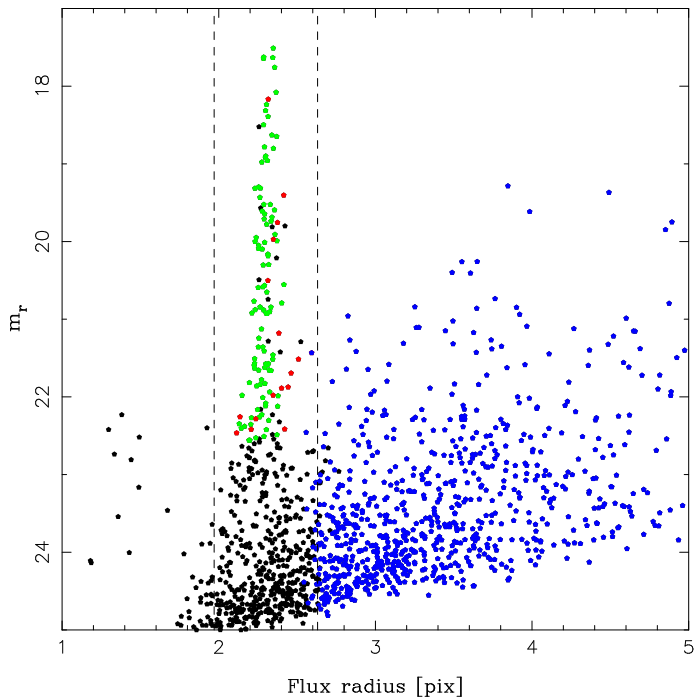}}
  \caption{The size-magnitude diagram of one of the chips in a randomly picked exposure. The black dots are the {\tt SExtractor} detections, the green dots are the selected stars, the red dots are the 3-sigma outliers, and the blue dots are the selected galaxies. The dashed lines indicate the location of the stellar branch. Thanks to the good image quality the stars are easily separated from the galaxies.}
  \label{plot_sizemag}
\end{figure}

\hspace{4mm} We retrieve the Elixir\footnotemark \ processed images from the Canadian Astronomy Data Centre (CADC) archive\footnotemark. 
\footnotetext[5]{http://www.cfht.hawaii.edu/Instruments/Elixir/}
\footnotetext[6]{http://www1.cadc-ccda.hia-iha.nrc-cnrc.gc.ca/cadc/}
We use the THELI pipeline \citep{Erben05,Erben09} to subtract the image backgrounds, to create weight maps that we use in the object detection phase, and to identify satellite and asteroid trails. To obtain accurate astrometry, we run {\tt SCAMP} \citep{Bertin06} on the images, which enables us to match our catalogues to the SDSS. The polynomial coefficients from {\tt SCAMP} describing the mapping from image to sky coordinates are used to calculate the camera distortion. We use the automated masking routines from the THELI pipeline to generate image masks and to combine them with the RCS2 masks in order to omit image regions that contaminate the lensing signal (e.g. saturated stars, satellite trails). All masks are inspected by eye, and manually improved where necessary. \\
\indent We use {\tt SExtractor} \citep{BertinA96} to detect the objects in the images. To select the stars for modelling the PSF variation across the images, we first identify the locus of the stellar branch in a size-magnitude diagram. We select the non-saturated objects close to the stellar branch with a signal to noise ratio larger than 30 and with no {\tt SExtractor} flags raised. To remove small galaxies that have been misidentified as stars, and stars that have been affected by cosmic rays, we fit a second-order polynomial to both the size and the ellipticity of these star-candidates, and discard all 3-sigma outliers. We clean the stellar selection even further in the shape measurement pipeline by removing shape parameter outliers. All objects larger than 1.2 times the local size of the PSF are classified as galaxies.\\
\indent In Figure \ref{plot_sizemag} we illustrate the star-galaxy separation. It has been fully automated, but as a precaution we inspect all size-magnitude diagrams by eye. The separation fails for a few chips that have either very few stars or a PSF with a large FWHM, and we manually adjust those. As neighbouring patches overlap by $\sim1$ arcminute, we remove all galaxies within 35 arcseconds from the image edges in order to avoid duplicating the lenses and sources in our analysis. \\
\indent Elixir provides approximate zeropoints for each pointing, which we use to measure the $r'$-band magnitudes of the objects in the images. We correct the magnitudes for galactic extinction using the dust maps from \citet{Schlegel98}. These magnitudes are not as accurately calibrated as those from \citet{Gilbank10}, and differ in the $r'$-band on average by $-0.01\pm0.32$. Our calibration is, however, sufficiently accurate to select the source galaxy sample. For the calculation of the luminosity overdensity, which is discussed in Section \ref{sec_lum_ml}, we use the catalogues from \citet{Gilbank10} instead. 

\begin{figure}
  \resizebox{\hsize}{!}{\includegraphics{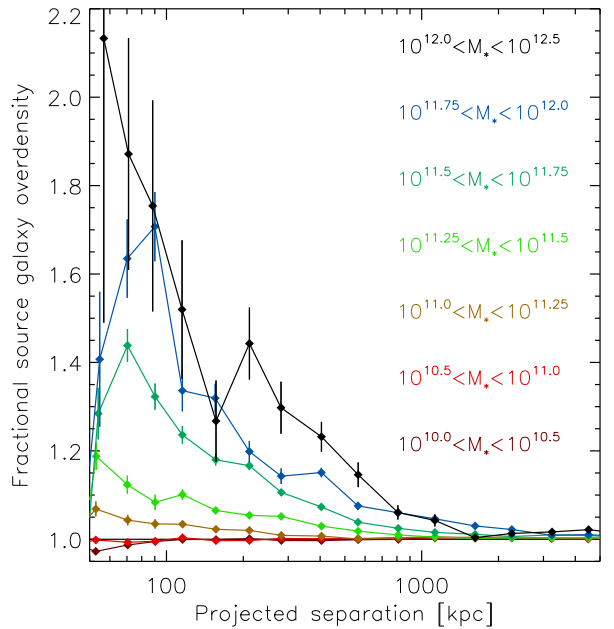}}
  \caption{The source galaxy overdensity as a function of distance from the lenses for the different stellar mass bins. The overdensity increases with stellar mass. Massive galaxies reside on average at higher redshifts and live in denser environments with more satellite galaxies.}
  \label{plot_galod}
\end{figure}
\subsection{Contamination correction \label{sec_contam}}
\hspace{4mm} A fraction of the galaxies in the source catalogue is physically associated with the lenses. Since we lack redshifts for the sources, we are unable to remove them. These objects are not lensed, and therefore dilute the lensing signal. To estimate this contamination we measure $f_{cg}(r)$, the excess source number density around the lenses. We show the overdensity around the lenses which have been divided into seven stellar mass bins (defined in Table \ref{tab_lens_mstel}) as a function of lens-source separation in Figure \ref{plot_galod}. The error bars are computed assuming that the number of source galaxies in each radial bin follows a Poisson distribution. The contamination increases with stellar mass, as massive galaxies reside in denser environments and therefore have more satellite galaxies. Although the overdensity is shown independently of the lens galaxy type in Figure \ref{plot_galod}, we measure it for the early- and late-types separately in the science analysis presented in Section \ref{sec_mdyn}, \ref{sec_lumi} and \ref{sec_mstel}. Assuming that the satellite galaxies have random orientations, we correct for the contamination by boosting the lensing signal with a factor $1+f_{cg}(r)$. Note, however, that the contamination correction may be too small if satellite galaxies are preferentially radially aligned in the direction of the lens. This type of intrinsic alignment has been studied with seemingly different results; some authors  \citep[e.g.][]{AgustssonB06,Faltenbacher07} who determined the galaxy orientation using the isophotal position angles, have observed a stronger alignment than others \citep[e.g.][]{Hirata04,Mandelbaum05sys} who used galaxy moments. \citet{Siverd09} and \citet{Hao11} attribute the discrepancy to the different definitions of the position angle of a galaxy. As we measure the shapes of source galaxies using galaxy moments, we expect that intrinsic alignment only has a minor impact on the correction factor and hence can be safely ignored. \\
\indent Gravitational lenses do not only shear the images of the source galaxies, but also magnify the background sky. As a result, the flux of the sources is magnified, and the source galaxy number density is diluted. These combined effects are known as magnification bias, and it changes the source density around the lenses. The effect is negligible for the lensing study presented here.


\subsection{Shape measurement}
\hspace{4mm} The measurement of the shapes of galaxies is central to any weak lensing analysis. The accuracy that is required depends on the science goal. For example, in cosmic shear studies aimed at constraining cosmological parameters, it is necessary to accurately correct the measured galaxy shapes for the anisotropic smearing of the PSF since the signal is small and very sensitive to any PSF residual systematic. In contrast, in the case of galaxy-galaxy lensing the signal is averaged over many lens-source pairs with random orientations, which removes most of the PSF systematics on small scales. \\
\indent For our lensing analysis we measure the shapes of galaxies with the KSB method \citep{Kaiser95,LuppinoK97,Hoekstra98}, using the implementation described by \citet{Hoekstra98,Hoekstra00}. The measured galaxy shapes are corrected for smearing by the PSF under the assumption that the brightness distribution of stars can be described by an isotropic profile convolved with a small anisotropic kernel. Generally, the PSF is more complicated which may lead to biases. The version of KSB we use has been tested on simulated images as part of the Shear Testing Programme (STEP) 1 and 2 (the 'HH' method in \citet{Heymans06} and \citet{Massey07} respectively). These tests have shown that the correction scheme works well for a variety of PSFs; in STEP2, the HH method underestimates the shear on average by 1-2\% only. \\
\begin{figure}
  \resizebox{\hsize}{!}{\includegraphics[angle=270]{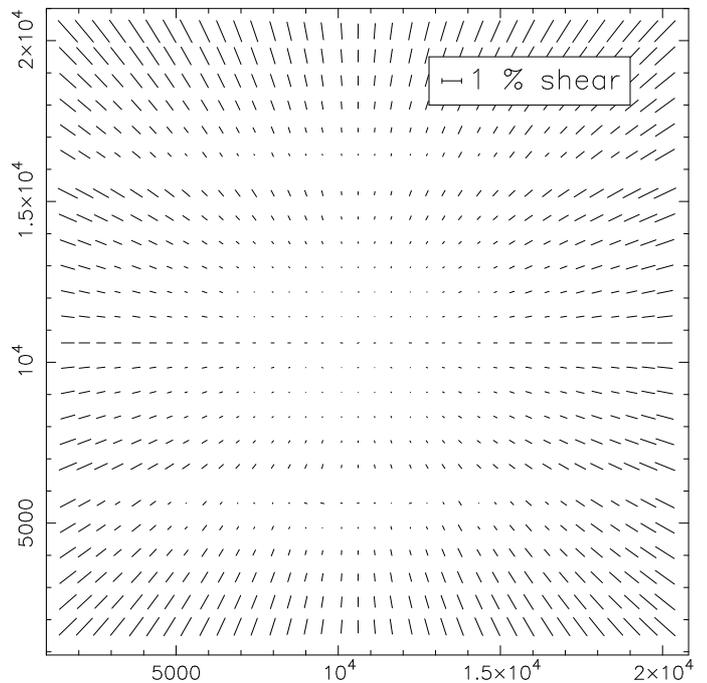}}
  \caption{Shear induced by camera distortion in the MegaCam imager. The camera shear is largest in the corners of the mosaic, with values up to 1.5\%. As the observed shear is the sum of the gravitational shear and the camera shear, we simply subtract the camera shear from the observed galaxy ellipticities to correct for it. }
  \label{plot_camdist}
\end{figure}
\indent  The mapping between the sky coordinates and the CCD pixels is slightly non-linear due to the camera optics, which causes an additional shear that needs to be corrected. We calculate the shear induced by this distortion using the polynomial coefficients from {\tt SCAMP} describing the mapping from image to sky coordinates. The camera shear of MegaCam is shown in Figure \ref{plot_camdist}. The images of both the stars and the galaxies are sheared, with a value reaching 1.5\% at the corners of the images. At large lens-source separations, where the gravitational lensing signal is small, the camera shear dominates the observed lensing signal. \citet{Hoekstra98,Hoekstra00} demonstrate that the observed shear is the sum of the gravitational shear and the camera shear. We therefore simply subtract the camera shear from the observed ellipticities of the galaxies to correct for it. \\ \\
\indent To demonstrate the excellence of the RCS2 as a lensing survey, we measure the galaxy-mass cross-correlation function in the exposures that significantly overlap with the SDSS (defined as having more than 30 matching objects). 301 exposures of the total overlapping 350 meet this requirement, which after masking and exclusion of the image boundaries leads to an effective area of approximately 260 square degrees. The galaxy-mass cross-correlation function measures the correlation between the galaxies and the surrounding distribution of (predominantly dark) matter. We compute it by measuring the azimuthally averaged tangential shear as a function of radial distance from the lens:
\begin{equation}
  \langle\gamma_t\rangle(r) = \frac{\Delta\Sigma(r)}{\Sigma_{\mathrm{crit}}},
\end{equation}
where $\Delta\Sigma(r)=\bar{\Sigma}(<r)-\bar{\Sigma}(r)$ is the difference between the mean projected surface density enclosed by $r$ and the mean projected surface density in an annulus at $r$, and $\Sigma_{\mathrm{crit}}$ is the critical surface density
\begin{equation}
  \Sigma_{\mathrm{crit}}=\frac{c^2}{4\pi G}\frac{D_s}{D_lD_{ls}},
\end{equation}
with $D_l$, $D_s$ and $D_{ls}$ the angular diameter distance to the lens, the source, and between the lens and the source respectively. \\
\indent Since we do not have redshifts for all galaxies we separate the lenses from the sources using magnitude cuts \citep[see e.g.][]{Hoekstra04}. Objects with $19.5 < m_{r'} < 21.5$ are defined as lenses, and objects with $22.0 < m_{r'} < 24.0$ are sources. We discard objects with ellipticities larger than 1, and objects that have a {\tt SExtractor} flag raised. Using these selection criteria we find 7.3$\times 10^5$ lenses and 5.9$\times 10^6$ sources. The corresponding effective source number density is 6.3 arcmin$^{-2}$, which is five times higher than the source density of 1.2 arcmin$^{-2}$ used in the SDSS analysis \citep{Mandelbaum05sys}. To obtain the approximate redshift distribution of the lenses and sources, we apply identical magnitude cuts to the photometric redshift catalogues of the Canada-France-Hawaii-Telescope Legacy Survey (CFHTLS) "Deep Survey" fields \citep{Ilbert06}. We stack the signals of all the lenses in the RCS2, and azimuthally average them in radial bins. To remove the contributions of systematic shear (from, e.g., the image masks), we subtract the signal computed around random lenses from the signal around the real lenses. We measure the source galaxy overdensity as a function of lens-source separation, and boost the signal to correct for the contamination as outlined in Section \ref{sec_contam}. Figure \ref{plot_gtall} shows the tangential shear, and the inset shows the signal at small scales using a linear vertical scale. \\
\begin{figure}
  \resizebox{\hsize}{!}{\includegraphics{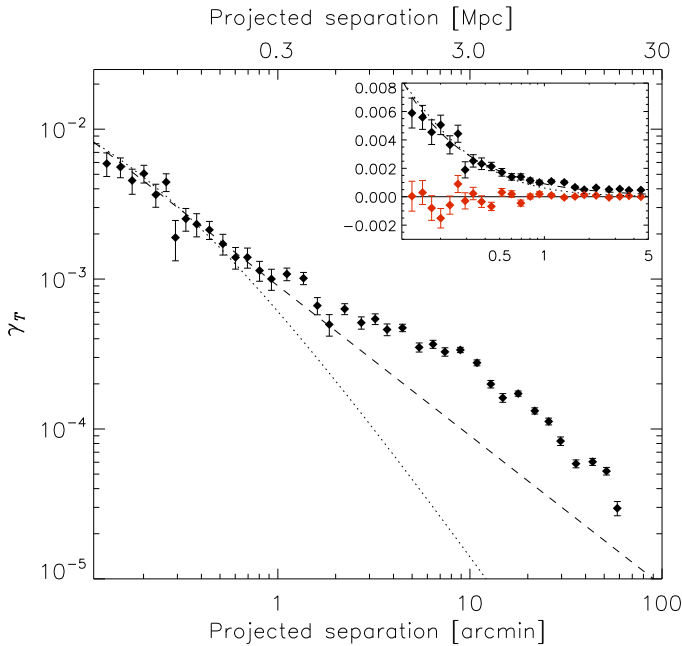}}
  \caption{The galaxy-mass cross-correlation function around 7.3$\times 10^5$ apparent magnitude selected lenses measured with 5.9$\times 10^6$ sources. The black symbols are the tangential shear, the red symbols are the cross shear. The top axis shows the projected separation in physical units for the median lens redshift $z_{med}$=0.34. The inset shows the signal on a linear scale for small separations. The signal has been corrected for contributions from systematic shear, and boosted to account for source galaxy contamination. The dashed (dotted) line shows the best fit SIS (NFW), fitted to the shear on scales between 0.2 and 0.6 arcminutes. The clustering of galaxies causes excess shear at scales $>$1 arcminutes.}
  \label{plot_gtall}
\end{figure}
\indent We also measure the cross shear around the lenses by rotating the background galaxies $45^\circ$ and repeating the measurement. Gravitational lensing does not produce cross shear, and a non-zero signal indicates the presence of residual systematics in the catalogues. We indicate the cross shear with the red symbols in the inset in Figure \ref{plot_gtall}, and note that it is consistent with zero on all scales. \\
\indent For reference, we fit a singular isothermal sphere (SIS) and a Navarro-Frenk-White (NFW) profile \citep{Navarro96} to the tangential shear on scales between 0.2 and 0.6 arcminutes ($\sim$60-180 kpc at the median lens redshift $z_{med}=0.34$). The SIS signal is given by
\begin{equation}
  \gamma_{t,\mathrm{SIS}}(r) = \frac{r_E}{2r}=\frac{4\pi \sigma^2}{c^2}\frac{D_{ls}}{D_s}\frac{1}{2r},
\end{equation}
where $r_E$ is the Einstein radius and $\sigma$ the velocity dispersion. We indicate the best fit SIS model with the dashed line in Figure \ref{plot_gtall}. The NFW density profile is given by
\begin{equation}
  \rho(r) = \frac{\delta_c \rho_c}{(r/r_s)(1+r/r_s)^2},
\end{equation}
with $\delta_c$ the characteristic overdensity of the halo, $\rho_c$ the critical density for closure of the universe, and $r_s=r_{200}/c_{\mathrm{NFW}}$ the scale radius, with $c_{\mathrm{NFW}}$ the concentration parameter. The NFW profile is specified by two free parameters: the mass and the concentration parameter. Since numerical simulations have shown that the concentration depends on the mass and redshift of the halo, we can reduce the number of free parameters in the fit by adopting a mass-concentration relation. We use the mass-concentration relation from \citet{Duffy08}, which is based on numerical simulations using the best fit parameters of the WMAP5 cosmology. It is given by
\begin{equation}
  c_{\mathrm{NFW}} = 5.71 \hspace{1mm} \Big( \frac{M_{200}}{2 \times 10^{12}h^{-1}M_{\odot}}\Big)^{-0.084} \hspace{1mm} (1+z)^{-0.47},
  \label{eq_mass_c}
\end{equation}
with $M_{200}$ the mass in units of $h^{-1}M_{\odot}$. $M_{200}$ is defined as the mass inside a sphere with radius $r_{200}$, the radius where the density is 200 times the critical density $\rho_c$. We use the median lens redshift $z_{med}=0.34$ for the stacked lenses in the NFW fit, and calculate the tangential shear profile using the analytical expressions provided by \citet{Bartelmann96} and \citet{WrightB00}. The best fit NFW profile is indicated by the dotted line in Figure \ref{plot_gtall}. \\
\indent It is clear that both the SIS and NFW profiles underestimate the signal at scales larger than $\sim$1 arcminute, which corresponds to $\sim$300 kpc at the median lens redshift. The majority of galaxies live in clustered environments, and with gravitational lensing we measure the shear induced by neighbouring galaxy haloes as well. This excess lensing signal complicates a straightforward analysis of the data. The problem could be avoided by studying the lensing signal on small scales around isolated galaxies (following \citet{Hoekstra05}), but this requires the availability of redshifts for all galaxies, which we do not have in the RCS2. Alternatively, the lensing signal can be modelled taking the clustering of the lenses into account, which enables the simultaneous study of the mass and of the clustering properties of the galaxies. This is inherent in the halo model \citep{Seljak00,CoorayS02}, which we will use here. \\
\indent The lenses in a bin generally have a range of masses. The correct interpretation of the signal therefore requires knowledge of the distribution of the masses of the lens galaxies, an issue we return to at the end of Section \ref{sec_halo}. \\


\section{HALO MODEL \label{sec_halo}}

\hspace{4mm} Galaxies form in the gravitational potential of dark matter haloes and therefore trace the large scale distribution of matter in the universe. The quantity that describes the relation between galaxies and dark matter is referred to as galaxy biasing. The description of galaxy biasing is non-trivial as the physics governing galaxy formation is complex, and the bias may depend on the dark matter halo mass, environment, scale and redshift \citep[e.g.][]{Cresswell09,Coupon11,Kovac11}. To gain insight into the relation between galaxies and dark matter the weak lensing signal around galaxies can be used, as it measures the correlation between the galaxies and the surrounding dark matter distribution. These lensing measurements provide constraints for models of the large scale distribution of matter, which are commonly described with the power spectrum of the density fluctuations \citep[e.g.][]{PeacockD96,Smith03}. For a given power spectrum, the lensing signal can be computed directly \citep{GuzikS01}:
\begin{equation}\begin{split}
  \gamma_t(\theta) = 6 \pi^2 \bigg(\frac{H_0}{c}\bigg)^2 \Omega_M \int_0^{\infty} \mathrm{d}\chi W_1(\chi) \frac{f(\chi)}{a(\chi)} \\
  \times \int \mathrm{d}k k P(k,\chi,\theta)J_2(k r(\chi)\theta),
  \label{eq_p2gt}
\end{split}\end{equation}
with $\chi$ the radial distance (in a flat universe, $\chi=a^{-1} \hspace{1mm} D_A$ with $a$ the scale factor and $D_A$ the angular diameter distance), $W_1(\chi)$ the normalized radial distribution of the lenses, $f(\chi)=\int_{\chi}^{\infty} \mathrm{d}\chi' g(\chi,\chi') W_2(\chi')$, with $W_2(\chi')$ the radial distribution of the sources, and 
\begin{equation}
g(\chi,\chi')=\frac{D_lD_{ls}}{D_sa(z_L)}.
\end{equation}
$P(k)$ is the power spectrum under consideration, and $J_2$ is the second Bessel function of the first kind. Instead of using a single power spectrum to describe the distribution of matter in the universe, it is beneficial to consider the various components that contribute, as is done in the halo model. This allows a simultaneous study of the halo masses of galaxies and of their clustering properties. \\

\indent In the halo model the mass distribution in the universe is described as a distinct number of dark matter haloes that are clustered. As the large scale spatial distribution of haloes is unlikely to affect the physics inside individual haloes, and vice versa, the description of the model can be separated into two steps: the halo mass function and the bias at large scales, and the halo occupation distribution at small scales. \\
\indent The large scale distribution of haloes can be described by the halo number density. In the Press-Schechter approach \citep{PressS74} the dark matter haloes are assumed to form by spherical collapse. This, however, leads to a halo number density that overestimates the abundance of galaxies below the non-linear mass scale. Better agreement with numerical simulations of hierarchical structure formation comes from the assumption of ellipsoidal rather that spherical collapse \citep{Sheth01}. The number density of bound objects is generally written as
\begin{equation}
  n_h(M,z) dM = \frac{\bar{\rho}}{M} f(\nu)d\nu,
  \label{eq_hmf}
\end{equation}
where $n_h(M,z)$ is the halo mass function which depends on the halo mass $M$ and redshift $z$, and $\bar{\rho}$ is the mean matter density of the universe at redshift $z$. Unless explicitly stated otherwise we use $M=M_{200}$. The peak height $\nu$ is given by
\begin{equation}
  \nu = \bigg(\frac{\delta_{sc}(z)}{\sigma(M,z)}\bigg)^2 ,
\end{equation}
with $\delta_{sc}(z)$ the critical overdensity required for spherical collapse at redshift $z$, and $\sigma(M,z)$ the rms of the density fluctuation field on the scale $R=(3M/4\pi\bar{\rho})^{1/3}$, extrapolated to $z$ using linear theory. In the case of ellipsoidal collapse, $f(\nu)$ is given by \citep{Sheth01}
\begin{equation}
  f(\nu) = A \hspace{1mm} (1+(a\nu)^{-p}) \hspace{1mm} \nu^{-1/2} e^{-a\nu/2},
\end{equation}
with $a=0.707$, $p=0.3$, and $A=0.13683$ a constant that is determined by requiring $\int f(\nu)d\nu =1$ (i.e. mass conservation). \\
\indent How the haloes trace the mass is given by the halo-to-mass bias, which is defined as the ratio of the power spectrum of the halo distribution to the power spectrum of the matter distribution. We use an analytical formula for the bias as given by \citet{Sheth01}, but incorporate the adjustments described in \citet{Tinker05}:
\begin{equation}\begin{split}
  b(\nu)&=1+\frac{1}{\sqrt{a}\delta_{sc}}\times\\
  &\Big[\sqrt{a}(a\nu)+\sqrt{a}b(a\nu)^{1-c}-\frac{(a\nu)^c}{(a\nu)^c+b(1-c)(1-c/2)}\Big],
  \label{eq_bias}
\end{split}\end{equation}
with $a=0.707$, $b=0.35$ and $c=0.80$. The scale dependence of the bias is given by
\begin{equation}
  b^2(\nu,r)=b^2(\nu)\frac{[1+1.17\xi_m(r)]^{1.49}}{[1+0.69\xi_m(r)]^{2.09}},
\end{equation}
where $\xi_m(r)$ is the matter correlation function, which in turn is the Fourier transform of the non-linear power spectrum $P_{\mathrm{NL}}(k)$ from \citet{Smith03}, and $r$ is the distance to the centre of the halo. \\
\indent To describe how the galaxies and dark matter are distributed within the haloes, we closely follow the approach outlined in \citet{GuzikS02} and \citet{Mandelbaum05HOD}. Galaxies living inside dark matter haloes are divided into two classes; they are either a central galaxy located in the central halo, or a satellite galaxy located in a subhalo inside the central halo. The fraction of satellites in a certain sample of galaxies is denoted by $\alpha$. The number of satellites in a central halo is described by the halo occupation distribution (HOD). Galaxy formation simulations \citep[e.g.][]{Zheng05,Kravtsov04} show that the HOD is well approximated by a powerlaw $N_s(M) \propto M^\epsilon$ with $\epsilon=1$, which is cut off below a certain minimal halo mass. Rather than this steep cut off, we follow \citet{Mandelbaum05HOD} and assume a more gradual transition, and use $\epsilon=2$ for halo masses smaller than $M_{\mathrm{char}}$, whilst $\epsilon=1$ for halo masses larger than $M_{\mathrm{char}}$, where $M_{\mathrm{char}}=3M_h$. $M_h$ is the typical halo mass of a certain set of galaxies (for example the galaxies selected in a luminosity bin). The amplitude is determined by normalizing to the total number of satellites in the set. \\
\subsection{Lensing signal from the halo model}

\hspace{4mm} We now proceed to explain how the lensing signal is computed. The ensemble averaged tangential shear is the sum of the signal around central galaxies and satellites, since we cannot distinguish between them. We compute each contribution separately, starting with the signal around central galaxies. It is assumed that the central galaxies are located at the centre of the dark matter haloes. Two terms contribute to the lensing signal around central galaxies: the signal coming from the halo where the galaxy resides ($\gamma_{t,\mathrm{cent}}^{1h}$), and the signal from nearby haloes ($\gamma_{t,\mathrm{cent}}^{2h}$). Hence the total signal around central galaxies is given by
\begin{equation}
  \gamma_{t,\mathrm{cent}}=\gamma_{t,\mathrm{cent}}^{1h} + \gamma_{t,\mathrm{cent}}^{2h}.
\end{equation}
The density profiles of the central haloes are assumed to be NFW, which we compute using the mass-concentration relation from \citet{Duffy08} given by Equation \ref{eq_mass_c}. By picking a central halo mass we can thus compute the tangential shear of the central halo term directly, as spectroscopic redshifts are available for all lenses. \\
\indent The calculation of $\gamma_{t,\mathrm{cent}}^{2h}$ requires the power spectrum describing the correlation between the galaxy in the central halo and the dark matter of nearby haloes:
\begin{equation}\begin{split}
  P_{\mathrm{cent}}^{2h}(k,M_h,r) = b_g(M_h,r) \frac{P_{\mathrm{NL}}(k)}{(2 \pi)^3} \\
  \times \int_0^{M_{\mathrm{lim}}} \mathrm{d}\nu f(\nu) b(\nu,r) y_{\mathrm{dm}}(k,M),
  \label{eq_pcent_2h}
\end{split}\end{equation}
with $b_g(M_h,r)$ the bias of the central galaxy, $P_\mathrm{{NL}}(k)$ the non-linear power spectrum from \citet{Smith03}, and $y_{\mathrm{dm}}(k,M)$ the radial Fourier transform of the central halo density profile divided by mass:
\begin{equation}
  y_{\mathrm{dm}}(k,M) = \frac{1}{M} \int_0^{r_{200}} {\mathrm d}r 4\pi r^2 \rho_{\mathrm{dm}}(r,M) \frac{\sin(kr)}{kr},
\end{equation}
which we calculate using the analytical formula given in \citet{Pielorz10}. \\
\indent The dark matter profiles of adjacent haloes cannot overlap, which is prevented by implementing halo exclusion. Different approaches to halo exclusion have been used in the literature. For example, \citet{Cacciato09} set the two-halo correlation function to zero below $r_{180}$, which leads to a sharp truncation in the halo models. We follow the approach of \citet{Tinker05}, which leads to a more natural smooth cut-off: the integral in Equation \ref{eq_pcent_2h} is cut off for masses greater than $M_{\mathrm{lim}}$ which is chosen such that the $r_{200}$ of the central halo does not overlap with the $r_{200}$ of nearby haloes: $r_{200}(M_h)+r_{200}(M_{\mathrm{lim}})=r$. It should be noted that this choice, as any other halo exclusion approach, is an approximation. Ultimately, numerical simulations should be used to provide improved estimates for $P^{2h}_{\mathrm{cent}}$. \\
\indent The contribution of the satellites to the lensing signal consists of three terms: the signal from the subhalo where the satellite resides ($\gamma_{t,\mathrm{sat}}^{\mathrm{trunc}}$), the signal from the central halo in which the subhalo resides ($\gamma_{t,\mathrm{sat}}^{1h}$), and the signal from nearby haloes ($\gamma_{t,\mathrm{sat}}^{2h}$). Hence the total signal around satellites is given by
\begin{equation}
  \gamma_{t,\mathrm{sat}}=\gamma_{t,\mathrm{sat}}^{\mathrm{trunc}} + \gamma_{t,\mathrm{sat}}^{1h} + \gamma_{t,\mathrm{sat}}^{2h}.
\end{equation}
First we compute the lensing signal of the subhalo, $\gamma_{t,\mathrm{sat}}^{\mathrm{trunc}}$, following \citet{Mandelbaum05HOD}. The density profile is assumed to follow an NFW profile in the inner regions. The outer regions of the subhalo are tidally stripped of its dark matter by the central halo. Due to this stripping the lensing signal is proportional to $r^{-2}$ at radii larger than the truncation radius. Based on good agreement with numerical simulations, \citet{Mandelbaum05HOD} chose a truncation radius of $0.4r_{200}$, and we use the same. This choice corresponds to roughly 50\% of the dark matter being stripped from the subhalo. \\
\indent To compute the lensing signal induced by the halo where the subhalo resides, we calculate the power spectrum describing the correlation between the subhalo and the dark matter profile of the central halo:
\begin{equation}\begin{split}
  P_{\mathrm{sat}}^{1h}(k,M_h) =\frac{1}{(2\pi)^3 \bar{n}} \int {\mathrm d}\nu f(\nu) N_s(M, M_h) \\
  \times y_{\mathrm{dm}}(k,M) y_{\mathrm{g}}(k,M),
\end{split}\end{equation}
with $\bar{n}$ the mean galaxy number density, which can be determined using $\bar{n}=\bar{\rho}\int {\mathrm d}\nu f(\nu)\frac{N_s(M, M_h)}{M}$, and $y_\mathrm{g}$ the radial Fourier transform of the radial distribution of satellites around the central halo. We assume that the radial distribution of satellites follows an NFW profile with a concentration $c_\mathrm{g}$, given by the mass-concentration relation from \citet{Duffy08}. To asses the sensitivity to the shape of the radial distribution of the satellites, we also calculate the $\gamma_{t,\mathrm{sat}}^{1h }$ term using a $c_\mathrm{g}$ that is varied by a factor of two. We find that this change mainly impacts the model signal at small scales: for a larger (smaller) concentration, the signal increases (decreases). At scales larger than a few hundred kpc, the change of the model signal is negligible. When we fit these adjusted models to the data, we find that the best fit model parameters do not change significantly. We conclude that the signal-to-noise of our data currently does not enable us to discriminate between halo models with different radial distributions of satellite galaxies. \\
\indent Finally we compute the contribution from nearby haloes to the lensing signal around satellite galaxies:
\begin{equation}\begin{split}
  P_{\mathrm{sat}}^{2h}(k,M_h,r) = \frac{P_{\mathrm{NL}}(k)}{(2 \pi)^3} \int_0^{M_{\mathrm{lim}}} {\mathrm d}\nu f(\nu)b(\nu,r)y_{\mathrm{dm}}(k,M) \\
  \times \frac{\bar{\rho}}{\bar{n}} \int  \mathrm{d}\nu f(\nu) b(\nu,r) \frac{N_s(M, M_h)}{M} y_{\mathrm{g}}(k,M).
\end{split}\end{equation}
The three power spectra are converted into their respective shear signals using Equation \ref{eq_p2gt}, and the contributions from the central galaxies and satellites are combined to yield
\begin{equation}
  \gamma_t = (1-\alpha)\hspace{1mm}\gamma_{t,\mathrm{cent}}+\alpha \hspace{1mm}\gamma_{t,\mathrm{sat}},
\end{equation}
where $\alpha$ is the fraction of satellites of the sample. The resulting model is compared to the data. \\ 

\indent The lens sample is selected to cover a range in an observable, such as luminosity or stellar mass, as the relation between the mean observable and the lensing mass is a useful constraint for simulations. The dark matter haloes of the lenses from such a sample have different masses, however, and it is therefore important to account for the scatter in the observable-halo mass relation. If the halo mass distribution is well-known, this can be done by integrating the models over the distribution of halo masses. Unfortunately, the distribution is generally not accurately known as the lenses span a considerable range in observable, redshift and environment. A simpler approach is to study how the lensing mass is related to the mean halo mass for a given halo mass distribution. This approach, which was proposed by \citet{Mandelbaum06}, provides the leading-order correction for the scatter, and we use it in this paper.


\section{COMPARISON WITH DYNAMICAL MASS \label{sec_mdyn}}
\begin{figure*}
\centering
  \includegraphics[width=17cm]{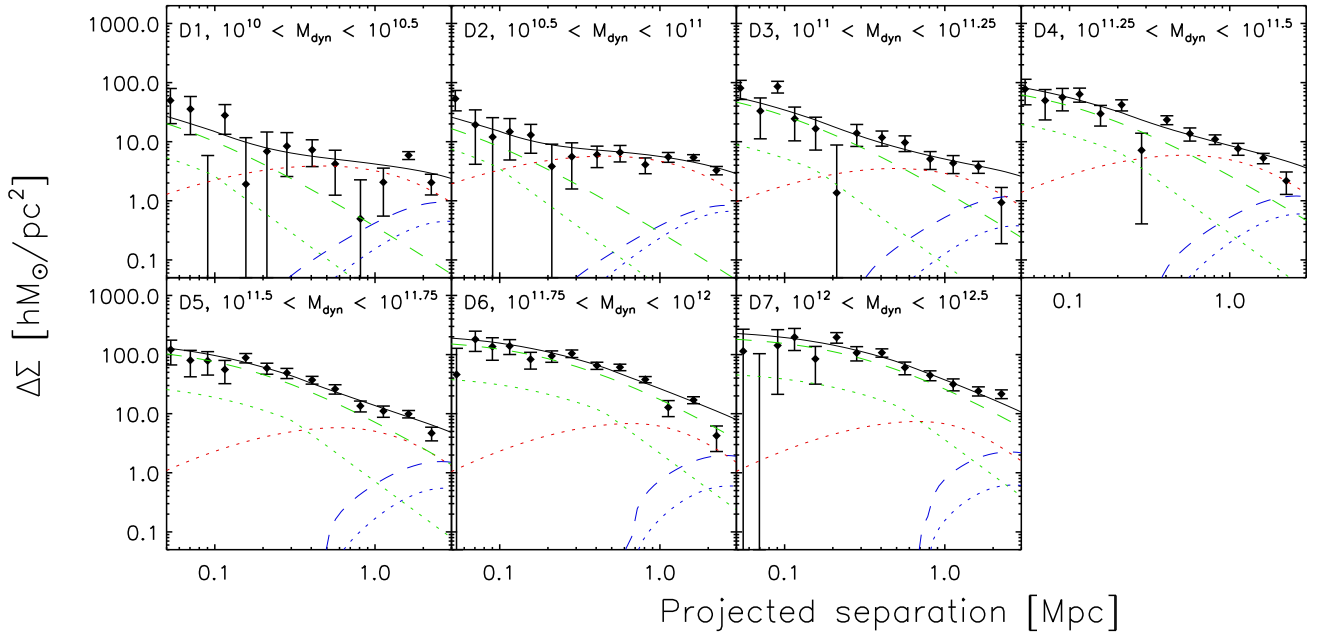}
  \caption{The lensing signal $\Delta \Sigma$ for each dynamical mass bin as a function of physical distance from the lens. The green dashed line shows the $\gamma_{t,\mathrm{cent}}^{1h}$ term, the blue dashed line the $\gamma_{t,\mathrm{cent}}^{2h}$ term, the green dotted line the $\gamma_{t,\mathrm{sat}}^{\mathrm{trunc}}$ term, the red dotted line the $\gamma_{t,\mathrm{sat}}^{\mathrm{1h}}$ term, the blue dotted line the $\gamma_{t,\mathrm{sat}}^{\mathrm{2h}}$ term, and the black line shows the sum of the terms. The $\gamma_{t,sat}^{1h}$ term causes a prominent bump for the two lowest dynamical mass bins, which indicates that a significant number of lenses in these bins are satellites. }
  \label{plot_mdyn_all}
\end{figure*}

\begin{figure}  
  \resizebox{\hsize}{!}{\includegraphics{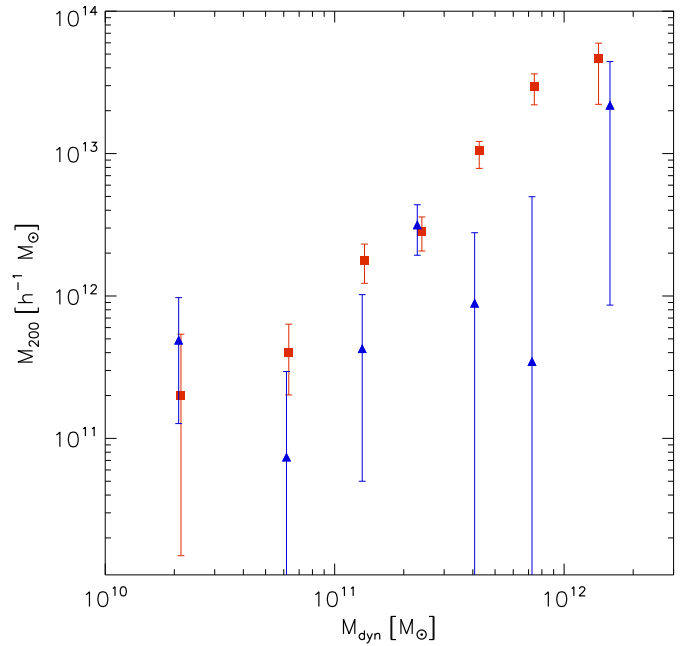}}
  \caption{The best fit halo mass as a function of the mean dynamical mass. The red squares (blue triangles) denote the halo mass for the early-types (late-types). The early-/late-type division is based on the brightness profiles of the lenses. The dynamical mass correlates well with the lensing mass for the early-type galaxies, but not for the late-type galaxies.}
  \label{plot_mdyn_mh}
\end{figure}

\hspace{4mm} The dynamical mass traces the gravitational potential of a galaxy at small scales, and typically provides estimates of the total mass enclosed by the effective radius, which is of the order of a few kpcs. Comparison to the mass derived from strong lensing shows that both estimates agree well for early-type galaxies \citep{Bolton08}. In contrast, weak lensing traces the gravitational potential at much larger scales, and the mass is usually determined within $r_{200}$, whose values range between a few tens to a few hundreds of kpc. To study how the dynamical mass is related to the weak lensing mass, we measure the lensing signal for galaxies divided into seven dynamical mass bins, as detailed in Table \ref{tab_lens_dyn}. The lensing signal of the stacked galaxies in each bin is shown in Figure \ref{plot_mdyn_all}. We fit our halo model to the lensing signal in the distance interval between 50 kpc and 2 Mpc. At scales smaller than 50 kpc the lensing signal is very noisy, since we do not have many sources at small separations, and lens light contamination might bias the shear signal. At scales larger than 2 Mpc we measure the lensing signal using mainly sources that reside at the edge of the images, where the PSF ellipticity is large for the data taken prior to a change in the MegaCam configuration\footnotemark (up to 15\%), and the residual PSF systematics noticeably bias the lensing signal. 
\footnotetext[7] {In November 2004, the lens L3 was accidentally mounted incorrectly after the wide-field corrector had been disassembled. As this surprisingly led to a significant improvement in the image quality for the $u^{*}$-, $g'$-,and $r'$-band, the new configuration was kept. About 20\% of the RCS2 survey was obtained prior to this change.}
We fit for the central halo mass and the satellite fraction, and use Equation \ref{eq_bias} to compute the bias because the lensing signal is not well constrained at scales $>3$Mpc. \\
\indent We impose two priors on the fits. Firstly, we do not fit halo masses that are lower than the mean stellar mass of the galaxies in the bin. This prior could introduce a bias if the assumed IMF is significantly different from the true one, leading to stellar mass estimates that are too high, but this is not expected to be the case. The second prior we impose is on the satellite fraction, which is not well constrained by the data for the most massive galaxies and is anti-correlated with the best fit halo mass (see Appendix \ref{ap_sat} for details). To prevent this from biasing the halo mass low, we limit the range of fitted satellite fractions to be less than 20\% in the three highest dynamical mass bins as they contain galaxies that are expected to be nearly exclusively centrals. The best fit halo model for each bin is also shown in Figure \ref{plot_mdyn_all}. We find that the model fits the data well. The resulting best fit halo masses for the early- and late-type galaxies are shown in Figure \ref{plot_mdyn_mh}, and detailed in Table \ref{tab_lens_dyn}. The error bars on the best fit halo mass (satellite fraction) indicate the 1$\sigma$ deviations determined by marginalizing over the satellite fraction (halo mass).\\ 
\begin{table*}
  \centering
  \caption{The dynamical mass results}             
  \begin{tabular}{c c c c c c c c c c } 
  \hline
  \vspace{1mm}
  Sample & log$(M_\mathrm{{dyn}})$ & $n_{\mathrm{lens}}$ & $\langle z \rangle$ & $\langle M_\mathrm{{dyn}} \rangle$ & $f_{\mathrm{late}}$ & $M_h^{\mathrm{early}}$ & $\alpha^{\mathrm{early}}$ & $M_h^{\mathrm{late}}$ &  $\alpha^{\rm{late}}$\\
   & (1) & (2) & (3) & (4) & (5) & (6) & (7) & (8) & (9) \\
  \hline\hline  \\
  D1  &    $[10.00,10.50]$ & $2 \hspace{0.5mm}011$ & 0.08 & 1.96 & 0.44 & $2.00^{+3.39}_{-1.85}$ & $0.24^{+0.11}_{-0.09}$ & $4.90^{+4.84}_{-3.63}$ & $0.17^{+0.11}_{-0.09}$ \\
  D2  &    $[10.50,11.00]$ & $4 \hspace{0.5mm}752$ & 0.10 & 5.91 & 0.35 & $3.98^{+2.36}_{-1.96}$ & $0.41^{+0.07}_{-0.06}$ & $0.74^{+2.21}_{-0.73}$ & $0.11^{+0.12}_{-0.08}$\\
  D3  &    $[11.00,11.25]$ & $2 \hspace{0.5mm}762$ & 0.13 & 13.2 & 0.25 & $17.8^{+5.36}_{-5.50}$ & $0.14^{+0.07}_{-0.07}$ & $4.27^{+5.95}_{-3.77}$ & $0.19^{+0.15}_{-0.12}$\\
  D4  &    $[11.25,11.50]$ & $2 \hspace{0.5mm}281$ & 0.16 & 23.6 & 0.16 & $28.2^{+7.72}_{-7.49}$ & $0.31^{+0.09}_{-0.08}$ & $31.6^{+12.1}_{-12.3}$ & $0.00^{+0.09}_{-0.00}$ \\
  D5  &    $[11.50,11.75]$ & $1 \hspace{0.5mm}715$ & 0.22 & 41.7 & 0.07 & $105^{+17.1}_{-26.0}$ & $0.20^{+0.00}_{-0.09}$ & $8.91^{+18.9}_{-8.90}$ & $0.00^{+0.39}_{-0.00}$\\
  D6  &    $[11.75,12.00]$ & 935 & 0.32 & 72.5 & 0.05 & $295^{+67.7}_{-75.1}$ & $0.20^{+0.00}_{-0.20}$ & $3.47^{+46.3}_{-3.46}$ & $0.00^{+0.20}_{-0.00}$\\
  D7  &    $[12.00,12.50]$ & 380 & 0.39 & 137.4 & 0.07 & $468^{+129}_{-246}$ & $0.20^{+0.00}_{-0.20}$ & $219^{+224}_{-210}$ & $0.20^{+0.00}_{-0.20}$ \\
  \hline
  \end{tabular}
  \tablefoot{(1) the dynamical mass range of the bin; (2) the number of lenses; (3) the mean redshift; (4) the mean dynamical mass in units of $10^{10}M_{\odot}$; (5) the fraction of late-type galaxies; (6) the best fit halo mass for the early-types in units of $10^{11}h^{-1}M_{\odot}$; (7) the best fit satellite fraction for the early-types; (8) the best fit halo mass for the late-types in units of $10^{11}h^{-1}M_{\odot}$; (9) the best fit satellite fraction for the late-types.}
  \label{tab_lens_dyn}
\end{table*}
\indent For the early-type galaxies, we find that the dynamical mass correlates well with the halo mass. The halo mass is $\sim$10 times larger than the mean dynamical mass for $M_{\mathrm{dyn}} < 1\times10^{11} M_{\odot}$, which increases to a factor $\sim$50 for the highest dynamical mass bins, as the galaxy dark matter haloes extend far beyond the effective radius. To establish whether we can scale the dynamical mass to the lensing mass, we replace $R_e$ with the best fit lensing $r_{200}$ in Equation \ref{eq_mdyn}. We find that the rescaled dynamical masses are 8 times larger than the best fit lensing masses for D1 and D2, but the difference decreases for the more massive bins: the rescaled dynamical mass is only 40\% larger than the best fit lensing mass for D7. We therefore cannot simply rescale the mean dynamical mass to the lensing mass. Note that at the high mass end, galaxies predominantly live in groups and clusters. With lensing we fit the halo mass of the entire structure, whereas the dynamical mass is determined for the individual galaxy only.  \\
\indent We observe that for the late-type galaxies the halo mass does not correlate well with the mean dynamical mass. In particular, the best fit halo masses of the D5 and D6 late-type bins are low. These low values may be explained if rotation constitutes a major part of the observed velocity dispersions of late-type galaxies, leading to an overestimation of the dynamical mass. Additionally, the effective radius for some late-type galaxies at high redshift may be overestimated, since a significant fraction consists of multiple objects with small separations as we observed in Section \ref{sec_lenssamp}. \\
\indent For early-type galaxies the dynamical mass is a useful tracer of the total mass at small scales, but it appears to be less reliable for late-type galaxies. How the dynamical mass changes for galaxies where rotation is important, or for galaxies that are populated over a large range of redshifts, may be studied with numerical simulations. In any case, it is not clear how to translate a dynamical mass estimate into a total mass estimate of the halo of a galaxy. With weak lensing we measure the total halo masses of galaxies directly, providing estimates that can easily be compared to simulations.


\section{LUMINOSITY RESULTS \label{sec_lumi}}

\begin{figure*}
   \centering
  \includegraphics[width=17cm]{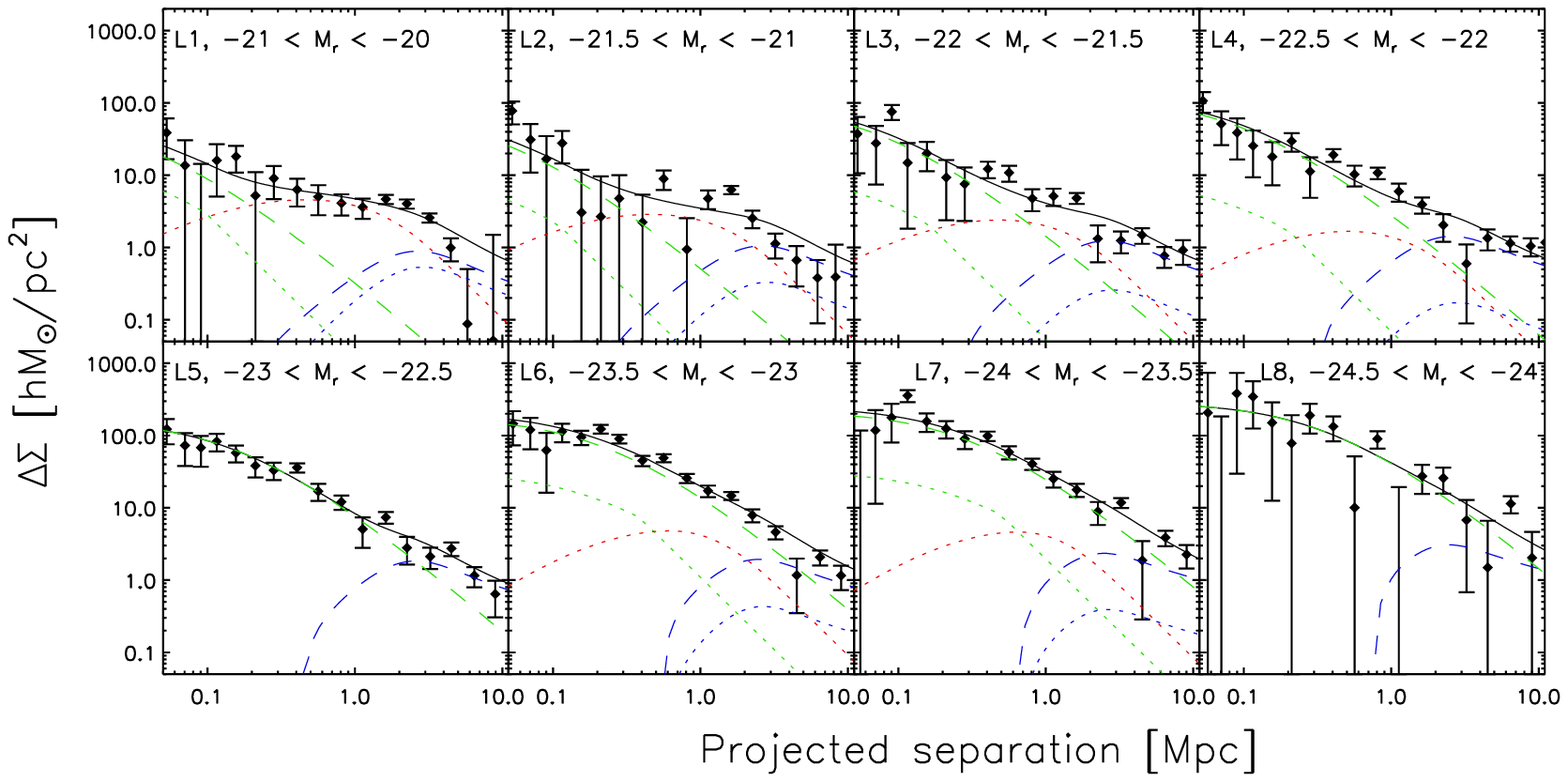}
  \caption{The lensing signal $\Delta \Sigma$ for each luminosity bin as a function of physical distance from the lens. The green dashed line shows the $\gamma_{t,\mathrm{cent}}^{1h}$ term, the blue dashed line the $\gamma_{t,\mathrm{cent}}^{2h}$ term, the green dotted line the $\gamma_{t,\mathrm{sat}}^{\mathrm{trunc}}$ term, the red dotted line the $\gamma_{t,\mathrm{sat}}^{\mathrm{1h}}$ term, the blue dotted line the $\gamma_{t,\mathrm{sat}}^{\mathrm{2h}}$ term, and the black line shows the sum of the terms. A significant fraction of the low luminosity lenses are satellites in larger haloes, as the $\gamma_{t,sat}^{1h}$ term causes a prominent bump at $\sim1$Mpc in the lensing signal.}
  \label{plot_lum_all}
\end{figure*}

\hspace{4mm} The optical luminosity is a readily measured quantity which is related to the stellar mass, and hence the baryonic content of a galaxy. Therefore, we continue by measuring the lensing signal as a function of luminosity. We divide our lens sample into eight luminosity bins, as detailed in Table \ref{tab_lens_lum}. We measure $\Delta \Sigma$ of the stacked lenses and show the results in Figure \ref{plot_lum_all}, together with the best fit halo model.
The amplitude of the lensing signal clearly increases for the brighter galaxies as expected. Furthermore, the shear from the $\gamma_{t,sat}^{1h}$ term causes a prominent bump for the fainter lenses, but not for the brighter ones. This indicates that a considerable fraction of the low luminosity lenses are satellites. We split the lenses into early- and late-types using the $frac\_deV$ parameter as before, and study the signals separately.\\ 

\begin{figure}
  \resizebox{\hsize}{!}{\includegraphics{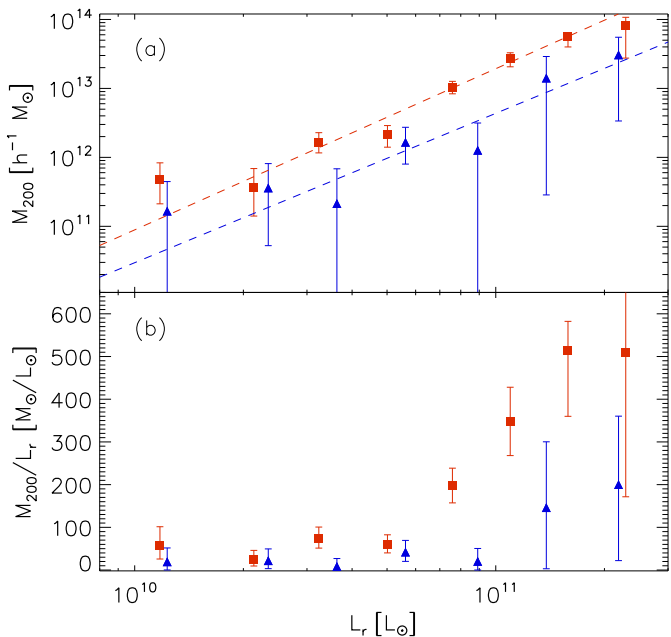}}
  \caption{The best fit halo mass ({\it top}), and the mass-to-light ratio ({\it bottom}) as a function of mean luminosity. The red squares (blue triangles) denote the early-type (late-type) results. The division in early-/late-types is based on the brightness profiles of the lenses. The dashed lines are the powerlaw fits, with values as indicated in the text. }
  \label{plot_lum_mh}
\end{figure}

\indent There are various issues we have to address before we can interpret the measurements. First of all, lens galaxies scatter between luminosity bins due to luminosity errors. If the luminosity errors are large compared to the width of the bins this could potentially introduce a bias. This bias is greatest at the highest luminosities, where the luminosity function is steep. In this case, on average more low luminosity (and mass) galaxies scatter into the higher luminosity bins, biasing the best fit halo mass low. 
\begin{table*}
  \caption{The luminosity results}             
  \centering
  \renewcommand{\tabcolsep}{0.13cm}
  \begin{tabular}{c c c c c c c c c c c c c} 
  \hline
  \vspace{1mm}
 Sample & $M_r$ & $n_{\rm{lens}}$ & $\langle z \rangle$ & $f_{\rm{late}}$ & $\langle L^{early}_r \rangle$ & $M_h^{\rm{early}}$ & $\alpha^{\rm{early}}$ & $L_{200}^{\rm{early}}$ & $\langle L^{late}_r \rangle$ & $M_h^{\rm{late}}$ & $\alpha^{\rm{late}}$ & $L_{200}^{\rm{late}}$ \\
  & (1) & (2) & (3) & (4) & (5) & (6) & (7) & (8) & (9) & (10) & (11) & (12) \\
  \hline\hline  \\
  L1  &  $[-21.0,-20.0]$ & $3 \hspace{0.5mm}563$ & 0.08 & 0.44 & 1.17 & $4.78^{+3.55}_{-2.66}$ & $0.37^{+0.08}_{-0.07}$ & 1.32$\pm$0.01 & 1.23 & $1.66^{+2.79}_{-1.65}$ & $0.07^{+0.11}_{-0.07}$ & 1.26$\pm$0.01 \\
  L2  &  $[-21.5,-21.0]$ & $2 \hspace{0.5mm}772$ & 0.10 & 0.35 & 2.14 & $3.67^{+3.21}_{-2.26}$ & $0.38^{+0.09}_{-0.08}$ & 2.36$\pm$0.01 & 2.34 & $3.60^{+4.51}_{-3.08}$ & $0.06^{+0.10}_{-0.06}$ & 2.52$\pm$0.01  \\
  L3  &  $[-22.0,-21.5]$ & $3 \hspace{0.5mm}064$ & 0.13 & 0.29 & 3.24 & $16.5^{+6.26}_{-4.91}$ & $0.19^{+0.07}_{-0.07}$ & 4.01$\pm$0.02 & 3.63 & $2.14^{+4.69}_{-2.13}$ & $0.35^{+0.28}_{-0.12}$ & 3.67$\pm$0.01 \\
  L4  &  $[-22.5,-22.0]$ & $2 \hspace{0.5mm}370$ & 0.16 & 0.20 & 5.01 & $21.3^{+7.58}_{-7.28}$ & $0.34^{+0.09}_{-0.09}$ & 6.57$\pm$0.03 & 5.62 & $16.6^{+10.7}_{-8.57}$ & $0.00^{+0.05}_{-0.00}$ & 6.50$\pm$0.04 \\
  L5  &  $[-23.0,-22.5]$ & $1 \hspace{0.5mm}658$ & 0.20 & 0.11 & 7.59 & $105^{+21.9}_{-21.3}$ & $0.05^{+0.11}_{-0.05}$ & 12.3$\pm$0.1 & 8.91 & $12.7^{+18.8}_{-12.7}$ & $0.20^{+0.00}_{-0.13}$ & 10.0$\pm$0.1 \\
  L6  &  $[-23.5,-23.0]$ & $1 \hspace{0.5mm}453$ & 0.32 & 0.03 & 11.0 & $267^{+61.1}_{-61.8}$ & $0.19^{+0.01}_{-0.19}$ & 21.2$\pm$0.2 & 13.8 & $141^{+149}_{-138}$ & $0.20^{+0.00}_{-0.20}$ & 23.4$\pm$1.7  \\
  L7  &  $[-24.0,-23.5]$ & 607 & 0.41 & 0.05 & 15.8 & $570^{+76.5}_{-170}$ & $0.00^{+0.20}_{-0.00}$ & 45.2$\pm$0.8 & 21.9 & $306^{+245}_{-273}$ & $0.00^{+0.20}_{-0.00}$ & - \\
  L8  &  $[-24.5,-24.0]$ & 83 & 0.47 & 0.04 & 22.9 & $818^{+257}_{-543}$ & $0.00^{+0.20}_{-0.00}$ & 68.0$\pm$3.6 & - & - & - & - \\
  \hline  
  \end{tabular}
  \tablefoot{(1) the magnitude range of the bin; (2) the number of lenses; (3) the mean redshift; (4) the fraction of late-type galaxies; (5)
the mean luminosity for the early-types in units of $10^{10}L_{\odot}$; (6) the best fit halo mass for the early-types in units of $10^{11}h^{-1}M_{\odot}$; (7) the best fit satellite fraction for the early-types; (8) the total luminosity within $r_{200}$ for the early-types in units of $10^{10}L_{\odot}$; (9) the mean luminosity for the late-types in units of $10^{10}L_{\odot}$; (10) the best fit halo mass for the late-types in units of $10^{11}h^{-1}M_{\odot}$; (11) the best fit satellite fraction for the late-types; (12) the total luminosity within $r_{200}$ for the late-types in units of $10^{10}L_{\odot}$ .}
  \label{tab_lens_lum}
\end{table*}
The average absolute magnitude error is $\sim$0.03 for $z<0.33$, and $\sim$0.07 for $z>0.33$, small compared to the minimal bin-width of 0.5. We find that the induced bias is relevant for the L7 and L8 bins of the early-types only, with corrections of 4\% and 7\% respectively. The corrections are smaller than the measurement errors on the halo mass for these bins. We detail the calculation of the correction factor in Appendix \ref{ap_scat}. \\ 
\indent When we fit a halo mass to the stacked shear signal of galaxies within a luminosity bin, the resulting mass is not equal to the mean halo mass, nor to the central mass of the original distribution \citep{Tasitsiomi04,Mandelbaum05HOD,Cacciato09,Leauthaud10} because the distribution in halo mass is not uniform (in addition, the NFW profile itself depends on mass). It is useful to convert the measured lensing mass to the mean halo mass to allow comparison with simulations. The correction we have to apply depends not only on the distribution of halo masses for a given luminosity, but also on the halo mass function. Since the halo mass function is a declining function --- steeply at the high mass end --- we will preferentially select lower mass haloes. Hence, the underlying function from which we draw our galaxies is the halo mass function convolved with the halo mass distribution. In Appendix \ref{ap_width} we discuss how we calculate the correction factor that we apply to obtain the mean of the halo mass in each luminosity bin. The values are given in Table \ref{tab_widthcorr}, and range between 5-30\%. \\
\indent The best fit halo mass for each luminosity bin, corrected for the scatter and the width of the halo mass distribution, is given in Table \ref{tab_lens_lum}, and is shown as a function of luminosity in Figure \ref{plot_lum_mh}a. The error bars on the halo masses are the 1$\sigma$ deviations determined by marginalizing over the satellite fraction.
We fit a powerlaw of the form $M_{200}=M_{0,L} (L/L_{0})^{\beta_L}$, with a pivot $L_{0}=10^{11}L_{r,\odot}$. As the errors of the best fit halo masses are asymmetric due to the constraints we impose on the halo model fits, we fit the powerlaw directly to the shear measurements (with symmetric error bars). Hence we do not fit for the halo mass for each bin, but determine the best fit $M_{0,L}$ and $\beta_L$ for all bins simultaneously, whilst fitting the satellite fraction for each bin separately. Note that the best fit satellite fractions from this approach are close to the values given in Table \ref{tab_lens_lum}. For the early-types, we find $M_{0,L}=1.93^{+0.13}_{-0.14}\times 10^{13}h^{-1}M_{\odot}$ and $\beta_L=2.34^{+0.09}_{-0.16}$, and for the late-types $M_{0,L}=0.43^{+0.20}_{-0.17}\times 10^{13}h^{-1}M_{\odot}$ and $\beta_L=2.2^{+0.7}_{-0.6}$, as shown in Figure \ref{plot_lum_mh}a.
\begin{figure}
  \resizebox{\hsize}{!}{\includegraphics[angle=-90]{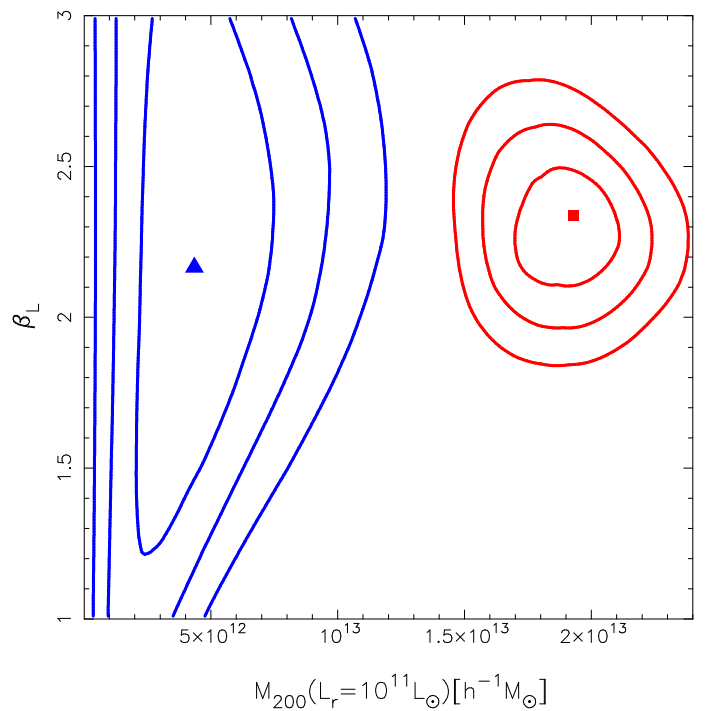}}
  \caption{The 67.8\%, 95.4\% and 99.7\% confidence limits of the powerlaw fits between luminosity and halo mass, in red (blue) for the early-type (late-type) galaxies. The red square (blue triangle) indicates the best fit of the early-types (late-types). The powerlaw fit for the early-types is better constrained than for the late-types, because the majority of galaxies in our lens sample are early-types. The early-types also reside in more massive haloes, and consequently produce a stronger lensing signal. }
  \label{plot_chisq_lum}
\end{figure}
The error on $M_{0,L}$ ($\beta_L$) is determined by marginalizing over $\beta_L$ ($M_{0,L}$). We show the 67.8\%, 95.4\% and 99.7\% confidence limits of the two powerlaw fits in Figure \ref{plot_chisq_lum}. The results for the early-types are better constrained because we have more early-type galaxies in our lensing sample. These are also more massive than the late-type galaxies and hence produce a stronger lensing signal. \\

\indent We compare our analysis to two previous weak lensing studies. \citet{Hoekstra05} measured the lensing signal of $\sim1.4\times10^5$ isolated galaxies with photometric redshift $0.2 < z < 0.4$ in the RCS. 
In the $R$-band, they found a virial mass of $M_{R}^{\mathrm{H05}}=7.5^{+1.2}_{-1.1}\times10^{11}h^{-1}M_{\odot}$ for a galaxy of luminosity $L_R=10^{10}h^{-2}L_{\odot}$, and a powerlaw index of $\beta_{R}^{\mathrm{H05}}=1.6\pm0.2$. We use the transformations from Lupton (2005)\footnotemark
\footnotetext[8]{http://www.sdss.org/dr7/algorithms/sdssUBVRITrans -\newline form.html\#Lupton2005}, and find that $r\approx R+0.24$ for the early-type galaxies in our sample, which make up the majority of the lenses. We convert $L_R$ to $L_r$, use our powerlaw fit to predict $M_{200}$, and convert that to the virial mass by increasing it by 30\%. We find that $M_{\mathrm{vir}}=(7.2\pm1.5)\times10^{11}h^{-1}M_{\odot}$ for a $L_R=10^{10}h^{-2}L_{\odot}$ galaxy, in good agreement with \citet{Hoekstra05}. The powerlaw index of \citet{Hoekstra05} is shallower than the $\beta_L=2.34^{+0.09}_{-0.16}$ that we find. A possible explanation is that a fraction of the low luminosity galaxies in \citet{Hoekstra05} are satellites, whose masses are biased high due to the added lensing signal of nearby galaxies, flattening the powerlaw index. We note two caveats: the lens sample of \citet{Hoekstra05} does not exclusively consist of early-types, and the lens samples we compare reside in different environments. \\
\indent \citet{Mandelbaum06} present results for $3.5 \times 10^5$ galaxies using SDSS data. Galaxies are divided into early-types and late-types based on their brightness profile (using the same selection criterium that we have applied to our lenses), and are studied in bins of absolute $r$-band magnitude. To compare the results, we convert our luminosities according to the definitions used in \citet{Mandelbaum06}: the absolute magnitude is calculated using a k-correction to z=0.1, the distance modulus is calculated using $h=1.0$ and a passive evolution term is included which is given by $1.6(z-0.1)$. As a result, we decrease the absolute magnitudes of our lenses by roughly one magnitude. Additionally, we increase our masses by 30\% since \citet{Mandelbaum06} define the halo mass using 180$\bar{\rho}$ instead of 200$\rho_c$. There are various other differences between the analyses, such as the use of a different correction factor for the width of the halo mass distribution, a different cosmology, a different mass-concentration relation for the NFW profiles, and differences in the modelling of the lensing signal. These differences are expected to have a minor impact on the best fit halo mass, but they limit the accuracy of a detailed comparison. \\
\indent Matching our luminosity bins to those of \citet{Mandelbaum06} closest in mean luminosity, we find that the best fit halo masses for the early- and late-type galaxies are generally in agreement. To quantify whether the results are consistent, we fit a powerlaw of the form $M_{180}=\tilde{M}_{0,L} (\tilde{L}/\tilde{L}_{0})^{\beta_{\tilde{L}}}$, where $\tilde{L}_{0}=1.2\times10^{10}h^{-2}L_{\odot}$. The tilde indicates that the luminosity is calculated following \citet{Mandelbaum06}. The powerlaw is fitted to the best fit halo mass directly, and the weights of the measurements are calculated from the error bars through which the model passes, i.e., if the model is larger (smaller) than the data point, we use the positive (negative) error bar. For the early-types we find $\tilde{M}_{0,L}=7.3^{+2.1}_{-1.7}\times 10^{11}h^{-1}M_{\odot}$ and $\beta_{\tilde{L}}=2.7\pm0.2$ for our data, while using \citet{Mandelbaum06} results we find $\tilde{M}_{0,L}=11.2^{+1.9}_{-1.8}\times 10^{11}h^{-1}M_{\odot}$ and $\beta_{\tilde{L}}=2.3\pm0.2$, in fair agreement with our findings. For the late-types we find $\tilde{M}_{0,L}=2.7^{+3.9}_{-1.8} \times 10^{11}h^{-1}M_{\odot}$ and $\beta_{\tilde{L}}=3.0^{+1.0}_{-1.6}$, while using the results of \citet{Mandelbaum06} we find $\tilde{M}_{0,L}=7.8\pm1.1 \times 10^{11}h^{-1}M_{\odot}$ and $\beta_{\tilde{L}}=1.1^{+0.3}_{-0.4}$. The results from \citet{Mandelbaum06} prefer a shallower slope and a higher offset, but the fits are consistent.  \\


\subsection{Mass-to-light ratio \label{sec_lum_ml}}
\begin{figure}
  \resizebox{\hsize}{!}{\includegraphics{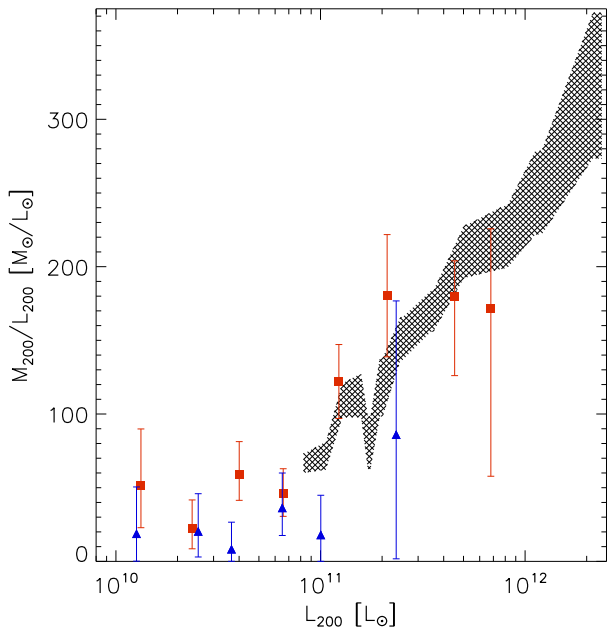}}
  \caption{The mass-to-light ratio using the total halo mass and luminosity within $r_{200}$, as a function of $L_{200}$. The red squares (blue triangles) denote the early-type (late-type) results. The hatched area indicates the converted $M_{200}/L_{200}$ of the maxBCG clusters from \citet{Sheldon09III}. The $M_{200}/L_{200}$ for individual galaxies at low luminosities are naturally extended to the ratios for the maxBCG clusters. }
  \label{plot_ml_l200}
\end{figure}

\indent A large number of the galaxies in our brightest luminosity bins reside in groups or small clusters. To identify those lenses, we cross-correlate our lens sample with the preliminary RCS2 cluster catalogue, to be presented in a future publication. We take galaxies with a separation $<250h^{-1}$kpc from the cluster centre, and within 0.05 from the cluster redshift, to be cluster members. Using these criteria, we find that from L5 to L7, 3\%, 26\%, 43\% of the late-type galaxies, and from L5 to L8, 12\%, 31\%, 48\% and 66\% of the early-type galaxies can be associated with clusters. The best fit halo mass of these galaxies is the mass of the group or cluster within $r_{200}$, while the luminosity is only measured for the lens galaxy. The resulting mass-to-light ratio, shown in Figure \ref{plot_lum_mh}b, is therefore higher than what we would measure for the individual galaxies, or for the clusters. \\
\indent To obtain the mass-to-light ratios of the groups and clusters, we estimate the amount of additional luminosity coming from other cluster members within $r_{200}$. We assume that the spectral energy distributions (SEDs) of the galaxies physically associated with the lens are similar to the SED of the lens, and convert their apparent magnitudes to absolute magnitudes using the same conversion that has been used for the lenses. The apparent magnitudes we use are those from the photometric catalogues from \citet{Gilbank10}. As these catalogues do not cover all fields (e.g. the fields in the uncompleted patch 1303), only $\sim$90\% of the lenses are used for the calculation of $L_{200}$. We measure the source galaxy overdensity as in Section \ref{sec_contam} using all the galaxies with $m_{\rm{low}}<m_r<24$, where $m_{\rm{low}}$ is the magnitude of the brightest galaxy that resides at the lens redshift, and calculate the mean luminosity overdensity as a function of lens-source separation. $m_{\rm{low}}$ is determined by selecting the brightest galaxy in the photometric redshift catalogues from \citet{Ilbert06} that resides at the redshift of the lens or higher. We sum the luminosity overdensity to $r_{200}$ and add it to the lens luminosity to obtain the total luminosity within $r_{200}$, $L_{200}$. To make sure that we do not miss a signicant fraction of $L_{200}$ from galaxies with $m_r>24$, we also calculate $L_{200}$ using an upper limit of 23.5, and find that the results do not change significantly. The values of $L_{200}$ are given in Table \ref{tab_lens_lum}. We show the mass-to-light ratio $M_{200}/L_{200}$ as a function of $L_{200}$ in Figure \ref{plot_ml_l200}. For $L_{200}<10^{11}L_{\odot}$ we calculate the weighted mean, and find a value of $M_{200}/L_{200}=42\pm10$ for early-type galaxies, whilst $M_{200}/L_{200}= 17\pm9$ for late-type galaxies. The total mass-to-light ratio increases with $L_{200}$ for the early-types to $\sim$180 at $L_{200}=5\times10^{11}L_{\odot}$. The total mass-to-light ratio is roughly a factor of two larger for early-types than for late-types. This suggests that the difference in the best fit halo mass between early- and late-types for a given luminosity is not solely due to the fact that early-types reside in denser environments, but is at least partly intrinsic. The value of $L_{200}$ for the L7 late-type bin could not be robustly determined, and is excluded from the results. \\
\indent We compare our results to the $M_{200}/L_{200}$ from \citet{Sheldon09III,Sheldon09I} which have been determined for the clusters in the maxBCG catalogue \citep{Koester07}. The quoted values of $L_{200}$ in their work have been measured in the $i$-band, and are calculated using a k-correction to $z=0.25$. We convert them to the $r$-band luminosities we use by accounting for the mean difference between $i$-band and $r$-band absolute magnitudes of early-type galaxies at $z=0.25$, the mean difference between the k-corrections to $z=0.25$ and $z=0.0$, and the difference between the $i$-band and $r$-band solar magnitudes. The final conversion factor is small as the corrections partly cancel each other, and we convert their luminosities to our definition by multiplying them by 1.06. Note that we do not account for differences in the redshift evolution of the luminosities, as it is not mentioned in \citet{Sheldon09III} which correction, if any, they have used. The converted $M_{200}/L_{200}$ from \citet{Sheldon09III} are indicated with the hatched area in Figure \ref{plot_ml_l200}. The mass-to-light ratios overlap, and the ratios we have determined, for individual galaxies at low luminosities, and for galaxy groups and small clusters at high luminosities, are naturally extended to the $M_{200}/L_{200}$ of clusters from the maxBCG cluster sample.

\subsection{Satellite fraction \label{sec_lum_sat}}
\begin{figure}
  \resizebox{\hsize}{!}{\includegraphics{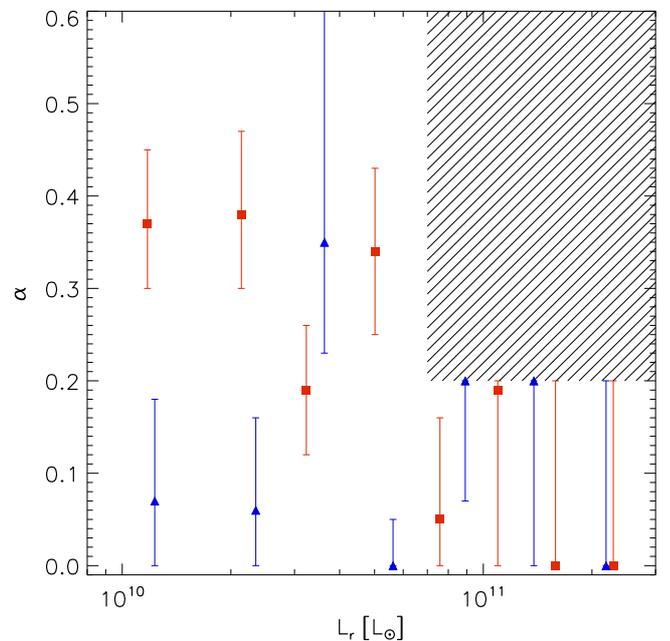}}
  \caption{The best fit satellite fraction as a function of mean luminosity. The red squares (blue triangles) denote the satellite fraction for the early-type (late-type) galaxies. The satellite fraction decreases with luminosity for the early-types, and no trend is observed for the late-types. The dashed area indicates the area excluded by the prior on the satellite fraction.}
  \label{plot_lum_alpha}
\end{figure}
\indent Figure \ref{plot_lum_alpha} shows the best fit satellite fraction as a function of luminosity. The satellite fraction is decreasing with increasing luminosity for the early-type galaxies, from $\sim40$\% at $L_r = 10^{10} L_\odot$ to $<10$\% at $L_r =10^{11} L_\odot$. For the late-type galaxies, no clear trend with luminosity is observed, and the satellite fraction has a value of 0-20\%. The satellite fractions are not well constrained for the highest luminosity bins. As demonstrated in Appendix \ref{ap_sat}, the sum of the halo model satellite terms has the same shape as the central term at the high halo mass end. As a result, the halo model fit cannot discriminate between the two profiles. The implementation of a more sophisticated description of the truncation of the subhaloes is necessary to improve the constraints on the satellite fraction at the high luminosity/stellar mass end. For instance, recent work by \citet{Limousin09} suggests that massive early-type satellite galaxies are stripped of a far larger fraction of their dark matter than the 50\% we have assumed so far, and we discuss the implications in Appendix \ref{ap_sat}. \\
\indent \citet{Mandelbaum06} find a satellite fraction of 10-15\% for late-type galaxies, independent of stellar mass or luminosity. The satellite fraction for early-types decreases with luminosity from 27\% at $\langle \tilde{L}/\tilde{L}_0\rangle=1.1$ to 15\% at $\langle \tilde{L}/\tilde{L}_0\rangle=4.9$, and both trends are consistent with our findings.

\section{STELLAR MASS RESULTS \label{sec_mstel}}

\begin{figure*}
  \centering
   \includegraphics[width=17cm]{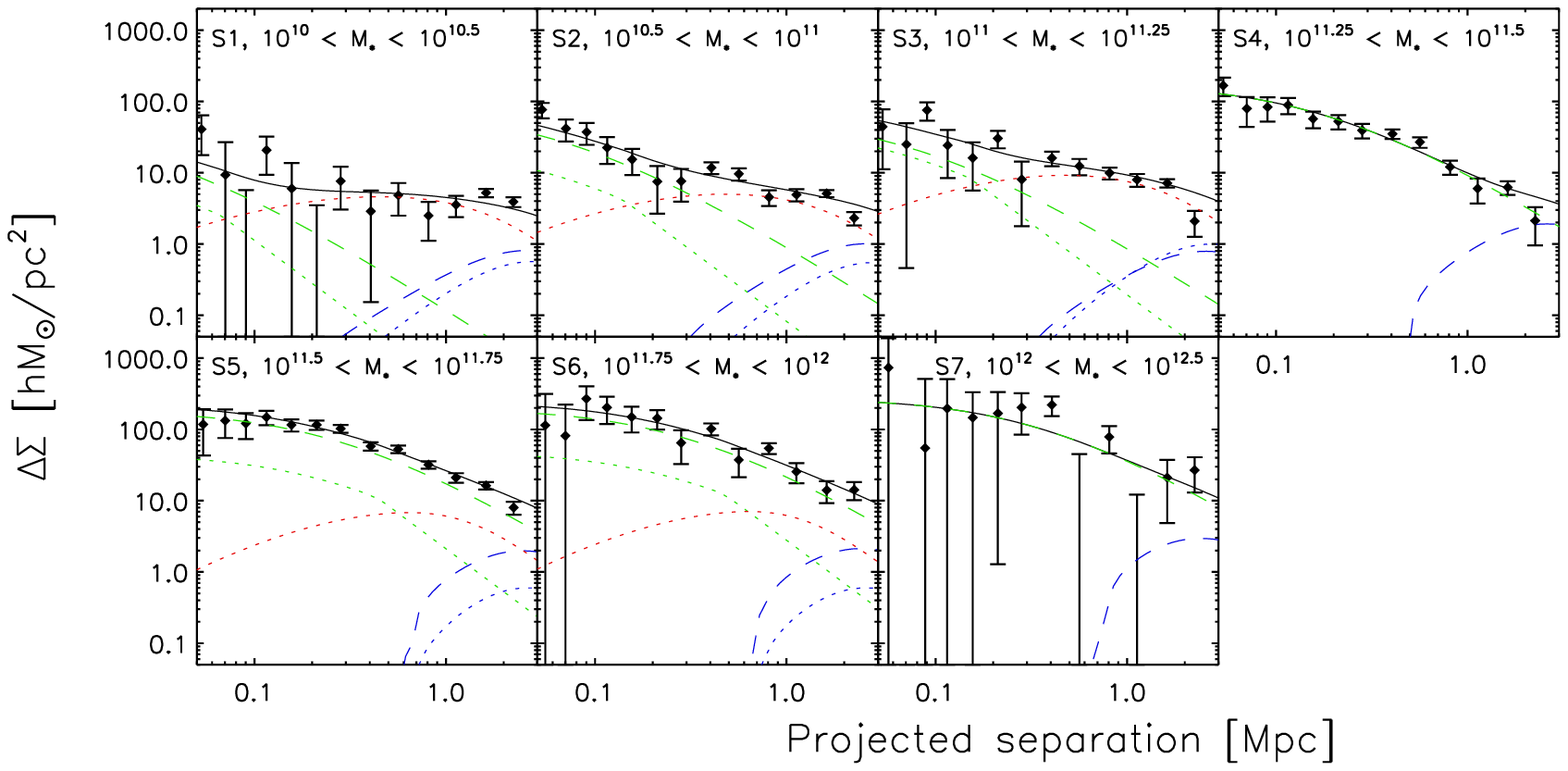}
   \caption{The lensing signal $\Delta \Sigma$ for each stellar mass bin as a function of physical distance from the lens. The green dashed line shows the $\gamma_{t,\mathrm{cent}}^{1h}$ term, the blue dashed line the $\gamma_{t,\mathrm{cent}}^{2h}$ term, the green dotted line the $\gamma_{t,\mathrm{sat}}^{\mathrm{trunc}}$ term, the red dotted line the $\gamma_{t,\mathrm{sat}}^{\mathrm{1h}}$ term, the blue dotted line the $\gamma_{t,\mathrm{sat}}^{\mathrm{2h}}$ term, and the black line shows the sum of the terms. Similar to the lensing signal of the luminosity bins, we find that the $\gamma_{t,sat}^{1h}$ term causes a clearly noticeable bump at $\sim1$Mpc in the lensing signal for the low stellar mass bins, which indicates that a significant fraction of these galaxies are satellites.}
  \label{plot_mstel_all}
\end{figure*}

\indent The stellar mass of a galaxy is believed to be a better tracer of the baryonic content of a galaxy than the luminosity, as it is less sensitive to recent star formation. Therefore, we divide our lens sample into seven stellar mass bins and study the lensing signal. The details of the samples are listed in Table \ref{tab_lens_mstel}. Figure \ref{plot_mstel_all} shows the lensing signal of the stacked lenses in each bin, together with the best fit halo model. Similar to the luminosity results, we find that the lensing signal increases with stellar mass, and observe the presence of the $\gamma^{1h}_{t,sat}$ bump for the lower stellar mass bins. We split the lens sample into early- and late-types using the $frac\_deV$ parameter as before, and study the signals separately. \\
\indent To interpret the results, we have to account for a number of issues. The random stellar mass errors are about 0.1 dex, independent of stellar mass, and do not include the systematic error. The random error determines the scattering of lenses amongst bins, and its value is large compared to the bin width. We calculate the bias resulting from this scatter in Appendix \ref{ap_scat}, and find that the best fit halo masses have to be corrected with a factor ranging between $0.9-1.4$. Once corrected for the scatter, we convert the lensing mass to the mean halo mass. This correction has already been introduced in Section \ref{sec_lumi}, and we discuss in Appendix \ref{ap_width} how we calculate it. We increase the corrected halo mass accordingly to obtain the mean halo mass (see Table \ref{tab_widthcorr} for details). \\
\begin{table*}
  \caption{The stellar mass results}             
  \centering
  \begin{tabular}{c c c c c c c c c c c c} 
  \hline \\ 
  \vspace{1mm}
  Sample & log$(M_*)$ & $n_{\rm{lens}}$ & $\langle z \rangle$ & $\langle M_* \rangle$ & $f_{\rm{late}}$ & $M_h^{\rm{early}}$ & $\alpha^{\rm{early}}$ & $M_{*,200}^{\rm{early}}$ &  $M_h^{\rm{late}}$ &  $\alpha^{\rm{late}}$ & $M_{*,200}^{\rm{late}}$\\
   & (1) & (2) & (3) & (4) & (5) & (6) & (7) & (8) & (9) & (10) & (11) \\
 \hline \hline\\ 
  S1  &    $[10.00,10.50]$ & $3 \hspace{0.5mm}359$ & 0.08 & 2.04 & 0.51 & $1.35^{+2.30}_{-1.34}$ & $0.44^{+0.21}_{-0.09}$ & 2.15$\pm$0.004 & $0.56^{+1.66}_{-0.55}$ & $0.10^{+0.11}_{-0.07}$ & 2.04$\pm$0.001 \\
  S2  &     $[10.50,11.00]$ & $5 \hspace{0.5mm}870$ & 0.11 & 6.03 & 0.28 & $8.39^{+2.83}_{-2.41}$ & $0.19^{+0.04}_{-0.04}$ & 7.66$\pm$0.02 & $10.7^{+5.11}_{-4.68}$ & $0.12^{+0.08}_{-0.08}$ & 6.75$\pm$0.01\\
  S3  &     $[11.00,11.25]$ & $2 \hspace{0.5mm}428$ & 0.15 & 13.2 & 0.10 & $14.0^{+6.21}_{-5.30}$ & $0.44^{+0.09}_{-0.08}$ & 16.8$\pm$0.1 & $1.49^{+17.3}_{-1.48}$ & $0.40^{+0.46}_{-0.31}$ & 14.4$\pm$0.1 \\
  S4  &    $[11.25,11.50]$ & $1 \hspace{0.5mm}631$ & 0.20 & 24.0 & 0.05 & $135^{+12.5}_{-21.1}$ & $0.00^{+0.07}_{-0.00}$ & 40.3$\pm$0.3 &  $2.66^{+31.5}_{-2.65}$ & $0.70^{+0.30}_{-0.51}$ & 27.3$\pm$0.4 \\
  S5  &     $[11.50,11.75]$ & $1 \hspace{0.5mm}505$ & 0.34 & 41.7 & 0.03 & $400^{+101}_{-79.6}$ & $0.20^{+0.00}_{-0.20}$ & 88.9$\pm$1.1 & $65.2^{+107}_{-65.2}$ & $0.20^{+0.00}_{-0.20}$ & 71.0$\pm$4.6 \\
  S6  &     $[11.75,12.00]$ & 396 & 0.41 & 69.2 & 0.05 & $640^{+139}_{-281}$ & $0.06^{+0.14}_{-0.06}$ & 230$\pm$6 & $221^{+353}_{-221}$ & $0.20^{+0.00}_{-0.20}$ & - \\
  S7  &    $[12.00,12.50]$ & 48 & 0.48 & 123 & 0.02 & $722^{+531}_{-517}$ & $0.20^{+0.00}_{-0.00}$ & 385$\pm$28 &  -  &  - & -\\
  \hline
  \end{tabular}
  \tablefoot{(1) the stellar mass range of the bin; (2) the number of lenses; (3) the mean redshift; (4) the mean stellar mass in units of $10^{10}M_{\odot}$; (5) the fraction of late-type galaxies; (6) the best fit halo mass for the early-types in units of $10^{11}h^{-1}M_{\odot}$; (7) the best fit satellite fraction for the early-types; (8) the total stellar mass within $r_{200}$ for the early-types in units of $10^{10}M_{\odot}$; (9) the best fit halo mass for the late-types in units of $10^{11}h^{-1}M_{\odot}$; (10) the best fit satellite fraction for the late-types; (11) the total stellar mass within $r_{200}$ for the late-types in units of $10^{10}M_{\odot}$ .}
  \label{tab_lens_mstel}
\end{table*}
\begin{figure}
   \resizebox{\hsize}{!}{\includegraphics{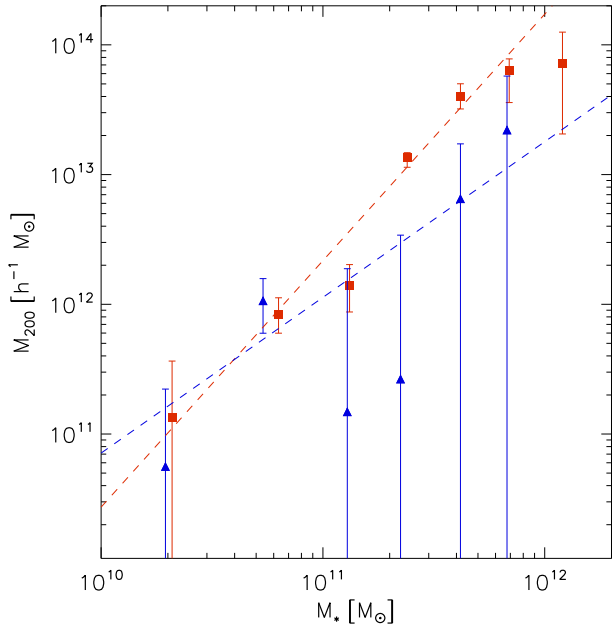}}
   \caption{The best fit halo mass as a function of mean stellar mass. The red squares (blue triangles) denote the early-type (late-type) galaxies. The separation of the lenses into early/late-types is based on their brightness profiles. The dashed lines are the powerlaw fits, with values as indicated in the text. For stellar masses lower than $10^{11}M_{\odot}$ the best fit halo masses of early- and late-type galaxies are similar, but for $M_*>10^{11}M_{\odot}$ we find that the best fit halo masses of early-types are greater. }
  \label{plot_mstel_mhalo}
\end{figure}
\indent The resulting halo masses are given in Table \ref{tab_lens_mstel}, and shown in Figure \ref{plot_mstel_mhalo}. This figure shows that the relation is different for early-types and late-types. Below a stellar mass of $10^{11}M_\odot$, the halo mass is similar for both galaxy types, but for stellar masses larger than $10^{11}M_\odot$ the halo masses of early-type galaxies are more massive for a given stellar mass than the halo masses of late-type galaxies, and increase more steeply with stellar mass. These trends in the stellar mass to halo mass relation are in agreement with those found by \citet{Mandelbaum06}.\\
\begin{figure}
  \resizebox{\hsize}{!}{\includegraphics[angle=-90]{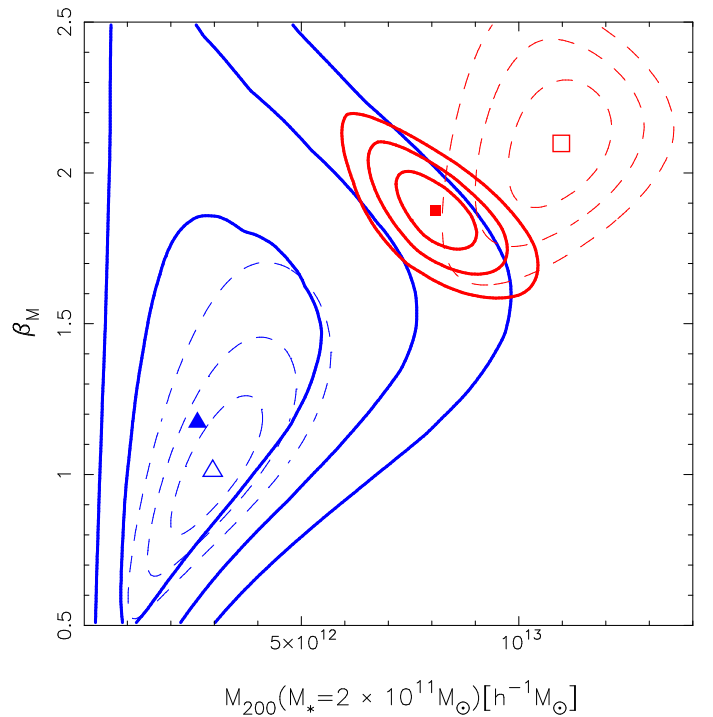}}
  \caption{The 67.8\%, 95.4\% and 99.7\% confidence limits of the powerlaw fits between stellar mass and halo mass, in red (blue) for the early-type (late-type) galaxies. The solid contour lines result from fitting the powerlaw to all the lensing data as described in the text. The dashed contours are the results from fitting the powerlaw between stellar mass and halo mass to the measurements in \citet{Mandelbaum06}. }
  \label{plot_chisq_mstel}
\end{figure}
\indent We fit a powerlaw of the form $M_{200}=M_{0,M} (M_*/M_{0})^{\beta_M}$, with $M_{0}=2\times10^{11}M_{\odot}$, fitting the lensing measurements simultaneously as we did for the luminosities. For the early-types, we find $M_{0,M}=8.1\pm0.6\times 10^{12}h^{-1}M_{\odot}$ and $\beta_M=1.9\pm0.1$, and for the late-types $M_{0,M}=2.6^{+1.8}_{-1.3}\times 10^{12}h^{-1}M_{\odot}$ and $\beta_M=1.2\pm0.4$. These fits are shown in Figure \ref{plot_mstel_mhalo} as the dashed red and blue lines for the early- and late-type galaxies respectively. We show the 67.8\%, 95.4\% and 99.7\% confidence limits of the two powerlaw fits in Figure \ref{plot_chisq_mstel}. \\
\indent In order to compare with the results of \citet{Mandelbaum06}, we lower their halo masses by 30\% to account for the difference between $M_{\mathrm{vir}}$ and $M_{200}$. We compare the halo masses of the bins with comparable mean stellar mass, and find that the best fit halo masses generally agree well. We fit a powerlaw between stellar mass and halo mass to their results, and the dashed contours in Figure \ref{plot_chisq_mstel} show the resulting best fit normalisation and slope. The results of the late-types agree, although the errors are large. For the early-types, \citet{Mandelbaum06} find a somewhat steeper slope and a higher offset. Note that this difference is mostly driven by their highest stellar mass bin, for which they fit a halo mass that is 50\% larger than what we find for our corresponding bin. If we exclude that point from the fit, the 1$\sigma$ contours overlap. \\
\indent \citet{Moster10} used numerical simulations to predict the relation between stellar mass and halo mass. We find that for $M_*<4\times 10^{11}M_{\odot}$, the halo masses we have determined are about 1-2 $\sigma$ lower than their models. At higher stellar masses, the discrepancy is significantly larger. Not only their model, but also the models of various other groups \cite[e.g.][]{Wang06,Croton06,Somerville08,Behroozi10,Neistein11} predict that the halo masses of galaxies with a stellar mass $>10^{11}M_{\odot}$ increases rapidly as a function of stellar mass, a trend we do not observe in our measurements. This would imply that the predicted relation between stellar mass and halo mass for galaxies with $M_*>4\times10^{11}M_{\odot}$ is too steep, possibly because the relation has not yet been well constrained by observations in this mass range. Although contamination of the high stellar mass bins by unresolved mergers may bias the best fit halo masses low, we estimate that this is not sufficient to explain the discrepancy.

\subsection{Baryon conversion efficiency}
\begin{figure}
   \resizebox{\hsize}{!}{\includegraphics{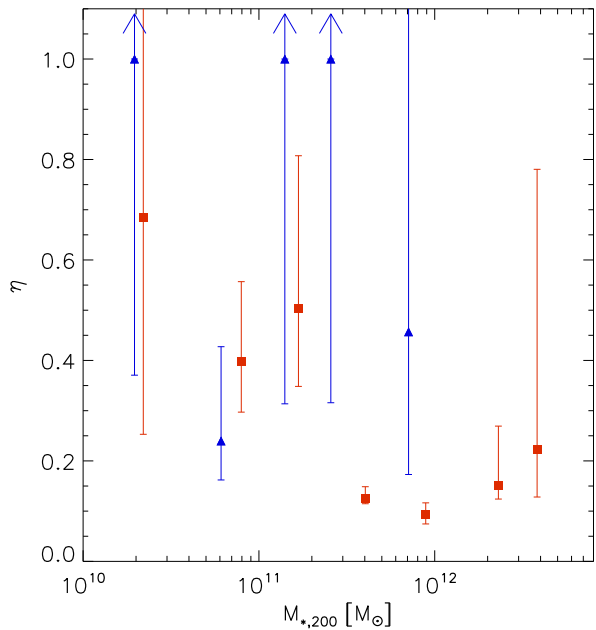}}
   \caption{The baryon conversion efficiency $\eta$ as a function of $M_{*,200}$, the stellar mass enclosed within $r_{200}$. The red squares (blue triangles) denote the early-type (late-type) galaxies. $\eta$ is smaller for the early-types than for the late-types for $M_{*,200}>10^{11}M_{\odot}$. }
  \label{plot_mstel_eta}
\end{figure}

To study the efficiency of star formation as a function of stellar mass, we measure the baryon conversion efficiency $\eta=M_*/(M_h \times f_b)$, where $f_b=\Omega_b / \Omega_M$ is the cosmological baryon fraction. We cannot simply use the mean stellar and halo mass, because we measure the halo mass of the environment where the galaxy resides. The mean stellar mass, however, is determined using the individual lenses only, which leads to an underestimation of $\eta$. To account for this, we estimate the additional amount of stellar mass within $r_{200}$ assuming that the SEDs of the cluster members are similar to that of the lens galaxy. Under that assumption we determine $M_{*,200}=\langle M_*\rangle \times(L_{200}/ L_r)$, where $L_{200}$ is the total luminosity within $r_{200}$ as discussed in Section \ref{sec_lum_ml}. The error bars assume that the number of source galaxies in each radial bin follows a Poisson distribution. We give the values of $M_{*,200}$ in Table \ref{tab_lens_mstel}, and plot $\eta$ as a function of $M_{*,200}$ in Figure \ref{plot_mstel_eta}. \\
\indent The stars that make up the diffuse intracluster light (ICL) also contribute to the total stellar mass. The ICL typically makes up 10--20\% of the stellar light in galaxy groups and clusters \citep[see][and references therein]{Giodini09}. We do not account for the additional stellar mass from the ICL, because our lens sample consists of a mixture of isolated galaxies and galaxies in groups and clusters. The average contribution from the ICL is hard to determine, particularly because the contribution for low mass structures is very uncertain. The ICL is expected to be of importance for the S6 and S7 bins only, as they contain the largest fraction of cluster associated galaxies, and the derived values of $\eta$ might at most increase with 10--20\%. \\
\indent We find that $\eta$ decreases from $\sim$40\% at $M_*\sim5\times10^{10} M_{\odot}$ to a minimum of $\sim$10\% for a stellar mass $M_{*,200}=10^{12}M_{\odot}$, and seems to increase again at higher stellar masses. The baryon conversion efficiency for the late-types is higher, and no clear trend is observable because of the large errors. For some bins $\eta$ is larger than unity, but the error bars cover the reasonable range of $\eta<1$. The value of $M_{*,200}$ for the S6 late-type bin could not be robustly determined, and is excluded from the results. \\
\indent \citet{Hoekstra05} divide the lens sample in red and blue galaxies based on their $B-V$ colour, and find that the baryon conversion efficiency for isolated blue galaxies in the magnitude range $18<R_C<24$ is about twice the value found for isolated red galaxies. Although we cannot compare the results in detail due to differences in the type selection and differences in the adopted IMF, our results also suggest a larger value for $\eta$ for late-type galaxies in the range $M_{*,200}>10^{11}M_{\odot}$. A similar trend has also been observed in \citet{Mandelbaum06} for stellar masses $M_*>10^{11}$, but note that the baryon conversion efficiencies were determined using $M_*$ instead of $M_{*,200}$, and the values are therefore lower limits. \\

\subsection{Satellite fraction}
\begin{figure}
   \resizebox{\hsize}{!}{\includegraphics{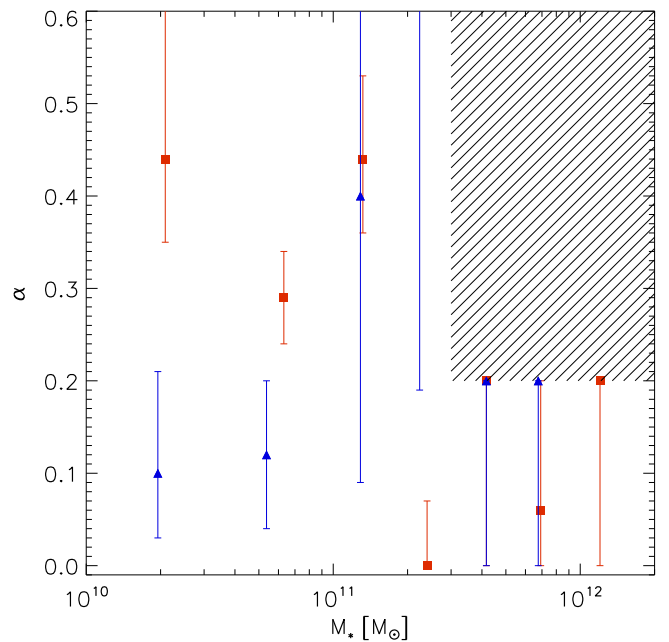}}
   \caption{The best fit satellite fraction as a function of the mean stellar mass. The red squares (blue triangles) denote the early-type (late-type) results. The satellite fraction decreases with stellar mass for the early-types, and no trend is observed for the late-types. The dashed area indicates the area excluded by the prior on the satellite fraction.}
  \label{plot_mstel_alpha}
\end{figure}

In Figure \ref{plot_mstel_alpha} we show the satellite fraction for the early- and late-type galaxies as a function of stellar mass. The satellite fraction of the late-types is only well determined for the S1 and S2 bins, and appears to be constant as a function of stellar mass, with a value of $\sim$10\%. The satellite fraction of the early-types is 45\% for the lowest stellar mass bin, but decreases to $<10$\% for $M_*\geq2\times10^{11} M_{\odot}$.
\citet{Mandelbaum06} find a satellite fraction of about 10-15\% for late-type galaxies, independent of stellar mass or luminosity. For the early-types, \citet{Mandelbaum06} find that the satellite fraction decreases with stellar mass from 50\% at $10^{10}M_{\odot}$ to roughly 10\% at $3\times10^{11}M_{\odot}$, consistent with our findings.

\subsection{Dependence on redshift}
\begin{figure}
  \resizebox{\hsize}{!}{\includegraphics{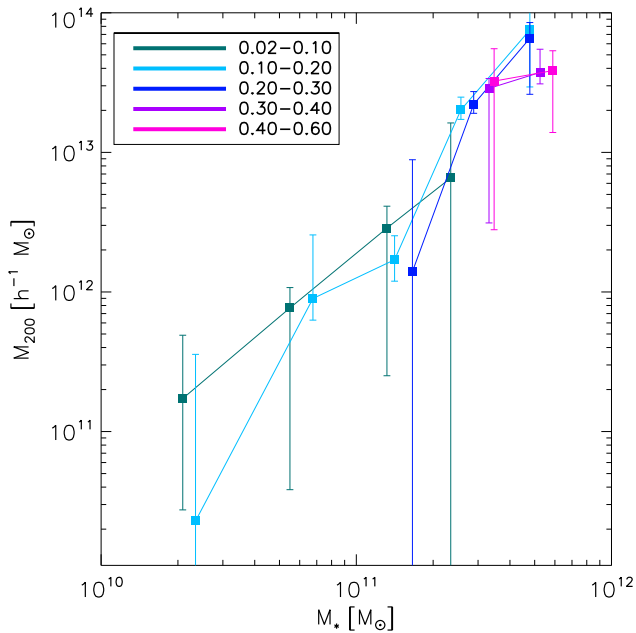}}
  \caption{The halo mass as a function of the mean stellar mass for early-type galaxies in different redshift slices. Although we lack the statistical precision to draw definite conclusions, the measurements support the view that at the high stellar mass end, galaxies at a higher redshift have lower halo masses.}
  \label{plot_mh_z}
\end{figure}

The stellar mass of a galaxy and the dark matter content of its halo evolve with time. The stellar mass increases as galaxies form stars and merge with satellites and other galaxies. Satellite galaxies residing in subhaloes are tidally stripped of their dark matter, whilst the dark matter content of central haloes increases due to mergers. The evolution of the relation between stellar mass and dark matter content of galaxies has been studied with numerical simulations \citep[e.g.][]{Moster10,ConroyW09}. These simulations predict that the dark matter content of haloes that host galaxies of $M_*>10^{11}M_{\odot}$ increases faster than the stellar mass, while the stellar mass grows faster for haloes hosting galaxies of $M_*<10^{11}M_{\odot}$. \\
\indent To study this, we bin the early-type galaxies in stellar mass and redshift, and measure their halo mass. To avoid the degeneracy between halo mass and satellite fraction affecting the results, we fix the satellite fraction to the value we find by fitting the halo model to all lenses in each stellar mass bin. We apply the various corrections (e.g. scattering of lenses between bins), and show the results in Figure \ref{plot_mh_z}. \\
\indent The errors on the best fit halo masses are large, and we therefore do not obtain tight constraints on the evolution of the halo masses for the low stellar mass bins. For the highest stellar mass bin, however, it appears that the halo mass is smaller by roughly a factor of two for the two highest redshift slices. The redshift dependent stellar-to-halo mass relation of \citet{Moster10} predicts that at $M_*=6\times10^{11}M_{\odot}$, the halo mass increases by $\sim35$\% between $z=0.5$ and $z=0.0$. In \citet{Leauthaud11}, the evolution of the stellar-to-halo mass relation from $z=1$ to $z=0.2$ is studied using a combined galaxy-galaxy weak lensing, galaxy spatial clustering, and galaxy number densities analysis in the COSMOS survey \citep{Scoville07}. At stellar masses $M_*>10^{11}$, the halo mass appears to decrease with redshift for a given stellar mass, but the small volume probed by COSMOS prevents a clear detection. \citet{Brown08} study the growth of the dark matter content of massive early-type galaxies between a redshift of 0.0 and 1.0 by measuring the space density and spatial clustering of the galaxies. They find that between redshift $z=1.0$ and $z=0.0$, the dark matter haloes grow with $\sim$100\%, while the stellar masses of these galaxies only grow with $\sim$30\%. \citet{Conroy07} utilizes the motions of satellite galaxies around isolated galaxies to constrain the evolution of the virial-to-stellar mass ratio, and they find that between $z\sim1$ and $z\sim0$ this ratio remains constant for host galaxies with a stellar mass below $1.5\times10^{11}M_{\odot}$, but increases by a factor $3.3\pm2.2$ for hosts with $M_*>1.5\times10^{11}M_{\odot}$. These findings are in qualitative agreement with our results.


\section{CONCLUSIONS \label{sec_conc}}
We measured the halo masses for early- and late-type galaxies and compared these to their luminosity and stellar mass. For this purpose, we measured the weak lensing signal induced by the galaxies with SDSS spectroscopy that overlap with the RCS2, and modelled the data with a halo model. This enabled us to improve the constraints on the lensing measurements for the most massive galaxies, which typically reside at redshifts where the SDSS is not very sensitive.  \\
\indent The halo mass and the dynamical mass correlate well for early-type galaxies, but not for late-type galaxies. A likely explanation is that late-type galaxies are rotating, resulting in an overestimation of the velocity dispersion, and hence of the dynamical mass. Furthermore, in contrast to the dynamical mass, the weak lensing mass can easily be related to numerical simulations, and provides constraints for the models that describe the relationship between baryons and dark matter. \\
\indent The halo masses of galaxies increase with luminosity and stellar mass. For a given luminosity, the halo mass of the early-types is on average about five times larger than the late-types. We fitted a powerlaw relation between the luminosity and halo mass, and find that in the range $10^{10}<L_r<10^{11.5} L_{\odot}$, the halo mass scales with luminosity as $M_h \propto L^{2.34^{+0.09}_{-0.16}}$ and $M_h \propto L^{2.2^{+0.7}_{-0.6}}$ for the early- and late-type galaxies respectively. For an early-type galaxy with a fiducal luminosity $L_0=10^{11}L_{r,\odot}$, we obtain a mass $M_{200}=(1.93^{+0.13}_{-0.14})\times10^{13}h^{-1}M_{\odot}$. We computed $L_{200}$, the additional luminosity around the lenses within $r_{200}$, and find that the $M_{200}/L_{200}$ ratio of the early-types is larger than for the late-types: for $L_{200}<10^{11}L_{\odot}$ we find $M_{200}/L_{200}=42\pm10$ for early-types, whilst $M_{200}/L_{200}= 17\pm9$ for late-types. This suggests that the difference in halo mass is not solely due to the fact that early-types reside in denser environments, but is at least partly intrinsic. \\
\indent Below a stellar mass of $10^{11}M_{\odot}$ the halo mass of early- and late-types are comparable. For larger stellar masses, the best fit halo masses of the early-types are larger than the late-types. We computed $M_{*,200}$, the total stellar mass within $r_{200}$, in order to calculate the baryon conversion efficiency $\eta$. Our results for early-type galaxies suggest a variation in efficiency with a minimum of $\sim$10\% for a stellar mass $M_{*,200}=10^{12}M_{\odot}$. The results for the late-type galaxies are not well constrained, but do suggest a larger value. \\ 
\indent The satellite fraction is $\sim$40\% for the low luminosity (stellar mass) early-type galaxies, and decreases rapidly to $<10$\% with increasing luminosity (stellar mass). The satellite fraction of the late-types has a value in the range 0-15\%, independent of luminosity or stellar mass. The satellite fraction is difficult to constrain at the high stellar mass/luminosity end, as the shape of the combined shear signal from the satellites mimics an NFW profile. Decreasing the truncation parameter leads to tighter constraints, and appears to be justified for the most massive early-type satellites based on the N-body simulation results of \citet{Limousin09}. Additional support comes from studying the shear signal of massive early-type galaxies that were selected to be satellites shown in Appendix \ref{ap_sat}, but the errors are currently too large to constrain the fraction of dark matter that is stripped. A more realistic description of the stripping of the haloes of massive satellite galaxies may result in an improvement of the constraints on the satellite fraction from weak lensing studies alone. \\
\indent The halo mass appears to decrease with redshift for the highest stellar mass bins, a trend that is qualitatively in agreement with predictions from numerical simulations. The signal-to-noise on the measurements is currently too low to provide a detailed view on the growth of dark matter haloes, but it shows that with future surveys weak lensing can be used to study in great detail the evolution of the relation between baryons and dark matter.

\paragraph{Acknowledgements \\ \\} 
We would like to thank Jarle Brinchman for his extensive help with the MPA/JHU catalogues, Alexie Leauthaud for useful discussions and suggestions concerning the halo model, and Thomas Erben and Hendrik Hildebrandt for their help with the THELI pipeline. Additionally, we would like to thank our colleagues at Leiden Observatory, in particular Konrad Kuijken and Elisabetta Semboloni, and Tim Schrabback, for useful suggestions and comments. Finally, we would like the anonymous referee for useful comments. \\
\indent H.H. acknowledges support from a Marie Curie International Reintegration Grant, and from a VIDI grant from the Nederlandse Organisatie voor Wetenschappelijk Onderzoek (NWO). The RCS2 project is supported in part by grants to H.K.C.Y. from the Canada Research Chairs program and the Natural Science and Engineering Research Council of Canada. \\
\indent This work is based on observations obtained with MegaPrime/MegaCam, a joint project of CFHT and CEA/DAPNIA, at the Canada-France-Hawaii Telescope (CFHT) which is operated by the National Research Council (NRC) of Canada, the Institute National des Sciences de l'Univers of the Centre National de la Recherche Scientifique of France, and the University of Hawaii. We used the facilities of the Canadian Astronomy Data Centre operated by the NRC with the support of the Canadian Space Agency. 

\bibliographystyle{aa}

\begin{thebibliography}{99}
\expandafter\ifx\csname natexlab\endcsname\relax\def\natexlab#1{#1}\fi

\bibitem[{{Abazajian} {et~al.}(2009){Abazajian}, {Adelman-McCarthy},
  {Ag{\"u}eros}, {Allam}, {Allende Prieto}, {An}, {Anderson}, {Anderson},
  {Annis}, {Bahcall}, \& et~al.}]{Abazajian09}
{Abazajian}, K.~N., {Adelman-McCarthy}, J.~K., {Ag{\"u}eros}, M.~A., {et~al.}
  2009, \apjs, 182, 543

\bibitem[{{Adelman-McCarthy} {et~al.}(2008){Adelman-McCarthy}, {Ag{\"u}eros},
  {Allam}, {Allende Prieto}, {Anderson}, {Anderson}, {Annis}, {Bahcall},
  {Bailer-Jones}, {Baldry}, {Barentine}, {Bassett}, {Becker}, {Beers}, {Bell},
  {Berlind}, {Bernardi}, {Blanton}, {Bochanski}, {Boroski}, {Brinchmann},
  {Brinkmann}, {Brunner}, {Budav{\'a}ri}, {Carliles}, {Carr}, {Castander},
  {Cinabro}, {Cool}, {Covey}, {Csabai}, {Cunha}, {Davenport}, {Dilday}, {Doi},
  {Eisenstein}, {Evans}, {Fan}, {Finkbeiner}, {Friedman}, {Frieman},
  {Fukugita}, {G{\"a}nsicke}, {Gates}, {Gillespie}, {Glazebrook}, {Gray},
  {Grebel}, {Gunn}, {Gurbani}, {Hall}, {Harding}, {Harvanek}, {Hawley},
  {Hayes}, {Heckman}, {Hendry}, {Hindsley}, {Hirata}, {Hogan}, {Hogg}, {Hyde},
  {Ichikawa}, {Ivezi{\'c}}, {Jester}, {Johnson}, {Jorgensen}, {Juri{\'c}},
  {Kent}, {Kessler}, {Kleinman}, {Knapp}, {Kron}, {Krzesinski}, {Kuropatkin},
  {Lamb}, {Lampeitl}, {Lebedeva}, {Lee}, {Leger}, {L{\'e}pine}, {Lima}, {Lin},
  {Long}, {Loomis}, {Loveday}, {Lupton}, {Malanushenko}, {Malanushenko},
  {Mandelbaum}, {Margon}, {Marriner}, {Mart{\'{\i}}nez-Delgado}, {Matsubara},
  {McGehee}, {McKay}, {Meiksin}, {Morrison}, {Munn}, {Nakajima}, {Neilsen},
  {Newberg}, {Nichol}, {Nicinski}, {Nieto-Santisteban}, {Nitta}, {Okamura},
  {Owen}, {Oyaizu}, {Padmanabhan}, {Pan}, {Park}, {Peoples}, {Pier}, {Pope},
  {Purger}, {Raddick}, {Re Fiorentin}, {Richards}, {Richmond}, {Riess}, {Rix},
  {Rockosi}, {Sako}, {Schlegel}, {Schneider}, {Schreiber}, {Schwope}, {Seljak},
  {Sesar}, {Sheldon}, {Shimasaku}, {Sivarani}, {Smith}, {Snedden}, {Steinmetz},
  {Strauss}, {SubbaRao}, {Suto}, {Szalay}, {Szapudi}, {Szkody}, {Tegmark},
  {Thakar}, {Tremonti}, {Tucker}, {Uomoto}, {Vanden Berk}, {Vandenberg},
  {Vidrih}, {Vogeley}, {Voges}, {Vogt}, {Wadadekar}, {Weinberg}, {West},
  {White}, {Wilhite}, {Yanny}, {Yocum}, {York}, {Zehavi}, \&
  {Zucker}}]{Adelman08}
{Adelman-McCarthy}, J.~K., {Ag{\"u}eros}, M.~A., {Allam}, S.~S., {et~al.} 2008,
  \apjs, 175, 297

\bibitem[{{Adelman-McCarthy} {et~al.}(2006){Adelman-McCarthy}, {Ag{\"u}eros},
  {Allam}, {Anderson}, {Anderson}, {Annis}, {Bahcall}, {Baldry}, {Barentine},
  {Berlind}, {Bernardi}, {Blanton}, {Boroski}, {Brewington}, {Brinchmann},
  {Brinkmann}, {Brunner}, {Budav{\'a}ri}, {Carey}, {Carr}, {Castander},
  {Connolly}, {Csabai}, {Czarapata}, {Dalcanton}, {Doi}, {Dong}, {Eisenstein},
  {Evans}, {Fan}, {Finkbeiner}, {Friedman}, {Frieman}, {Fukugita}, {Gillespie},
  {Glazebrook}, {Gray}, {Grebel}, {Gunn}, {Gurbani}, {de Haas}, {Hall},
  {Harris}, {Harvanek}, {Hawley}, {Hayes}, {Hendry}, {Hennessy}, {Hindsley},
  {Hirata}, {Hogan}, {Hogg}, {Holmgren}, {Holtzman}, {Ichikawa}, {Ivezi{\'c}},
  {Jester}, {Johnston}, {Jorgensen}, {Juri{\'c}}, {Kent}, {Kleinman}, {Knapp},
  {Kniazev}, {Kron}, {Krzesinski}, {Kuropatkin}, {Lamb}, {Lampeitl}, {Lee},
  {Leger}, {Lin}, {Long}, {Loveday}, {Lupton}, {Margon},
  {Mart{\'{\i}}nez-Delgado}, {Mandelbaum}, {Matsubara}, {McGehee}, {McKay},
  {Meiksin}, {Munn}, {Nakajima}, {Nash}, {Neilsen}, {Newberg}, {Newman},
  {Nichol}, {Nicinski}, {Nieto-Santisteban}, {Nitta}, {O'Mullane}, {Okamura},
  {Owen}, {Padmanabhan}, {Pauls}, {Peoples}, {Pier}, {Pope}, {Pourbaix},
  {Quinn}, {Richards}, {Richmond}, {Rockosi}, {Schlegel}, {Schneider},
  {Schroeder}, {Scranton}, {Seljak}, {Sheldon}, {Shimasaku}, {Smith}, {Smol{\v
  c}i{\'c}}, {Snedden}, {Stoughton}, {Strauss}, {SubbaRao}, {Szalay},
  {Szapudi}, {Szkody}, {Tegmark}, {Thakar}, {Tucker}, {Uomoto}, {Vanden Berk},
  {Vandenberg}, {Vogeley}, {Voges}, {Vogt}, {Walkowicz}, {Weinberg}, {West},
  {White}, {Xu}, {Yanny}, {Yocum}, {York}, {Zehavi}, {Zibetti}, \&
  {Zucker}}]{Adelman06}
{Adelman-McCarthy}, J.~K., {Ag{\"u}eros}, M.~A., {Allam}, S.~S., {et~al.} 2006,
  \apjs, 162, 38

\bibitem[{{Agustsson} \& {Brainerd}(2006)}]{AgustssonB06}
{Agustsson}, I. \& {Brainerd}, T.~G. 2006, \apjl, 644, L25

\bibitem[{{Ann} {et~al.}(2008){Ann}, {Park}, \& {Choi}}]{Ann08}
{Ann}, H.~B., {Park}, C., \& {Choi}, Y. 2008, \mnras, 389, 86

\bibitem[{{Bartelmann}(1996)}]{Bartelmann96}
{Bartelmann}, M. 1996, \aap, 313, 697

\bibitem[{{Behroozi} {et~al.}(2010){Behroozi}, {Conroy}, \&
  {Wechsler}}]{Behroozi10}
{Behroozi}, P.~S., {Conroy}, C., \& {Wechsler}, R.~H. 2010, \apj, 717, 379

\bibitem[{{Bell} \& {de Jong}(2001)}]{BelldJ01}
{Bell}, E.~F. \& {de Jong}, R.~S. 2001, \apj, 550, 212

\bibitem[{{Bertin}(2006)}]{Bertin06}
{Bertin}, E. 2006, in Astronomical Society of the Pacific Conference Series,
  Vol. 351, Astronomical Data Analysis Software and Systems XV, ed.
  {C.~Gabriel, C.~Arviset, D.~Ponz, \& S.~Enrique}, 112--+

\bibitem[{{Bertin} \& {Arnouts}(1996)}]{BertinA96}
{Bertin}, E. \& {Arnouts}, S. 1996, \aaps, 117, 393

\bibitem[{{Bertin} {et~al.}(2002){Bertin}, {Ciotti}, \& {Del
  Principe}}]{Bertin02}
{Bertin}, G., {Ciotti}, L., \& {Del Principe}, M. 2002, \aap, 386, 149

\bibitem[{{Blanton} {et~al.}(2003){Blanton}, {Brinkmann}, {Csabai}, {Doi},
  {Eisenstein}, {Fukugita}, {Gunn}, {Hogg}, \& {Schlegel}}]{Blanton03}
{Blanton}, M.~R., {Brinkmann}, J., {Csabai}, I., {et~al.} 2003, \aj, 125, 2348

\bibitem[{{Blanton} {et~al.}(2001){Blanton}, {Dalcanton}, {Eisenstein},
  {Loveday}, {Strauss}, {SubbaRao}, {Weinberg}, {Anderson}, {Annis}, {Bahcall},
  {Bernardi}, {Brinkmann}, {Brunner}, {Burles}, {Carey}, {Castander},
  {Connolly}, {Csabai}, {Doi}, {Finkbeiner}, {Friedman}, {Frieman}, {Fukugita},
  {Gunn}, {Hennessy}, {Hindsley}, {Hogg}, {Ichikawa}, {Ivezi{\'c}}, {Kent},
  {Knapp}, {Lamb}, {Leger}, {Long}, {Lupton}, {McKay}, {Meiksin}, {Merelli},
  {Munn}, {Narayanan}, {Newcomb}, {Nichol}, {Okamura}, {Owen}, {Pier}, {Pope},
  {Postman}, {Quinn}, {Rockosi}, {Schlegel}, {Schneider}, {Shimasaku},
  {Siegmund}, {Smee}, {Snir}, {Stoughton}, {Stubbs}, {Szalay}, {Szokoly},
  {Thakar}, {Tremonti}, {Tucker}, {Uomoto}, {Vanden Berk}, {Vogeley},
  {Waddell}, {Yanny}, {Yasuda}, \& {York}}]{Blanton01}
{Blanton}, M.~R., {Dalcanton}, J., {Eisenstein}, D., {et~al.} 2001, \aj, 121,
  2358

\bibitem[{{Blanton} \& {Roweis}(2007)}]{BlantonR07}
{Blanton}, M.~R. \& {Roweis}, S. 2007, \aj, 133, 734

\bibitem[{{Blanton} {et~al.}(2005){Blanton}, {Schlegel}, {Strauss},
  {Brinkmann}, {Finkbeiner}, {Fukugita}, {Gunn}, {Hogg}, {Ivezi{\'c}}, {Knapp},
  {Lupton}, {Munn}, {Schneider}, {Tegmark}, \& {Zehavi}}]{Blanton05}
{Blanton}, M.~R., {Schlegel}, D.~J., {Strauss}, M.~A., {et~al.} 2005, \aj, 129,
  2562

\bibitem[{{Bolton} {et~al.}(2008){Bolton}, {Treu}, {Koopmans}, {Gavazzi},
  {Moustakas}, {Burles}, {Schlegel}, \& {Wayth}}]{Bolton08}
{Bolton}, A.~S., {Treu}, T., {Koopmans}, L.~V.~E., {et~al.} 2008, \apj, 684,
  248

\bibitem[{{Bower} {et~al.}(2010){Bower}, {Vernon}, {Goldstein}, {Benson},
  {Lacey}, {Baugh}, {Cole}, \& {Frenk}}]{Bower10}
{Bower}, R.~G., {Vernon}, I., {Goldstein}, M., {et~al.} 2010, \mnras, 407, 2017

\bibitem[{{Brainerd} {et~al.}(1996){Brainerd}, {Blandford}, \&
  {Smail}}]{Brainerd96}
{Brainerd}, T.~G., {Blandford}, R.~D., \& {Smail}, I. 1996, \apj, 466, 623

\bibitem[{{Brown} {et~al.}(2008){Brown}, {Zheng}, {White}, {Dey}, {Jannuzi},
  {Benson}, {Brand}, {Brodwin}, \& {Croton}}]{Brown08}
{Brown}, M.~J.~I., {Zheng}, Z., {White}, M., {et~al.} 2008, \apj, 682, 937

\bibitem[{{Bruzual} \& {Charlot}(2003)}]{BruzualC03}
{Bruzual}, G. \& {Charlot}, S. 2003, \mnras, 344, 1000

\bibitem[{{Cacciato} {et~al.}(2009){Cacciato}, {van den Bosch}, {More}, {Li},
  {Mo}, \& {Yang}}]{Cacciato09}
{Cacciato}, M., {van den Bosch}, F.~C., {More}, S., {et~al.} 2009, \mnras, 394,
  929

\bibitem[{{Conroy} {et~al.}(2007){Conroy}, {Prada}, {Newman}, {Croton}, {Coil},
  {Conselice}, {Cooper}, {Davis}, {Faber}, {Gerke}, {Guhathakurta}, {Klypin},
  {Koo}, \& {Yan}}]{Conroy07}
{Conroy}, C., {Prada}, F., {Newman}, J.~A., {et~al.} 2007, \apj, 654, 153

\bibitem[{{Conroy} \& {Wechsler}(2009)}]{ConroyW09}
{Conroy}, C. \& {Wechsler}, R.~H. 2009, \apj, 696, 620

\bibitem[{{Cooray} \& {Sheth}(2002)}]{CoorayS02}
{Cooray}, A. \& {Sheth}, R. 2002, \physrep, 372, 1

\bibitem[{{Coupon} {et~al.}(2011){Coupon}, {Kilbinger}, {McCracken}, {Ilbert},
  {Arnouts}, {Mellier}, {Abbas}, {de la Torre}, {Goranova}, {Hudelot}, {Kneib},
  \& {Lefevre}}]{Coupon11}
{Coupon}, J., {Kilbinger}, M., {McCracken}, H.~J., {et~al.} 2011, ArXiv
  e-prints

\bibitem[{{Cresswell} \& {Percival}(2009)}]{Cresswell09}
{Cresswell}, J.~G. \& {Percival}, W.~J. 2009, \mnras, 392, 682

\bibitem[{{Croton} {et~al.}(2006){Croton}, {Springel}, {White}, {De Lucia},
  {Frenk}, {Gao}, {Jenkins}, {Kauffmann}, {Navarro}, \& {Yoshida}}]{Croton06}
{Croton}, D.~J., {Springel}, V., {White}, S.~D.~M., {et~al.} 2006, \mnras, 365,
  11

\bibitem[{{Duffy} {et~al.}(2008){Duffy}, {Schaye}, {Kay}, \& {Dalla
  Vecchia}}]{Duffy08}
{Duffy}, A.~R., {Schaye}, J., {Kay}, S.~T., \& {Dalla Vecchia}, C. 2008,
  \mnras, 390, L64

\bibitem[{{Eisenstein} {et~al.}(2001){Eisenstein}, {Annis}, {Gunn}, {Szalay},
  {Connolly}, {Nichol}, {Bahcall}, {Bernardi}, {Burles}, {Castander},
  {Fukugita}, {Hogg}, {Ivezi{\'c}}, {Knapp}, {Lupton}, {Narayanan}, {Postman},
  {Reichart}, {Richmond}, {Schneider}, {Schlegel}, {Strauss}, {SubbaRao},
  {Tucker}, {Vanden Berk}, {Vogeley}, {Weinberg}, \& {Yanny}}]{Eisenstein01}
{Eisenstein}, D.~J., {Annis}, J., {Gunn}, J.~E., {et~al.} 2001, \aj, 122, 2267

\bibitem[{{Emsellem} {et~al.}(2007){Emsellem}, {Cappellari}, {Krajnovi{\'c}},
  {van de Ven}, {Bacon}, {Bureau}, {Davies}, {de Zeeuw}, {Falc{\'o}n-Barroso},
  {Kuntschner}, {McDermid}, {Peletier}, \& {Sarzi}}]{Emsellem07}
{Emsellem}, E., {Cappellari}, M., {Krajnovi{\'c}}, D., {et~al.} 2007, \mnras,
  379, 401

\bibitem[{{Erben} {et~al.}(2009){Erben}, {Hildebrandt}, {Lerchster}, {Hudelot},
  {Benjamin}, {van Waerbeke}, {Schrabback}, {Brimioulle}, {Cordes}, {Dietrich},
  {Holhjem}, {Schirmer}, \& {Schneider}}]{Erben09}
{Erben}, T., {Hildebrandt}, H., {Lerchster}, M., {et~al.} 2009, \aap, 493, 1197

\bibitem[{{Erben} {et~al.}(2005){Erben}, {Schirmer}, {Dietrich}, {Cordes},
  {Haberzettl}, {Hetterscheidt}, {Hildebrandt}, {Schmithuesen}, {Schneider},
  {Simon}, {Deul}, {Hook}, {Kaiser}, {Radovich}, {Benoist}, {Nonino}, {Olsen},
  {Prandoni}, {Wichmann}, {Zaggia}, {Bomans}, {Dettmar}, \&
  {Miralles}}]{Erben05}
{Erben}, T., {Schirmer}, M., {Dietrich}, J.~P., {et~al.} 2005, Astronomische
  Nachrichten, 326, 432

\bibitem[{{Faltenbacher} {et~al.}(2007){Faltenbacher}, {Li}, {Mao}, {van den
  Bosch}, {Yang}, {Jing}, {Pasquali}, \& {Mo}}]{Faltenbacher07}
{Faltenbacher}, A., {Li}, C., {Mao}, S., {et~al.} 2007, \apjl, 662, L71

\bibitem[{{Fischer} {et~al.}(2000){Fischer}, {McKay}, {Sheldon}, {Connolly},
  {Stebbins}, {Frieman}, {Jain}, {Joffre}, {Johnston}, {Bernstein}, {Annis},
  {Bahcall}, {Brinkmann}, {Carr}, {Csabai}, {Gunn}, {Hennessy}, {Hindsley},
  {Hull}, {Ivezi{\'c}}, {Knapp}, {Limmongkol}, {Lupton}, {Munn}, {Nash},
  {Newberg}, {Owen}, {Pier}, {Rockosi}, {Schneider}, {Smith}, {Stoughton},
  {Szalay}, {Szokoly}, {Thakar}, {Vogeley}, {Waddell}, {Weinberg}, {York}, \&
  {The SDSS Collaboration}}]{Fischer00}
{Fischer}, P., {McKay}, T.~A., {Sheldon}, E., {et~al.} 2000, \aj, 120, 1198

\bibitem[{{Gallazzi} {et~al.}(2005){Gallazzi}, {Charlot}, {Brinchmann},
  {White}, \& {Tremonti}}]{Gallazzi05}
{Gallazzi}, A., {Charlot}, S., {Brinchmann}, J., {White}, S.~D.~M., \&
  {Tremonti}, C.~A. 2005, \mnras, 362, 41

\bibitem[{{Gilbank} {et~al.}(2011){Gilbank}, {Gladders}, {Yee}, \&
  {Hsieh}}]{Gilbank10}
{Gilbank}, D.~G., {Gladders}, M.~D., {Yee}, H.~K.~C., \& {Hsieh}, B.~C. 2011,
  \aj, 141, 94

\bibitem[{{Giodini} {et~al.}(2009){Giodini}, {Pierini}, {Finoguenov}, {Pratt},
  {Boehringer}, {Leauthaud}, {Guzzo}, {Aussel}, {Bolzonella}, {Capak}, {Elvis},
  {Hasinger}, {Ilbert}, {Kartaltepe}, {Koekemoer}, {Lilly}, {Massey},
  {McCracken}, {Rhodes}, {Salvato}, {Sanders}, {Scoville}, {Sasaki}, {Smolcic},
  {Taniguchi}, {Thompson}, \& {the COSMOS Collaboration}}]{Giodini09}
{Giodini}, S., {Pierini}, D., {Finoguenov}, A., {et~al.} 2009, \apj, 703, 982

\bibitem[{{Gladders} \& {Yee}(2005)}]{Gladders05}
{Gladders}, M.~D. \& {Yee}, H.~K.~C. 2005, \apjs, 157, 1

\bibitem[{{Gr{\"u}tzbauch} {et~al.}(2011){Gr{\"u}tzbauch}, {Conselice},
  {Varela}, {Bundy}, {Cooper}, {Skibba}, \& {Willmer}}]{Grutzbauch11}
{Gr{\"u}tzbauch}, R., {Conselice}, C.~J., {Varela}, J., {et~al.} 2011, \mnras,
  411, 929

\bibitem[{{Guo} {et~al.}(2009){Guo}, {McIntosh}, {Mo}, {Katz}, {van den Bosch},
  {Weinberg}, {Weinmann}, {Pasquali}, \& {Yang}}]{Guo09}
{Guo}, Y., {McIntosh}, D.~H., {Mo}, H.~J., {et~al.} 2009, \mnras, 398, 1129

\bibitem[{{Guzik} \& {Seljak}(2001)}]{GuzikS01}
{Guzik}, J. \& {Seljak}, U. 2001, \mnras, 321, 439

\bibitem[{{Guzik} \& {Seljak}(2002)}]{GuzikS02}
{Guzik}, J. \& {Seljak}, U. 2002, \mnras, 335, 311

\bibitem[{{Hao} {et~al.}(2011){Hao}, {Kubo}, {Feldmann}, {Annis}, {Johnston},
  {Lin}, \& {McKay}}]{Hao11}
{Hao}, J., {Kubo}, J.~M., {Feldmann}, R., {et~al.} 2011, ArXiv e-prints

\bibitem[{{Heymans} {et~al.}(2006){Heymans}, {Van Waerbeke}, {Bacon}, {Berge},
  {Bernstein}, {Bertin}, {Bridle}, {Brown}, {Clowe}, {Dahle}, {Erben}, {Gray},
  {Hetterscheidt}, {Hoekstra}, {Hudelot}, {Jarvis}, {Kuijken}, {Margoniner},
  {Massey}, {Mellier}, {Nakajima}, {Refregier}, {Rhodes}, {Schrabback}, \&
  {Wittman}}]{Heymans06}
{Heymans}, C., {Van Waerbeke}, L., {Bacon}, D., {et~al.} 2006, \mnras, 368,
  1323

\bibitem[{{Hirata} {et~al.}(2004){Hirata}, {Mandelbaum}, {Seljak}, {Guzik},
  {Padmanabhan}, {Blake}, {Brinkmann}, {Bud{\'a}vari}, {Connolly}, {Csabai},
  {Scranton}, \& {Szalay}}]{Hirata04}
{Hirata}, C.~M., {Mandelbaum}, R., {Seljak}, U., {et~al.} 2004, \mnras, 353,
  529

\bibitem[{{Hoekstra} {et~al.}(2000){Hoekstra}, {Franx}, \&
  {Kuijken}}]{Hoekstra00}
{Hoekstra}, H., {Franx}, M., \& {Kuijken}, K. 2000, \apj, 532, 88

\bibitem[{{Hoekstra} {et~al.}(1998){Hoekstra}, {Franx}, {Kuijken}, \&
  {Squires}}]{Hoekstra98}
{Hoekstra}, H., {Franx}, M., {Kuijken}, K., \& {Squires}, G. 1998, \apj, 504,
  636

\bibitem[{{Hoekstra} {et~al.}(2005){Hoekstra}, {Hsieh}, {Yee}, {Lin}, \&
  {Gladders}}]{Hoekstra05}
{Hoekstra}, H., {Hsieh}, B.~C., {Yee}, H.~K.~C., {Lin}, H., \& {Gladders},
  M.~D. 2005, \apj, 635, 73

\bibitem[{{Hoekstra} {et~al.}(2004){Hoekstra}, {Yee}, \&
  {Gladders}}]{Hoekstra04}
{Hoekstra}, H., {Yee}, H.~K.~C., \& {Gladders}, M.~D. 2004, \apj, 606, 67

\bibitem[{{Hsieh} {et~al.}(2005){Hsieh}, {Yee}, {Lin}, \& {Gladders}}]{Hsieh05}
{Hsieh}, B.~C., {Yee}, H.~K.~C., {Lin}, H., \& {Gladders}, M.~D. 2005, \apjs,
  158, 161

\bibitem[{{Ilbert} {et~al.}(2006){Ilbert}, {Arnouts}, {McCracken},
  {Bolzonella}, {Bertin}, {Le F{\`e}vre}, {Mellier}, {Zamorani}, {Pell{\`o}},
  {Iovino}, {Tresse}, {Le Brun}, {Bottini}, {Garilli}, {Maccagni}, {Picat},
  {Scaramella}, {Scodeggio}, {Vettolani}, {Zanichelli}, {Adami}, {Bardelli},
  {Cappi}, {Charlot}, {Ciliegi}, {Contini}, {Cucciati}, {Foucaud}, {Franzetti},
  {Gavignaud}, {Guzzo}, {Marano}, {Marinoni}, {Mazure}, {Meneux}, {Merighi},
  {Paltani}, {Pollo}, {Pozzetti}, {Radovich}, {Zucca}, {Bondi}, {Bongiorno},
  {Busarello}, {de La Torre}, {Gregorini}, {Lamareille}, {Mathez}, {Merluzzi},
  {Ripepi}, {Rizzo}, \& {Vergani}}]{Ilbert06}
{Ilbert}, O., {Arnouts}, S., {McCracken}, H.~J., {et~al.} 2006, \aap, 457, 841

\bibitem[{{Kaiser} {et~al.}(2002){Kaiser}, {Aussel}, {Burke}, {Boesgaard},
  {Chambers}, {Chun}, {Heasley}, {Hodapp}, {Hunt}, {Jedicke}, {Jewitt},
  {Kudritzki}, {Luppino}, {Maberry}, {Magnier}, {Monet}, {Onaka}, {Pickles},
  {Rhoads}, {Simon}, {Szalay}, {Szapudi}, {Tholen}, {Tonry}, {Waterson}, \&
  {Wick}}]{Kaiser02}
{Kaiser}, N., {Aussel}, H., {Burke}, B.~E., {et~al.} 2002, in Presented at the
  Society of Photo-Optical Instrumentation Engineers (SPIE) Conference, Vol.
  4836, Society of Photo-Optical Instrumentation Engineers (SPIE) Conference
  Series, ed. {J.~A.~Tyson \& S.~Wolff}, 154--164

\bibitem[{{Kaiser} {et~al.}(1995){Kaiser}, {Squires}, \&
  {Broadhurst}}]{Kaiser95}
{Kaiser}, N., {Squires}, G., \& {Broadhurst}, T. 1995, \apj, 449, 460

\bibitem[{{Kauffmann} {et~al.}(2003){Kauffmann}, {Heckman}, {White}, {Charlot},
  {Tremonti}, {Brinchmann}, {Bruzual}, {Peng}, {Seibert}, {Bernardi},
  {Blanton}, {Brinkmann}, {Castander}, {Cs{\'a}bai}, {Fukugita}, {Ivezic},
  {Munn}, {Nichol}, {Padmanabhan}, {Thakar}, {Weinberg}, \&
  {York}}]{Kauffmann03}
{Kauffmann}, G., {Heckman}, T.~M., {White}, S.~D.~M., {et~al.} 2003, \mnras,
  341, 33

\bibitem[{{Kim} {et~al.}(2009){Kim}, {Baugh}, {Cole}, {Frenk}, \&
  {Benson}}]{Kim09}
{Kim}, H., {Baugh}, C.~M., {Cole}, S., {Frenk}, C.~S., \& {Benson}, A.~J. 2009,
  \mnras, 400, 1527

\bibitem[{Koester {et~al.}(2007)Koester, McKay, Annis, Wechsler, Evrard, Bleem,
  Becker, Johnston, Sheldon, Nichol, Miller, Scranton, Bahcall, Barentine,
  Brewington, Brinkmann, Harvanek, Kleinman, Krzesinski, Long, Nitta,
  Schneider, Sneddin, Voges, \& York}]{Koester07}
Koester, B.~P., McKay, T.~A., Annis, J., {et~al.} 2007, \apj, 660, 239

\bibitem[{{Komatsu} {et~al.}(2009){Komatsu}, {Dunkley}, {Nolta}, {Bennett},
  {Gold}, {Hinshaw}, {Jarosik}, {Larson}, {Limon}, {Page}, {Spergel},
  {Halpern}, {Hill}, {Kogut}, {Meyer}, {Tucker}, {Weiland}, {Wollack}, \&
  {Wright}}]{Komatsu09}
{Komatsu}, E., {Dunkley}, J., {Nolta}, M.~R., {et~al.} 2009, \apjs, 180, 330

\bibitem[{{Kova{\v c}} {et~al.}(2011){Kova{\v c}}, {Porciani}, {Lilly},
  {Marinoni}, {Guzzo}, {Cucciati}, {Zamorani}, {Iovino}, {Oesch}, {Bolzonella},
  {Peng}, {Meneux}, {Zucca}, {Bardelli}, {Carollo}, {Contini}, {Kneib}, {Le
  F{\`e}vre}, {Mainieri}, {Renzini}, {Scodeggio}, {Bongiorno}, {Caputi},
  {Coppa}, {de la Torre}, {de Ravel}, {Finoguenov}, {Franzetti}, {Garilli},
  {Kampczyk}, {Knobel}, {Lamareille}, {Le Borgne}, {Le Brun}, {Maier},
  {Mignoli}, {Pello}, {Perez-Montero}, {Pozzetti}, {Ricciardelli}, {Silverman},
  {Tanaka}, {Tasca}, {Tresse}, {Vergani}, {Abbas}, {Bottini}, {Cappi},
  {Cassata}, {Cimatti}, {Fumana}, {Koekemoer}, {Leauthaud}, {Maccagni},
  {McCracken}, {Memeo}, {Scaramella}, \& {Scoville}}]{Kovac11}
{Kova{\v c}}, K., {Porciani}, C., {Lilly}, S.~J., {et~al.} 2011, \apj, 731, 102

\bibitem[{{Kravtsov} {et~al.}(2004){Kravtsov}, {Berlind}, {Wechsler}, {Klypin},
  {Gottl{\"o}ber}, {Allgood}, \& {Primack}}]{Kravtsov04}
{Kravtsov}, A.~V., {Berlind}, A.~A., {Wechsler}, R.~H., {et~al.} 2004, \apj,
  609, 35

\bibitem[{{Kroupa}(2001)}]{Kroupa01}
{Kroupa}, P. 2001, \mnras, 322, 231

\bibitem[{{Leauthaud} {et~al.}(2010){Leauthaud}, {Finoguenov}, {Kneib},
  {Taylor}, {Massey}, {Rhodes}, {Ilbert}, {Bundy}, {Tinker}, {George}, {Capak},
  {Koekemoer}, {Johnston}, {Zhang}, {Cappelluti}, {Ellis}, {Elvis}, {Giodini},
  {Heymans}, {Le F{\`e}vre}, {Lilly}, {McCracken}, {Mellier},
  {R{\'e}fr{\'e}gier}, {Salvato}, {Scoville}, {Smoot}, {Tanaka}, {Van
  Waerbeke}, \& {Wolk}}]{Leauthaud10}
{Leauthaud}, A., {Finoguenov}, A., {Kneib}, J., {et~al.} 2010, \apj, 709, 97

\bibitem[{{Leauthaud} {et~al.}(2011){Leauthaud}, {Tinker}, {Bundy}, {Behroozi},
  {Massey}, {Rhodes}, {George}, {Kneib}, {Benson}, {Wechsler}, {Busha},
  {Capak}, {Cortes}, {Ilbert}, {Koekemoer}, {Le Fevre}, {Lilly}, {McCracken},
  {Salvato}, {Schrabback}, {Scoville}, {Smith}, \& {Taylor}}]{Leauthaud11}
{Leauthaud}, A., {Tinker}, J., {Bundy}, K., {et~al.} 2011, ArXiv e-prints

\bibitem[{{Limousin} {et~al.}(2009){Limousin}, {Sommer-Larsen}, {Natarajan}, \&
  {Milvang-Jensen}}]{Limousin09}
{Limousin}, M., {Sommer-Larsen}, J., {Natarajan}, P., \& {Milvang-Jensen}, B.
  2009, \apj, 696, 1771

\bibitem[{{Luppino} \& {Kaiser}(1997)}]{LuppinoK97}
{Luppino}, G.~A. \& {Kaiser}, N. 1997, \apj, 475, 20

\bibitem[{{Mandelbaum} {et~al.}(2005{\natexlab{a}}){Mandelbaum}, {Hirata},
  {Seljak}, {Guzik}, {Padmanabhan}, {Blake}, {Blanton}, {Lupton}, \&
  {Brinkmann}}]{Mandelbaum05sys}
{Mandelbaum}, R., {Hirata}, C.~M., {Seljak}, U., {et~al.} 2005{\natexlab{a}},
  \mnras, 361, 1287

\bibitem[{{Mandelbaum} {et~al.}(2006){Mandelbaum}, {Seljak}, {Kauffmann},
  {Hirata}, \& {Brinkmann}}]{Mandelbaum06}
{Mandelbaum}, R., {Seljak}, U., {Kauffmann}, G., {Hirata}, C.~M., \&
  {Brinkmann}, J. 2006, \mnras, 368, 715

\bibitem[{{Mandelbaum} {et~al.}(2005{\natexlab{b}}){Mandelbaum}, {Tasitsiomi},
  {Seljak}, {Kravtsov}, \& {Wechsler}}]{Mandelbaum05HOD}
{Mandelbaum}, R., {Tasitsiomi}, A., {Seljak}, U., {Kravtsov}, A.~V., \&
  {Wechsler}, R.~H. 2005{\natexlab{b}}, \mnras, 362, 1451

\bibitem[{{Massey} {et~al.}(2007){Massey}, {Heymans}, {Berg{\'e}}, {Bernstein},
  {Bridle}, {Clowe}, {Dahle}, {Ellis}, {Erben}, {Hetterscheidt}, {High},
  {Hirata}, {Hoekstra}, {Hudelot}, {Jarvis}, {Johnston}, {Kuijken},
  {Margoniner}, {Mandelbaum}, {Mellier}, {Nakajima}, {Paulin-Henriksson},
  {Peeples}, {Roat}, {Refregier}, {Rhodes}, {Schrabback}, {Schirmer}, {Seljak},
  {Semboloni}, \& {van Waerbeke}}]{Massey07}
{Massey}, R., {Heymans}, C., {Berg{\'e}}, J., {et~al.} 2007, \mnras, 376, 13

\bibitem[{{More} {et~al.}(2011){More}, {van den Bosch}, {Cacciato}, {Skibba},
  {Mo}, \& {Yang}}]{More11}
{More}, S., {van den Bosch}, F.~C., {Cacciato}, M., {et~al.} 2011, \mnras, 410,
  210

\bibitem[{{Moster} {et~al.}(2010){Moster}, {Somerville}, {Maulbetsch}, {van den
  Bosch}, {Macci{\`o}}, {Naab}, \& {Oser}}]{Moster10}
{Moster}, B.~P., {Somerville}, R.~S., {Maulbetsch}, C., {et~al.} 2010, \apj,
  710, 903

\bibitem[{{Napolitano} {et~al.}(2009){Napolitano}, {Romanowsky}, {Coccato},
  {Capaccioli}, {Douglas}, {Noordermeer}, {Gerhard}, {Arnaboldi}, {de Lorenzi},
  {Kuijken}, {Merrifield}, {O'Sullivan}, {Cortesi}, {Das}, \&
  {Freeman}}]{Napolitano09}
{Napolitano}, N.~R., {Romanowsky}, A.~J., {Coccato}, L., {et~al.} 2009, \mnras,
  393, 329

\bibitem[{{Navarro} {et~al.}(1996){Navarro}, {Frenk}, \& {White}}]{Navarro96}
{Navarro}, J.~F., {Frenk}, C.~S., \& {White}, S.~D.~M. 1996, \apj, 462, 563

\bibitem[{{Neistein} {et~al.}(2011){Neistein}, {Li}, {Khochfar}, {Weinmann},
  {Shankar}, \& {Boylan-Kolchin}}]{Neistein11}
{Neistein}, E., {Li}, C., {Khochfar}, S., {et~al.} 2011, ArXiv e-prints

\bibitem[{{Padmanabhan} {et~al.}(2008){Padmanabhan}, {Schlegel}, {Finkbeiner},
  {Barentine}, {Blanton}, {Brewington}, {Gunn}, {Harvanek}, {Hogg},
  {Ivezi{\'c}}, {Johnston}, {Kent}, {Kleinman}, {Knapp}, {Krzesinski}, {Long},
  {Neilsen}, {Nitta}, {Loomis}, {Lupton}, {Roweis}, {Snedden}, {Strauss}, \&
  {Tucker}}]{Padmanabhan08}
{Padmanabhan}, N., {Schlegel}, D.~J., {Finkbeiner}, D.~P., {et~al.} 2008, \apj,
  674, 1217

\bibitem[{{Peacock} \& {Dodds}(1996)}]{PeacockD96}
{Peacock}, J.~A. \& {Dodds}, S.~J. 1996, \mnras, 280, L19

\bibitem[{{Pielorz} {et~al.}(2010){Pielorz}, {R{\"o}diger}, {Tereno}, \&
  {Schneider}}]{Pielorz10}
{Pielorz}, J., {R{\"o}diger}, J., {Tereno}, I., \& {Schneider}, P. 2010, \aap,
  514, A79+

\bibitem[{{Pozzetti} {et~al.}(2010){Pozzetti}, {Bolzonella}, {Zucca},
  {Zamorani}, {Lilly}, {Renzini}, {Moresco}, {Mignoli}, {Cassata}, {Tasca},
  {Lamareille}, {Maier}, {Meneux}, {Halliday}, {Oesch}, {Vergani}, {Caputi},
  {Kova{\v c}}, {Cimatti}, {Cucciati}, {Iovino}, {Peng}, {Carollo}, {Contini},
  {Kneib}, {Le F{\'e}vre}, {Mainieri}, {Scodeggio}, {Bardelli}, {Bongiorno},
  {Coppa}, {de la Torre}, {de Ravel}, {Franzetti}, {Garilli}, {Kampczyk},
  {Knobel}, {Le Borgne}, {Le Brun}, {Pell{\`o}}, {Perez Montero},
  {Ricciardelli}, {Silverman}, {Tanaka}, {Tresse}, {Abbas}, {Bottini}, {Cappi},
  {Guzzo}, {Koekemoer}, {Leauthaud}, {Maccagni}, {Marinoni}, {McCracken},
  {Memeo}, {Porciani}, {Scaramella}, {Scarlata}, \& {Scoville}}]{Pozzetti10}
{Pozzetti}, L., {Bolzonella}, M., {Zucca}, E., {et~al.} 2010, \aap, 523, A13+

\bibitem[{{Press} \& {Schechter}(1974)}]{PressS74}
{Press}, W.~H. \& {Schechter}, P. 1974, \apj, 187, 425

\bibitem[{{Salim} {et~al.}(2007){Salim}, {Rich}, {Charlot}, {Brinchmann},
  {Johnson}, {Schiminovich}, {Seibert}, {Mallery}, {Heckman}, {Forster},
  {Friedman}, {Martin}, {Morrissey}, {Neff}, {Small}, {Wyder}, {Bianchi},
  {Donas}, {Lee}, {Madore}, {Milliard}, {Szalay}, {Welsh}, \& {Yi}}]{Salim07}
{Salim}, S., {Rich}, R.~M., {Charlot}, S., {et~al.} 2007, \apjs, 173, 267

\bibitem[{{Schlegel} {et~al.}(1998){Schlegel}, {Finkbeiner}, \&
  {Davis}}]{Schlegel98}
{Schlegel}, D.~J., {Finkbeiner}, D.~P., \& {Davis}, M. 1998, \apj, 500, 525

\bibitem[{{Scoville} {et~al.}(2007){Scoville}, {Abraham}, {Aussel}, {Barnes},
  {Benson}, {Blain}, {Calzetti}, {Comastri}, {Capak}, {Carilli}, {Carlstrom},
  {Carollo}, {Colbert}, {Daddi}, {Ellis}, {Elvis}, {Ewald}, {Fall},
  {Franceschini}, {Giavalisco}, {Green}, {Griffiths}, {Guzzo}, {Hasinger},
  {Impey}, {Kneib}, {Koda}, {Koekemoer}, {Lefevre}, {Lilly}, {Liu},
  {McCracken}, {Massey}, {Mellier}, {Miyazaki}, {Mobasher}, {Mould}, {Norman},
  {Refregier}, {Renzini}, {Rhodes}, {Rich}, {Sanders}, {Schiminovich},
  {Schinnerer}, {Scodeggio}, {Sheth}, {Shopbell}, {Taniguchi}, {Tyson}, {Urry},
  {Van Waerbeke}, {Vettolani}, {White}, \& {Yan}}]{Scoville07}
{Scoville}, N., {Abraham}, R.~G., {Aussel}, H., {et~al.} 2007, \apjs, 172, 38

\bibitem[{{Seljak}(2000)}]{Seljak00}
{Seljak}, U. 2000, \mnras, 318, 203

\bibitem[{{S\'ersic}(1968)}]{Sersic68}
{S\'ersic}, J.~L. 1968, {Atlas de galaxias australes}, ed. {S\'ersic, J.~L.}

\bibitem[{{Sheldon} {et~al.}(2009{\natexlab{a}}){Sheldon}, {Johnston},
  {Masjedi}, {McKay}, {Blanton}, {Scranton}, {Wechsler}, {Koester}, {Hansen},
  {Frieman}, \& {Annis}}]{Sheldon09III}
{Sheldon}, E.~S., {Johnston}, D.~E., {Masjedi}, M., {et~al.}
  2009{\natexlab{a}}, \apj, 703, 2232

\bibitem[{{Sheldon} {et~al.}(2009{\natexlab{b}}){Sheldon}, {Johnston},
  {Scranton}, {Koester}, {McKay}, {Oyaizu}, {Cunha}, {Lima}, {Lin}, {Frieman},
  {Wechsler}, {Annis}, {Mandelbaum}, {Bahcall}, \& {Fukugita}}]{Sheldon09I}
{Sheldon}, E.~S., {Johnston}, D.~E., {Scranton}, R., {et~al.}
  2009{\natexlab{b}}, \apj, 703, 2217

\bibitem[{{Sheth} {et~al.}(2001){Sheth}, {Mo}, \& {Tormen}}]{Sheth01}
{Sheth}, R.~K., {Mo}, H.~J., \& {Tormen}, G. 2001, \mnras, 323, 1

\bibitem[{{Siverd} {et~al.}(2009){Siverd}, {Ryden}, \& {Gaudi}}]{Siverd09}
{Siverd}, R.~J., {Ryden}, B.~S., \& {Gaudi}, B.~S. 2009, ArXiv e-prints

\bibitem[{{Smith} {et~al.}(2003){Smith}, {Peacock}, {Jenkins}, {White},
  {Frenk}, {Pearce}, {Thomas}, {Efstathiou}, \& {Couchman}}]{Smith03}
{Smith}, R.~E., {Peacock}, J.~A., {Jenkins}, A., {et~al.} 2003, \mnras, 341,
  1311

\bibitem[{{Somerville} {et~al.}(2008){Somerville}, {Hopkins}, {Cox},
  {Robertson}, \& {Hernquist}}]{Somerville08}
{Somerville}, R.~S., {Hopkins}, P.~F., {Cox}, T.~J., {Robertson}, B.~E., \&
  {Hernquist}, L. 2008, \mnras, 391, 481

\bibitem[{{Springel} {et~al.}(2005){Springel}, {White}, {Jenkins}, {Frenk},
  {Yoshida}, {Gao}, {Navarro}, {Thacker}, {Croton}, {Helly}, {Peacock}, {Cole},
  {Thomas}, {Couchman}, {Evrard}, {Colberg}, \& {Pearce}}]{Springel05}
{Springel}, V., {White}, S.~D.~M., {Jenkins}, A., {et~al.} 2005, \nat, 435, 629

\bibitem[{{Strateva} {et~al.}(2001){Strateva}, {Ivezi{\'c}}, {Knapp},
  {Narayanan}, {Strauss}, {Gunn}, {Lupton}, {Schlegel}, {Bahcall}, {Brinkmann},
  {Brunner}, {Budav{\'a}ri}, {Csabai}, {Castander}, {Doi}, {Fukugita}, {Gy{\H
  o}ry}, {Hamabe}, {Hennessy}, {Ichikawa}, {Kunszt}, {Lamb}, {McKay},
  {Okamura}, {Racusin}, {Sekiguchi}, {Schneider}, {Shimasaku}, \&
  {York}}]{Strateva01}
{Strateva}, I., {Ivezi{\'c}}, {\v Z}., {Knapp}, G.~R., {et~al.} 2001, \aj, 122,
  1861

\bibitem[{{Strauss} {et~al.}(2002){Strauss}, {Weinberg}, {Lupton}, {Narayanan},
  {Annis}, {Bernardi}, {Blanton}, {Burles}, {Connolly}, {Dalcanton}, {Doi},
  {Eisenstein}, {Frieman}, {Fukugita}, {Gunn}, {Ivezi{\'c}}, {Kent}, {Kim},
  {Knapp}, {Kron}, {Munn}, {Newberg}, {Nichol}, {Okamura}, {Quinn}, {Richmond},
  {Schlegel}, {Shimasaku}, {SubbaRao}, {Szalay}, {Vanden Berk}, {Vogeley},
  {Yanny}, {Yasuda}, {York}, \& {Zehavi}}]{Strauss02}
{Strauss}, M.~A., {Weinberg}, D.~H., {Lupton}, R.~H., {et~al.} 2002, \aj, 124,
  1810

\bibitem[{{Tasitsiomi} {et~al.}(2004){Tasitsiomi}, {Kravtsov}, {Wechsler}, \&
  {Primack}}]{Tasitsiomi04}
{Tasitsiomi}, A., {Kravtsov}, A.~V., {Wechsler}, R.~H., \& {Primack}, J.~R.
  2004, \apj, 614, 533

\bibitem[{{Tinker} {et~al.}(2005){Tinker}, {Weinberg}, {Zheng}, \&
  {Zehavi}}]{Tinker05}
{Tinker}, J.~L., {Weinberg}, D.~H., {Zheng}, Z., \& {Zehavi}, I. 2005, \apj,
  631, 41

\bibitem[{{Vulcani} {et~al.}(2011){Vulcani}, {Poggianti},
  {Arag{\'o}n-Salamanca}, {Fasano}, {Rudnick}, {Valentinuzzi}, {Dressler},
  {Bettoni}, {Cava}, {D'Onofrio}, {Fritz}, {Moretti}, {Omizzolo}, \&
  {Varela}}]{Vulcani10}
{Vulcani}, B., {Poggianti}, B.~M., {Arag{\'o}n-Salamanca}, A., {et~al.} 2011,
  \mnras, 412, 246

\bibitem[{{Wang} {et~al.}(2006){Wang}, {Li}, {Kauffmann}, \& {De
  Lucia}}]{Wang06}
{Wang}, L., {Li}, C., {Kauffmann}, G., \& {De Lucia}, G. 2006, \mnras, 371, 537

\bibitem[{{Wright} \& {Brainerd}(2000)}]{WrightB00}
{Wright}, C.~O. \& {Brainerd}, T.~G. 2000, \apj, 534, 34

\bibitem[{{York} {et~al.}(2000){York}, {Adelman}, {Anderson}, {Anderson},
  {Annis}, {Bahcall}, {Bakken}, {Barkhouser}, {Bastian}, {Berman}, {Boroski},
  {Bracker}, {Briegel}, {Briggs}, {Brinkmann}, {Brunner}, {Burles}, {Carey},
  {Carr}, {Castander}, {Chen}, {Colestock}, {Connolly}, {Crocker}, {Csabai},
  {Czarapata}, {Davis}, {Doi}, {Dombeck}, {Eisenstein}, {Ellman}, {Elms},
  {Evans}, {Fan}, {Federwitz}, {Fiscelli}, {Friedman}, {Frieman}, {Fukugita},
  {Gillespie}, {Gunn}, {Gurbani}, {de Haas}, {Haldeman}, {Harris}, {Hayes},
  {Heckman}, {Hennessy}, {Hindsley}, {Holm}, {Holmgren}, {Huang}, {Hull},
  {Husby}, {Ichikawa}, {Ichikawa}, {Ivezi{\'c}}, {Kent}, {Kim}, {Kinney},
  {Klaene}, {Kleinman}, {Kleinman}, {Knapp}, {Korienek}, {Kron}, {Kunszt},
  {Lamb}, {Lee}, {Leger}, {Limmongkol}, {Lindenmeyer}, {Long}, {Loomis},
  {Loveday}, {Lucinio}, {Lupton}, {MacKinnon}, {Mannery}, {Mantsch}, {Margon},
  {McGehee}, {McKay}, {Meiksin}, {Merelli}, {Monet}, {Munn}, {Narayanan},
  {Nash}, {Neilsen}, {Neswold}, {Newberg}, {Nichol}, {Nicinski}, {Nonino},
  {Okada}, {Okamura}, {Ostriker}, {Owen}, {Pauls}, {Peoples}, {Peterson},
  {Petravick}, {Pier}, {Pope}, {Pordes}, {Prosapio}, {Rechenmacher}, {Quinn},
  {Richards}, {Richmond}, {Rivetta}, {Rockosi}, {Ruthmansdorfer}, {Sandford},
  {Schlegel}, {Schneider}, {Sekiguchi}, {Sergey}, {Shimasaku}, {Siegmund},
  {Smee}, {Smith}, {Snedden}, {Stone}, {Stoughton}, {Strauss}, {Stubbs},
  {SubbaRao}, {Szalay}, {Szapudi}, {Szokoly}, {Thakar}, {Tremonti}, {Tucker},
  {Uomoto}, {Vanden Berk}, {Vogeley}, {Waddell}, {Wang}, {Watanabe},
  {Weinberg}, {Yanny}, \& {Yasuda}}]{York00}
{York}, D.~G., {Adelman}, J., {Anderson}, Jr., J.~E., {et~al.} 2000, \aj, 120,
  1579

\bibitem[{{Zheng} {et~al.}(2005){Zheng}, {Berlind}, {Weinberg}, {Benson},
  {Baugh}, {Cole}, {Dav{\'e}}, {Frenk}, {Katz}, \& {Lacey}}]{Zheng05}
{Zheng}, Z., {Berlind}, A.~A., {Weinberg}, D.~H., {et~al.} 2005, \apj, 633, 791

\end{thebibliography}


\begin{appendix}
\section{Scatter of lenses between bins \label{ap_scat}}
\begin{figure}
  \resizebox{\hsize}{!}{\includegraphics{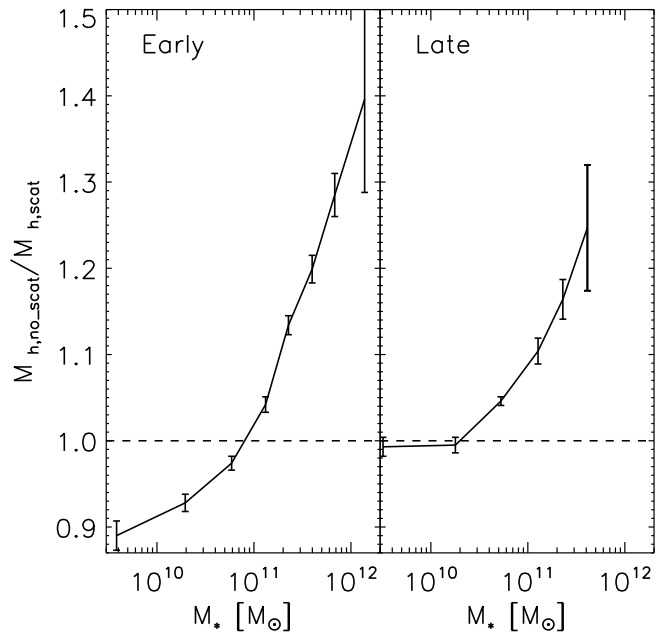}}
  \caption{The ratio of the best fit halo mass for the unscattered lens catalogue to the best fit halo mass for the lenses to which a log-normal scatter of 0.1 in stellar mass has been applied, for the early-type galaxies ({\it left}) and the late-type galaxies ({\it right}). The halo masses are underestimated at the high stellar mass end due to low mass objects scattering into the high mass bins.}
  \label{plotscatmstel}
\end{figure}

In this Appendix we describe how we calculate the bias that results from the scatter of galaxies between lensing bins due to the stellar mass errors. The bias that results from the scatter due to the luminosity errors has been calculated in a similar fashion. To begin, we create a large set of simulated lens catalogues. We construct the stellar mass function from the MPA/JHU catalogue, randomly draw stellar masses from this distribution and assign these to our lenses. We fit a powerlaw of the form $M_{200}=\alpha_*M_*^{\beta_*}$ to our initial observations, and calculate the halo mass of each galaxy. Assuming that the density profile of each lens follows an NFW profile, we calculate the ellipticities of the source galaxies under the assumption that they are intrinsically round. Next we create 20 new lens catalogues by applying a log-normal scatter with a width of 0.1 to the stellar masses. We use the stellar mass bins from Table \ref{tab_lens_mstel} to stack the lensing signal of the scattered lenses, and measure the tangential shear using the original source catalogue. We fit the lensing signal between 30 and 200 kpc with the halo model, imposing a satellite fraction of 0\% as the lenses were randomly inserted in the images. The ratio of the best fit halo masses for the original lenses and the lenses with scattered stellar masses gives the bias.
\\ 
\indent As the stellar mass function and the best fit powerlaw are different for the two galaxy types, we make two sets of simulations to study the bias for early- and late-type galaxies separately. We do not account for evolution with redshift, although the stellar mass function evolves between $z=0.0$ and $z=1.0$, most strongly for $M_*<10^{11}M_{\odot}$ \citep[e.g.][]{Vulcani10,Pozzetti10}. We are only sensitive to the change of the shape of the stellar mass function, which is most noticeable for $10^{10.5}<M_*<10^{11}M_{\odot}$. However, the bias in this regime is small, and we do not expect the change in shape to strongly affect our results. The relation between stellar mass and halo mass may also evolve between $z=0.0$ and $z=0.5$ \citep[e.g.][]{Moster10,Leauthaud11}. We currently lack sufficient signal-to-noise to study this in detail. As we will demonstrate, the bias is not very sensitive to changes in the powerlaw slope, and a mild evolution does not significantly alter the results. \\
\indent The ratio of the input halo masses to the best fit halo mass measured for the lenses that have been scattered is shown in Figure \ref{plotscatmstel}. The error bars indicate the standard deviation of the simulations. 
We find that the bias is highest for the early-types at the high mass end. This is due to the steepness of the stellar mass function, which leads to the net effect that low stellar mass objects scatter into and contaminate the high stellar mass bins. The bias for early-types at the low mass end is slightly smaller than 1, as the stellar mass function turns over at $\sim 5\times 10^{10} M_*$ and becomes smaller with decreasing stellar mass. The stellar mass function of the late-types is monotonically decreasing, and consequently the bias does not become smaller than unity. At the high mass end, the stellar mass function of the late-types is poorly determined due to the lack of objects. We cannot reliably determine the bias for the S6 late-type bin, and therefore apply the correction factor of the S5 bin to this bin as well. \\
\indent The observed stellar masses have already been scattered, and the best fit powerlaw is therefore too shallow. To investigate how this affects the bias, we correct our initial halo masses for the scattering, and again fit a powerlaw between stellar mass and halo mass. We repeat our simulations with these new powerlaw slopes, and find that the correction factors change by at most 4\%. The correction we apply is obtained using the corrected powerlaw slopes. \\
\indent The intrinsic stellar mass function is steeper than the observed one as on average more low stellar mass objects have scattered upward. Although we cannot retrieve the intrinsic stellar mass function, we can obtain an estimate of the level of contamination. For this purpose, we draw $1\times10^8$ objects from the observed stellar mass function, apply the log-normal scatter, and compare the number of objects in the stellar mass bins before and after the scatter. The number of lenses in the three lowest stellar mass bins does not change much after the scatter, but it increases with stellar mass for the more massive bins, reaching a maximum of 36\% more lenses in the S7 early-type bin. The increase in the number of objects may be even larger, as the observed stellar masses have already been scattered, and therefore the observed stellar mass function is smoother than the intrinsic one. As the stellar mass function at the high mass end is already very uncertain, we do not attempt to retrieve the intrinsic stellar mass function. However, the bias correction is sensitive to the slope at the high mass end, and the correction factors may actually be larger. \\


\section{Mean versus fitted halo mass \label{ap_width}}
\begin{figure}
   \resizebox{\hsize}{!}{\includegraphics{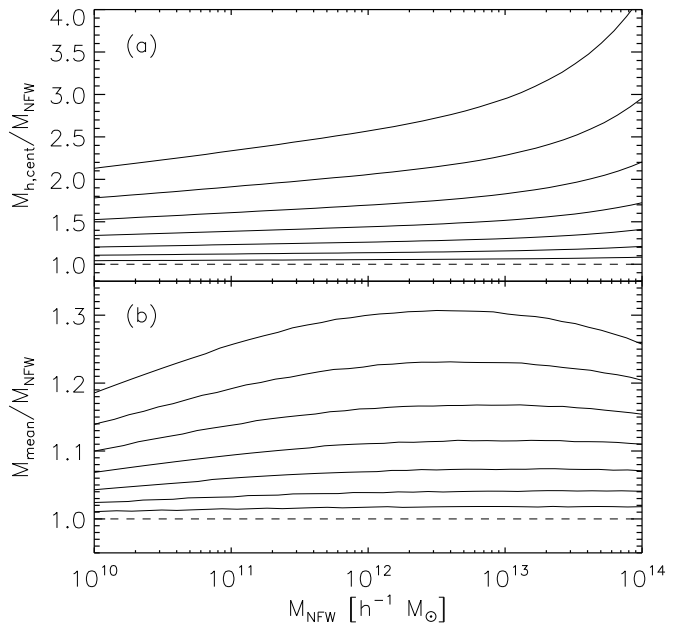}}
   \caption{The ratio of the central mass of the halo mass distribution, $m_{h,cent}$, and the best fit NFW mass ({\it top}) and the ratio of the mean halo mass and the best fit NFW mass ({\it bottom}) as a function of best fit NFW mass. Different lines correspond to values of $\sigma_{m_h}$ 0.10 (bottom line), 0.15, 0.20, 0.25, 0.30, 0.35 and 0.40 (top line). The lensing mass is converted to the mean halo mass using the corrections from the bottom panel.}
  \label{PlotWidth}
\end{figure}

\indent The distribution of halo masses for a certain luminosity (or stellar mass) is given by the conditional probability function, which is usually described by a log-normal function of the form
\begin{equation}
  P(m_h|l) \propto \exp\Big(-\frac{(m_h-m_{h,cent})^2}{2\sigma^2_{m_h}}\Big)
  \label{eq_mhdist}
\end{equation}
where $l=\log(L)$, $m_h=\log(M_h)$ and $\sigma_{m_h}$ is the scatter in $m_h$. In this Appendix we study how the best fit lensing mass is related to either the mean halo mass or to the centre of the halo mass distribution, $m_{h,cent}$. To mimic the selection of real galaxies, we assign a value to $m_{h,cent}$ and $\sigma_{m_h}$, and randomly draw 1000 galaxies from the conditional probability function which has been convolved with the halo mass function (Equation \ref{eq_hmf}). We calculate the NFW shear profiles of these galaxies, average their signals to simulate the usual lensing procedure, and fit an NFW profile to the stacked shear. Figure \ref{PlotWidth} shows the ratio of $M_{h,cent}$ to the best fit NFW mass in the top panel, and the ratio of the mean halo mass to the best fit NFW mass in the lower panel. The lines correspond to different values of $\sigma_{m_h}$, ranging from 0.10 to 0.40 from bottom to top. Note that the scale of the vertical axes in the two panels is different.\\
\indent In Figure \ref{PlotWidth}a we see that the best fit NFW mass is considerably lower than the central mass of the distribution. This is mainly the result of the declining halo mass function, which leads us to preferentially pick lower mass haloes. The shape of an NFW profile changes with halo mass because the NFW concentration parameter depends on halo mass. The shape and amplitude of the stacked shear signal is therefore not equal to the profile of an NFW with a corresponding mean halo mass. Therefore, the best fit NFW mass underestimates the mean halo mass, as demonstrated in Figure \ref{PlotWidth}b.  \\
\indent The ratios in Figure \ref{PlotWidth}a and \ref{PlotWidth}b are sensitive to the value of $\sigma_{m_h}$. We use the results from \citet{More11}, who studied the distribution of halo masses as a function of luminosity and stellar mass using the kinematics of satellite galaxies orbiting central galaxies. As only central galaxies are considered in their work, the actual scatter for a sample of galaxies consisting of both centrals and satellites may be larger. On the other hand, part of the scatter may be introduced through uncertainties in the determination of the halo masses, which would imply a lower intrinsic scatter. \\
\indent We use Figures 4 and 9 from \citet{More11} to read off the values we assign to $\sigma_{m_h}$ for the luminosity and stellar mass bins. We list these values, and the corresponding correction factor to the mean halo mass, in Table \ref{tab_widthcorr}. The luminosities and stellar masses in our sample extend to higher values than \citet{More11} use, but their figures suggest that $\sigma_{m_h}$ does not change rapidly at the high mass/luminosity end, and we therefore assume that the values remain constant. For the stellar masses we use the NFW masses that have been corrected for the scattering of objects between the bins.\\
\begin{table}
  \caption{The values of $\sigma_{m_h}$ assigned to the luminosity and stellar mass  bins, and the correction factors $f_{corr}$ we apply to convert the measured lensing mass into the mean halo mass.}   
  \centering
  \begin{tabular}{c c c c c } 
  \hline
  Sample & $\sigma_{m_h}$(early) & $f_{corr}$(early) & $\sigma_{m_h}$(late) &  $f_{corr}$(late) \\
  & &  &  & \\
  \hline\hline  \\
    L1 & 0.20 & 1.07 & 0.25 & 1.10 \\
    L2 & 0.25 & 1.11 & 0.29 & 1.14 \\
    L3 & 0.30 & 1.17 & 0.30 & 1.15 \\
    L4 & 0.33 & 1.20 & 0.33 & 1.20 \\
    L5 & 0.37 & 1.26 & 0.34 & 1.21 \\
    L6 & 0.39 & 1.28 & 0.35 & 1.23 \\
    L7 & 0.40 & 1.28 & 0.35 & 1.22 \\
    L8 & 0.40 & 1.27 & 0.35 & 1.23 \\
    & & & & \\
    S1 & 0.15 & 1.03 & 0.10 & 1.01 \\
    S2 & 0.18 & 1.06 & 0.10 & 1.02 \\
    S3 & 0.26 & 1.12 & 0.10 & 1.02 \\
    S4 & 0.32 & 1.19 & 0.10 & 1.02 \\
    S5 & 0.36 & 1.24 & 0.10 & 1.02 \\
    S6 & 0.40 & 1.28 & 0.10 & 1.02 \\
    S7 & 0.40 & 1.27 & - & - \\
  \hline \\
  \end{tabular}
  \label{tab_widthcorr}
\end{table}
\indent There are further sources of uncertainty to consider in future studies, and we list a few of them: luminosity bins have a certain width, the luminosity function is not constant inside a luminosity bin, and lens galaxies are located at a range of redshifts. We expect that these complications further broaden the conditional probability function, which means that the correction factors we use may be too low. These complications should be taken into account to enable a detailed comparison between observations and simulations. \\


\section{Constraints on the satellite fraction at high halo masses \label{ap_sat}}
\begin{figure}
  \resizebox{\hsize}{!}{\includegraphics{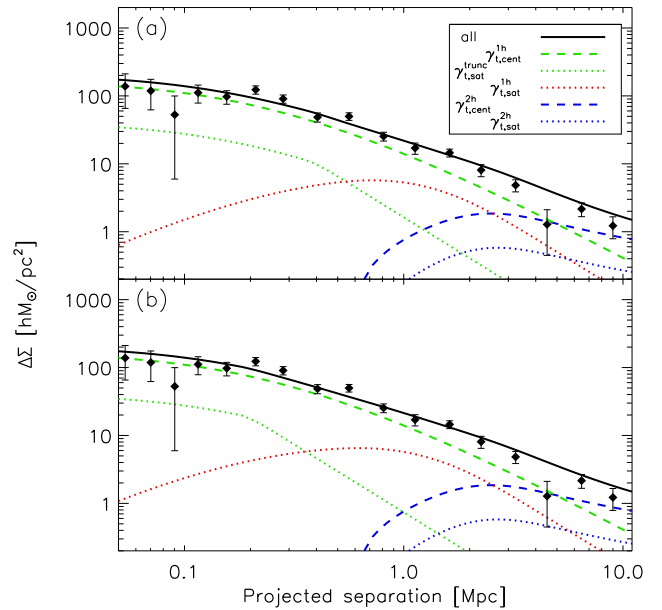}}
  \caption{The lensing signal of the L6 early-type bin, shown together with the five components of the best fit halo model. In the upper panel the truncation radius of the stripped satellites is 0.4$r_{200}$, and the shape of the combined satellite 1-halo terms mimicks the shape of the central NFW term. In the lower panel the truncation radius is 0.2$r_{200}$, changing the shape of the combined satellite 1-halo terms. Note that the halo model in the lower panel is not a fit, but serves to illustrate the effect of choosing a different truncation radius.}
  \label{plot_sat_strip}
\end{figure}

\begin{figure}
  \resizebox{\hsize}{!}{\includegraphics[angle=270]{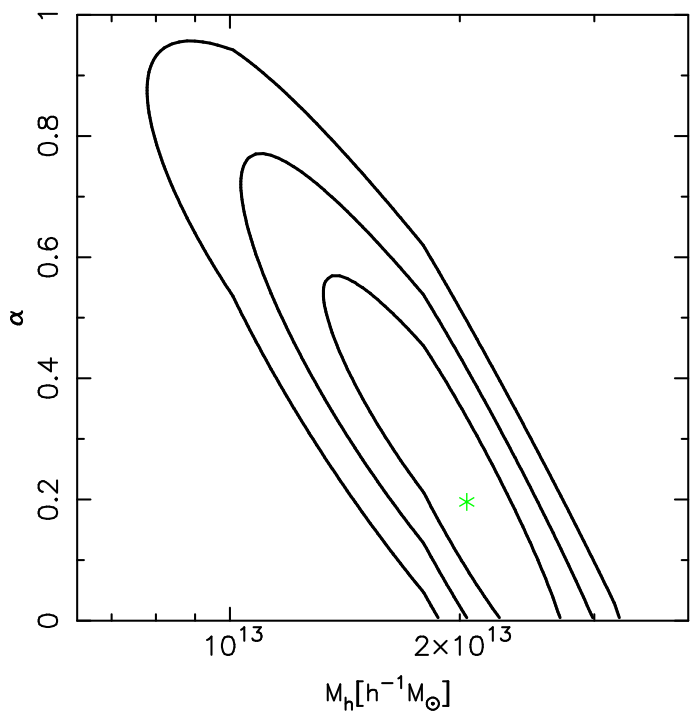}}
  \caption{The $\chi^2$ values of the halo model fits to the L6 early-type bin. The green star indicates the best fit. The three contours show the 67.8\%, 95.4\% and 99.7\% confidence intervals ($\Delta \chi^2$ of 2.3, 6.2 and 11.8 respectively). The best fit halo mass is anti-correlated with the best fit satellite fraction. }
  \label{plot_chisq}
\end{figure}

\begin{figure}
  \resizebox{\hsize}{!}{\includegraphics{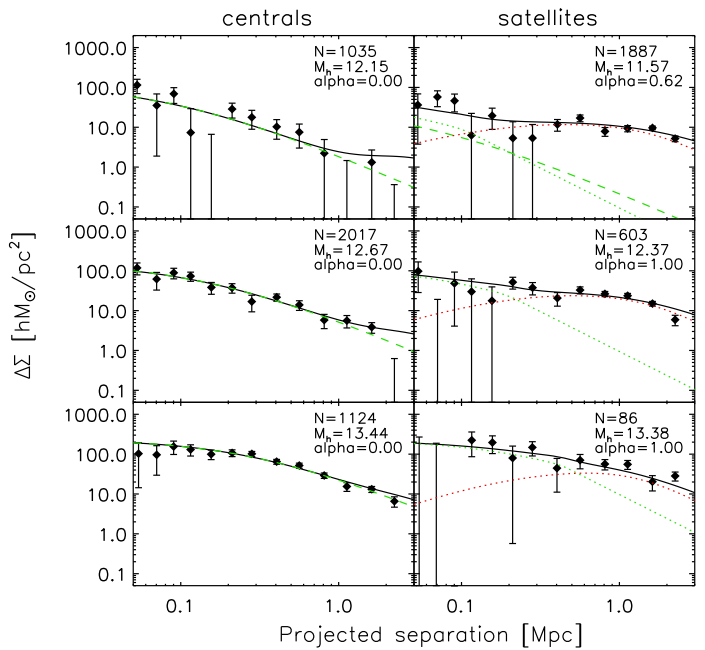}}
  \caption{The lensing signal $\Delta \Sigma$ as a function of physical distance from the lens. The lensing signal is measured for central galaxies ({\it left}) and for the satellite galaxies ({\it right}), for the $10^{10.5}<M_{*}<10^{11.0}M_{\odot}$ bin ({\it top}), $10^{11.0} <M_{*}<10^{11.5} M_{\odot}$ bin ({\it middle}) and $10^{11.5} <M_{*}<10^{11.75} M_{\odot}$ bin ({\it bottom}). Indicated in each plot is the number of lenses, the logarithm of the best fit halo mass and the best fit satellite fraction. The shear signal is reduced at large lens-source separations for the central galaxies, indicating that they are isolated. At small lens-source separations the shear signal of the satellite sample appears to be reduced compared to the central sample. Note that the 2-halo terms are not shown for clarity. }
  \label{plot_envir}
\end{figure}

\begin{figure}
  \resizebox{\hsize}{!}{\includegraphics{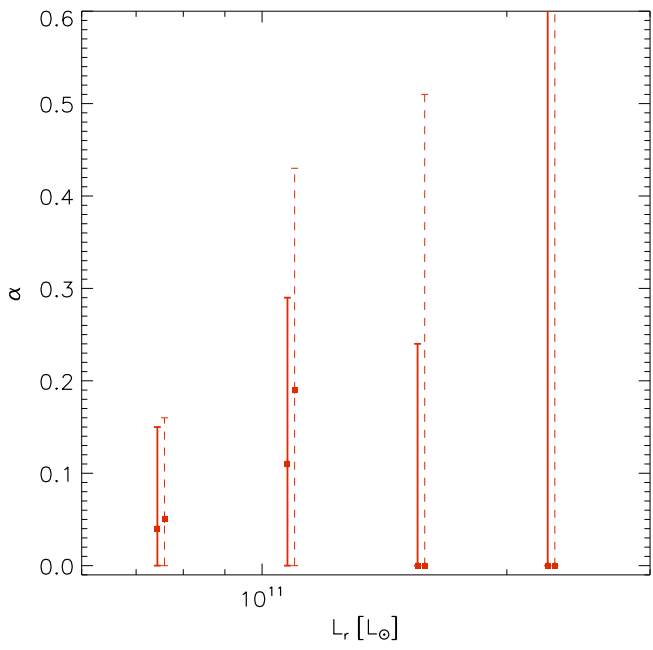}}
  \caption{The best fit satellite fraction for the four highest luminosity bins of early-types. The thick solid (thin dashed) lines indicate the results determined using a truncation radius of 0.2$r_{200}$ (0.4$r_{200}$). Decreasing the truncation radius tightens the constraints on the satellite fraction. }
  \label{plot_cons_alpha}
\end{figure}

The satellite fraction is not well constrained at the high luminosity/stellar mass end. The reason for this is illustrated in Figure \ref{plot_sat_strip}. In Figure \ref{plot_sat_strip}a we show the lensing signal of the L6 luminosity bin, together with the five terms of the halo model, using the standard truncation radius of 0.4$r_{200}$ for the satellite galaxies. The satellite shear signal on scales $<1h^{-1}$Mpc in the halo model is the sum of stripped satellite term and the $\gamma_{t,sat}^{1h}$ term. It is clear that the shape of the combined signal is very similar to the shape of the shear signal coming from the central halo. As a result the error on the satellite fraction is large. The satellite fraction and the halo mass are anti-correlated, as we can see from Figure \ref{plot_chisq}. 
The model either prefers a large mass and small satellite fraction, or a small mass and large satellite fraction. To reduce any bias in the best fit halo mass, we decrease the allowed range for the satellite fractions to a uniform prior between 0\% and 20\% for the highest stellar mass and luminosity bins, as almost all of the galaxies in these bins are expected to be centrals. \\
\indent Recent work by \citet{Limousin09} shows that the half mass radius of a subhalo is a strongly decreasing function of projected cluster-centric distance. Furthermore, the radial distribution of early-type satellites is more peaked around the cluster centre than the radial distribution of late-type satellites \citep[e.g.][]{Ann08}. Hence we expect that the massive elliptical satellite galaxies, which practically always reside close to the centre of a cluster, are stripped of a far larger fraction of their dark matter. \\
\indent To determine whether we can observe a change in the truncation radius of massive early-type satellite galaxies, we make a selection of galaxies that are likely to be satellites and study their shear profile. We consider early-types in the mass range $10^{10.5}<M_{*}<10^{11.75}M_{\odot}$, and divide them in three mass bins; galaxies more massive than $10^{11.75}M_{\odot}$ will almost exclusively be central galaxies and hence not significantly stripped. To determine whether the galaxies are satellites or centrals, we use the SDSS DR7 photometric redshift catalogue $Photoz$, which contains the photometric redshifts of 260 million galaxies, and match them to our source galaxy catalogue. The lenses that have a neighbouring galaxy of the same luminosity or brighter within 750 kpc, and lie within the $1\sigma$ errors of the photometric redshift of the source, are selected for the satellite sample. The galaxies that do not have brighter neighbours within 1 Mpc and within the $1\sigma$ errors of the photometric redshift are selected for the central sample. Note that we do not aim to obtain samples that are complete, but we strive to make a selection that enables us to quantitatively study the differences in the lensing signal. \\
\indent In Figure \ref{plot_envir} we show the stacked shear signal of the galaxies, for the central sample and for the satellite sample, together with their halo model components.
The shear signals of the central sample are indeed described well by an NFW profile. The galaxies preferentially live in isolated environments, and consequently the $\gamma_{t,\mathrm{cent}}^{2h}$ term is overestimated. The halo model fits of the satellite sample are dominated by the satellite terms, as can be observed from the best fit satellite fractions indicated in the plot. The shear signal around 100 kpc is different from the signal of the central sample at the same scale, and is lower than the halo model fit. This suggests that additional stripping of dark matter occurs at small scales. The measurements are too noisy, however, to constrain which fraction of the dark matter haloes is stripped. \\
\indent To illustrate the impact the choice of truncation radius has on the best fit satellite fractions, we also consider stripped satellite profiles with a truncation radius of $0.2r_{200}$. In Figure \ref{plot_sat_strip}b we show the shear signal of the same bin, but with this smaller truncation radius. Note that the halo model parameters are identical in both panels for illustrative purposes, and that the model in the lower panel is not a fit. The shear signal of the satellites at small scales is now clearly different from the central halo term, and the satellite fraction can be better constrained. We have also fit halo models with a truncation radius of 0.2$r_{200}$ to the four most luminous early-type bins. The constraints on the satellite fraction for both models are shown in Figure \ref{plot_cons_alpha}. The satellite fraction is better constrained for the models with a truncation radius of 0.2$r_{200}$. Setting the truncation radius to 0.2$r_{200}$ is a rather arbitrary choice, however, and in future studies it is necessary to include a more realistic prescription for the stripping of the satellites.\\
\indent \citet{Mandelbaum06} study the environmental dependence of the shear profile as a function of luminosity. They distinguish galaxies residing in a high-density environment and in a low-density environment. The brightest galaxies of their low-density sample are almost exclusively centrals, whilst in the high-density sample they are a mixture of centrals and satellites. As the lensing signal is then an average of the shear profiles from satellites and centrals, this may explain why they do not observe a reduction of the signal at small scales. Note that the satellite galaxies we study are more massive, and hence are expected to reside in denser environments where the haloes are stripped of a larger fraction of their dark matter content.
\end{appendix}

\end{document}